\documentclass[aps,pra,twocolumn,showpacs,nobibnotes,nofootinbib]{revtex4-2}

\usepackage[utf8]{inputenc}
\usepackage[T1]{fontenc}
\usepackage{lmodern}
\usepackage{physics}
\usepackage{xcolor}
\usepackage[colorlinks,citecolor=blue,linkcolor=magenta,hypertexnames=false]{hyperref}
\usepackage[nameinlink,capitalize]{cleveref}
\usepackage{graphicx}
\usepackage{epsfig}
\usepackage{color}
\usepackage{bm}
\usepackage{amsmath,amssymb,amsfonts,bbm,physics,mathtools}

\newcommand{\prlrunin}[1]{\textit{#1}---\ignorespaces}

\makeatletter
\let\original@bib@device\bib@device
\renewcommand{\bib@device}[2]{%
  \begingroup
    \let\addcontentsline\@gobblethree
    \original@bib@device{#1}{#2}%
  \endgroup
}
\makeatother

\begin{document}

\title{Wigner-Negative Magnon Steady States from Incoherent Qubit Pumping}
\author{Yan Liu}\email{neonorth@foxmail.com}
\author{Jiahua Li}\email{Contact author: huajia\_li@163.com}
\affiliation{School of Physics, Huazhong University of Science and
Technology, Wuhan 430074, People's Republic of China}

\date{\today}

\begin{abstract}
We show that incoherently pumped qubits can realize a cascaded dissipative mechanism for stabilizing Wigner-negative magnon steady states.
The mechanism combines qubit pumping with dispersive magnon-number selectivity to direct the steady-state population toward selected magnon Fock states.
In the single-qubit case, the single-magnon population can approach unity, accompanied by strong antibunching and pronounced Wigner negativity.
Extending the same principle to multiple qubits yields Wigner-negative steady states dominated by higher magnon Fock components.
We further derive an analytical birth--death model that captures the mechanism and agrees with numerical results.
These results establish incoherent qubit pumping as a controllable dissipative resource for generating nonclassical magnon states in hybrid quantum systems.
\end{abstract}

\maketitle

\prlrunin{Introduction}
Preparing and controlling nonclassical magnon states remains challenging~\cite{LachanceQuirion2017,LachanceQuirion2020,PhysRevLett.130.193603,Kounalakis2022,PhysRevA.108.053715,PhysRevA.108.063703,zhgm-p3ss,Weng2026,Zhang2016CavityMagnomechanics,Li2018MagnonPhotonPhononEntanglement,Li2019SqueezedMagnonsPhonons,Lu2025NonGaussianMagnon} because collective magnon modes in high-quality ferrimagnets are nearly harmonic~\cite{Serga2010,Chumak2015,Yuan2022,ZareRameshti2022,LachanceQuirion2019,TABUCHI2016729,Li2020,Zhang2023}.
Although the magnon Kerr effect has been observed in cavity-magnon systems~\cite{Wang2016Kerr,Wang2018Bistability,Wu2022NonlinearityHeating}, the intrinsic Kerr anharmonicity is weak at the single-quantum level and becomes appreciable mainly for strongly driven, many-magnon states~\cite{Zheng2023NonlinearMagnonics,Zhang2019Kerr,Xiong2022KerrMagnonInterface,Jiang2022KerrNonclassicality}.
Consequently, blockade schemes based solely on the bare magnon Kerr nonlinearity face severe constraints, and recent magnon-blockade studies have instead explored qubit-induced anharmonicity, quantum interference, nonreciprocal dissipation, magnon squeezing, spin-magnon coupling, and enhanced or effective Kerr-type nonlinearities~\cite{PhysRevA.101.042331,Xu2021,Wang2022,PhysRevA.108.053702,Hou2024KerrBlockade,Wang2020MagnonBlockadePT,Wu2021MultimagnonBlockade,Wang2022HybridMagnonAtomBlockade,Huang:24,Yan2024RemoteMagnonBlockade,Wang2024GiantSpinBlockade,Amazioug2024SqueezingBlockade,Gupta2024InterferenceBlockade,Ding2024NonlinearCavityMagnonBlockade,Xie2025AnisotropicMagnonBlockade,Zhao2025SpinMagnonDetuning,Falch2025AntiferromagnetBlockade,Zheng2026HybridMagnonSuperconductingQubit}.
In particular, recent optimized magnon-blockade theory has shown that extremely strong antibunching is possible, although the corresponding steady-state single-magnon population remains limited by the vacuum component~\cite{PhysRevA.110.012459}.
Achieving a steady state that combines strong antibunching with a high single-magnon population approaching unity therefore remains an important challenge for quantum magnonics~\cite{Yuan2022,ZareRameshti2022,LachanceQuirion2019,TABUCHI2016729,LachanceQuirion2017,LachanceQuirion2020,PhysRevLett.130.193603,Li2022,Kounalakis2022,PhysRevA.108.053715,PhysRevA.108.063703,Weng2026,Lu2025NonGaussianMagnon}.

This motivates coupling magnons to superconducting quantum circuits, where the Josephson nonlinearity of superconducting qubits provides a much stronger artificial anharmonicity than the intrinsic Kerr nonlinearity of a bare magnon mode~\cite{Blais2004CircuitQED,Wallraff2004,Clarke2008,RevModPhys.93.025005,Schuster2007PhotonNumber,Hofheinz2008,Hofheinz2009,Vlastakis2013,Ofek2016Nature,GrimmNature2020}. Strong magnon-photon coupling has been demonstrated in cavity-magnon systems~\cite{Huebl2013,Zhang2014,Goryachev2014,PhysRevLett.113.083603,Zhang2015,Morris2017,Baity2021,PhysRevApplied.20.024039}, establishing hybrid microwave platforms for coherently interfacing collective spin excitations with superconducting circuits~\cite{LachanceQuirion2019,ZareRameshti2022,Yuan2022,TABUCHI2016729,Li2020,Zhang2023}. Building on this platform, coherent coupling between a ferromagnetic magnon and a superconducting qubit has been achieved in the single-magnon regime~\cite{Tabuchi2015,TABUCHI2016729}. In the strong-dispersive regime, the same hybrid architecture has resolved magnon-number states through qubit spectroscopy~\cite{LachanceQuirion2017} and enabled single-shot detection of a single magnon~\cite{LachanceQuirion2020}.
More recently, superconducting-qubit–magnon systems have enabled, or been proposed for, quantum control and nonclassical-state engineering of magnons~\cite{PhysRevLett.130.193603,Kounalakis2022,PhysRevA.108.063703,zhgm-p3ss,Weng2026}.
These advances establish superconducting qubits as both nonlinear elements and quantum probes for magnon systems.
An autonomous, number-selective route to high-population Wigner-negative magnon steady states dominated by selected Fock components, however, remains to be established.
Unlike coherent pulsed protocols that prepare bosonic Fock states at selected times~\cite{PhysRevLett.130.193603,Hofheinz2008}, the present mechanism autonomously stabilizes a Wigner-negative magnon steady state under continuous incoherent qubit pumping.

In this Letter, we propose a cascaded dissipative scheme that uses incoherently pumped qubits to stabilize Wigner-negative magnon steady states with populations concentrated near selected Fock states, inspired by reservoir engineering and autonomous dissipative state stabilization in quantum optical and superconducting-circuit systems~\cite{Poyatos1996,Verstraete2009_DissipativeQE,PhysRevLett.109.183602,Shankar2013,LeghtasScience2015,Harrington2022,PhysRevLett.132.203602,Li2024}.
Instead of relying on the weak intrinsic Kerr nonlinearity of the magnon mode, we use dispersive magnon-qubit coupling to resolve adjacent magnon-number transitions and use incoherent qubit pumping to generate directed steady-state injection, building on the number-resolved and qubit-assisted control capabilities developed in circuit quantum electrodynamics (QED) and quantum magnonics~\cite{Blais2004CircuitQED,Schuster2007PhotonNumber,Hofheinz2008,Hofheinz2009,Tabuchi2015,LachanceQuirion2017,LachanceQuirion2020,PhysRevLett.130.193603}.
This combination stabilizes a high single-magnon population approaching unity, together with strong antibunching and pronounced Wigner negativity, and it can be extended to higher Fock states by assigning different qubits to successive transitions in the ladder.
We further derive an analytical birth--death model that explains the operating window and agrees with numerical results, thereby identifying incoherent qubit pumping as a controllable dissipative resource for nonclassical magnon-state preparation.

\prlrunin{Model and mechanism}
We consider a single magnon mode coupled to $N_Q$ independently and incoherently pumped qubits, as motivated by hybrid quantum magnonic platforms based on superconducting circuits~\cite{LachanceQuirion2019,ZareRameshti2022,Yuan2022,TABUCHI2016729,Tabuchi2015,LachanceQuirion2017,LachanceQuirion2020,PhysRevLett.130.193603}.
The magnon mode has effective resonance frequency $\omega_m$ and annihilation (creation) operator $\hat a$ ($\hat a^\dagger$); we denote its number operator by $\hat n=\hat a^\dagger\hat a$.
The qubits are labeled by $j=0,1,\ldots,N_{Q}-1$.
For qubit $j$, with effective transition frequency $\omega_j$, the ground and excited states are denoted by $\ket{g_j}$ and $\ket{e_j}$, respectively.
For this qubit, we define the raising and lowering operators as $\hat\sigma_j^+=\ket{e_j}\bra{g_j}$ and $\hat\sigma_j^-=\ket{g_j}\bra{e_j}$.
The corresponding population-inversion, or Pauli-$z$, operator is $\hat\sigma_{z,j}=\ket{e_j}\bra{e_j}-\ket{g_j}\bra{g_j}$, with eigenvalues $+1$ and $-1$ for $\ket{e_j}$ and $\ket{g_j}$, respectively.
In a frame rotating at the effective magnon frequency $\omega_m$ with respect to the total excitation number, and setting $\hbar=1$, the effective exchange--dispersive Hamiltonian takes a Jaynes--Cummings form supplemented by magnon-number-dependent dispersive shifts~\cite{Jaynes1963,Blais2004CircuitQED,Wallraff2004,Schuster2007PhotonNumber,Hofheinz2008,Hofheinz2009,LachanceQuirion2017}
\begin{equation}
\label{eq:model_H}
\hat H
=
\sum_{j=0}^{N_Q-1}
\Delta_j
\hat\sigma_j^+\hat\sigma_j^-
+
\sum_{j=0}^{N_Q-1}
\chi_j
\hat n\hat\sigma_{z,j}
+
\sum_{j=0}^{N_Q-1}
g_j(
\hat a^\dagger\hat\sigma_j^-
+
\hat a\hat\sigma_j^+).
\end{equation}
Here $\Delta_j=\omega_j-\omega_m$ is the effective detuning of qubit $j$ from the magnon mode.
The parameter $g_j$ denotes the effective Jaynes--Cummings exchange coupling between the magnon mode and qubit $j$, while $\chi_j$ is the magnon-number-dependent dispersive shift associated with qubit $j$~\cite{Jaynes1963,Blais2004CircuitQED,Wallraff2004,Schuster2007PhotonNumber,Tabuchi2015,LachanceQuirion2017,LachanceQuirion2020,PhysRevLett.130.193603}.
The density operator $\hat\rho$ obeys the Markovian master equation~\cite{Gorini1976,Lindblad1976,Carmichael1999,GardinerZoller2004,Breuer2007}
\begin{equation}
\label{eq:model_master}
\frac{d\hat\rho}{dt}
=
-i[\hat H,\hat\rho]
+
\kappa\mathcal D[\hat a]\hat\rho
+
\sum_{j=0}^{N_Q-1}
\gamma_j
\mathcal D[\hat\sigma_j^-]\hat\rho
+
\sum_{j=0}^{N_Q-1}
\wp_j
\mathcal D[\hat\sigma_j^+]\hat\rho ,
\end{equation}
with
$\mathcal D[\hat o]\hat\rho=\hat o\hat\rho\hat o^\dagger-\frac{1}{2}\hat o^\dagger\hat o\hat\rho-\frac{1}{2}\hat\rho\hat o^\dagger\hat o$, as in the standard Lindblad form of Markovian open-system dynamics~\cite{Gorini1976,Lindblad1976,Carmichael1999,GardinerZoller2004,Breuer2007}.
Here $\kappa$ is the magnon loss rate, while $\gamma_j$ and $\wp_j$ are the relaxation and incoherent pump rates of qubit $j$, respectively~\cite{Poyatos1996,PhysRevLett.109.183602,Shankar2013,LeghtasScience2015,PhysRevA.84.043816}.
Equation~\eqref{eq:model_H} is an effective exchange--dispersive Hamiltonian, with $g_j$ and $\chi_j$ treated as independently tunable effective parameters, as can in principle be engineered in driven or parametrically controlled circuit-QED-type hybrid systems~\cite{Blais2004CircuitQED,Wallraff2004,Schuster2007PhotonNumber,PhysRevLett.104.177004,PhysRevLett.113.193601,Tabuchi2015,LachanceQuirion2017,LachanceQuirion2020,PhysRevLett.130.193603}; see Sec.\,S9 of the Supplemental Material (SM) for details~\cite{supplementary2025}.
Unless otherwise stated, we take $g_j=g$, $\chi_j=\chi$, $\gamma_j=\gamma$, and $\wp_j=\wp$.
The robustness of the mechanism against parameter inhomogeneity and additional pure-dephasing channels is analyzed in Secs.\,S6 and S7 of the SM~\cite{supplementary2025}.

The full product basis is written as $\ket{n;s_0,\ldots,s_{N_Q-1}}\equiv\ket{n}
\bigotimes_{j=0}^{N_Q-1}\ket{s_j}$, with $s_j\in\{g_j,e_j\}$.
Here $n$ labels the magnon Fock state, and $s_j$ specifies the state of qubit $j$.
When no qubit labels are displayed, $\ket{n}$ denotes the magnon Fock state alone.
To identify the number-selective resonant exchange processes associated with the Jaynes--Cummings interaction and dispersive state-dependent frequency shifts, we compare the energies of the two product states connected by $\hat a^\dagger\hat\sigma_j^-+\hat a\hat\sigma_j^+$.
During such a transition, only qubit $j$ changes its state, whereas the other qubits remain as spectators.
We therefore write $\ket{n,s_j;v_j}$, where $v_j$ denotes the configuration of all spectator qubits $k\neq j$.

The role of the incoherent pump is crucial.
The Jaynes--Cummings term alone does not provide directionality: on resonance it coherently hybridizes $\ket{n,e_j;v_j}$ and $\ket{n+1,g_j;v_j}$, leading to reversible exchange
rather than irreversible population transfer.
Directionality is introduced by the pump, which pumps $\ket{n+1,g_j;v_j}$ to $\ket{n+1,e_j;v_j}$.
If the subsequent transition $\ket{n+1,e_j;v_j}\leftrightarrow\ket{n+2,g_j;v_j}$ is detuned by the number-dependent dispersive shift, the same qubit cannot efficiently transfer population from the magnon level $\ket{n+1}$ to $\ket{n+2}$.
The resonant exchange followed by incoherent pumping therefore acts as an effective one-step injection process, $\ket{n}\rightarrow\ket{n+1}$, rather than as an unrestricted pump to higher magnon numbers.
In this way, a pumped qubit acts as a number-selective single-magnon injector.

This mechanism motivates a cascaded injection scheme in which the $j$th qubit is assigned to
the $j$th injection step, $\ket{j,e_j;v_j}\leftrightarrow\ket{j+1,g_j;v_j}$, with $j=0,1,\ldots,N_Q-1$.
If all these $N_Q$ number-selective injection steps are simultaneously
realized, the magnon mode is driven along the sequence
\begin{equation}
\ket{0}\rightarrow\ket{1}\rightarrow\cdots\rightarrow\ket{N_Q}.
\label{eq4}
\end{equation}
Since no qubit is assigned to the next step $\ket{N_Q}\rightarrow\ket{N_Q+1}$, the steady-state magnon population is dynamically concentrated near the selected Fock state $\ket{N_Q}$.

The above scheme requires the $j$th injection step to be resonant.
Thus, for a given spectator configuration $v_j$, one has to compare the energies of the two states connected by the exchange term, $\ket{n,e_j;v_j}\leftrightarrow\ket{n+1,g_j;v_j}$.
In a multi-qubit system this resonance condition is shifted by the spectator qubits: although qubits $k\neq j$ do not change their states during the exchange, the dispersive terms $\chi_k\hat n\hat\sigma_{z,k}$ shift the magnon transition frequency.
For a given $v_j$, we denote by $z_k(v_j)=+1$ $(-1)$ the eigenvalue of $\hat\sigma_{z,k}$ when spectator qubit $k\neq j$ is in $\ket{e_k}$ $(\ket{g_k})$.
The corresponding conditional detuning is then $\delta_j(n;v_j)
=-\Delta_j-(2n+1)\chi_j+\sum_{\substack{k=0\\k\neq j}}^{N_Q-1}z_k(v_j)\chi_k$.
To make qubit $j$ resonant with its assigned step $n=j$, we impose
$\delta_j(j;\bar v_j)=0$, where $\bar v_j$ is the spectator configuration
used for compensation. This gives $\Delta_j
=-(2j+1)\chi_j+\sum_{\substack{k=0\\k\neq j}}^{N_Q-1}\bar z_k\chi_k$,
with $\bar z_k=z_k(\bar v_j)$. In the strongly pumped regime
$\wp_k\gg\gamma_k$, the spectator qubits are predominantly excited, so that
$\bar z_k=+1$. For identical dispersive shifts, $\chi_j=\chi$, this reduces
to $\Delta_j=(N_Q-2j-2)\chi$.
Details of the energy-mismatch calculation, including the dependence of the
conditional detuning on the spectator eigenvalues $z_k(v_j)$, are given in Sec.\,S1\,B of the SM~\cite{supplementary2025}.

\begin{figure}[htb]
\centerline{\includegraphics[width=1.0\linewidth]{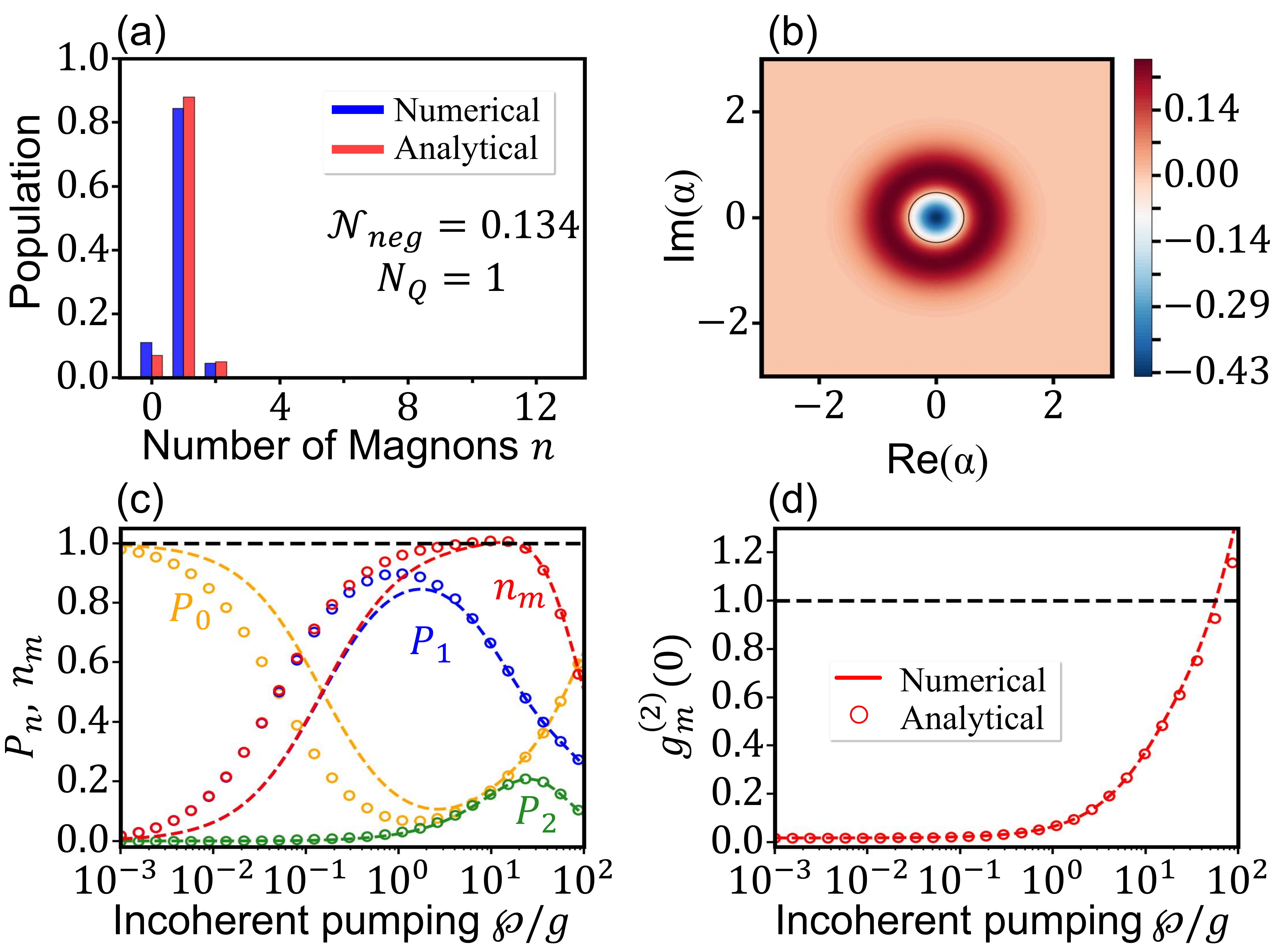}}
\caption{Single-magnon steady state for one incoherently pumped qubit ($N_Q=1$).
Common parameters are $\chi/g=10$, $\kappa/g=0.1$, $\gamma/g=0.05$, and $\Delta_0/g=-10$, such that $|0,e\rangle\leftrightarrow |1,g\rangle$ is resonant.
(a, b) Results at $\wp/g=2$: (a) magnon populations $P_n$, with blue and red bars denoting numerical and analytical results, respectively; (b) numerical Wigner function $W(\alpha)$, with $\mathcal{N}_{\rm neg}=0.134$.
(c) $P_0$, $P_1$, $P_2$, and $n_m$ versus $\wp/g$.
(d) $g_m^{(2)}(0)$ versus $\wp/g$.
In (c, d), dashed curves and open circles denote numerical and analytical results, respectively.
The black dashed lines in (c) and (d) mark $n_m=1$ and $g_m^{(2)}(0)=1$, respectively.
}
\label{fig1}
\end{figure}

To characterize the nonclassicality of the magnon state in phase space, we first trace out the qubits and define the reduced density matrix of the magnon mode $\hat\rho_m={\rm Tr}_{q}[\hat\rho]$.
The Wigner function of the magnon mode is then written in the displaced-parity form~\cite{Wigner1932,Cahill1969,PhysRevA.15.449,Banaszek1996,Lutterbach1997,PhysRevA.60.674,Schleich2001,lvovsky2009} as $W(\alpha)=\frac{2}{\pi}{\rm Tr}_m[\hat\rho_m\hat D(\alpha)\hat\Pi\hat D^\dagger(\alpha)]$, where $\alpha$ is the phase-space coordinate, $\hat D(\alpha)=\exp(\alpha\hat a^\dagger-\alpha^*\hat a)$ is the magnon displacement operator and $\hat\Pi=(-1)^{\hat a^\dagger\hat a}$ is the magnon-number parity operator.
Negative values of $W(\alpha)$ constitute a direct phase-space witness of nonclassicality~\cite{Hudson1974,Kenfack2004,PhysRevLett.89.200402,Deleglise2008,PhysRevLett.105.253603,PhysRevA.87.062104,PhysRevA.98.052350,Veitch_2012,PhysRevLett.109.230503,Chabaud2021WitnessingWN}.
We quantify the total negative volume by the integrated Wigner negativity~\cite{Kenfack2004,Arkhipov2018}
\begin{equation}
\mathcal{N}_{\rm neg}
=
\frac{1}{2}\int
\left[
|W(\alpha)|-W(\alpha)
\right]d^2\alpha ,
\end{equation}
which is positive if and only if $W(\alpha)$ has negative regions.

\prlrunin{Single-magnon steady state}
The steady magnon state is characterized by the Fock-state populations $P_n=\langle n|\hat\rho_m|n\rangle$, with mean magnon number $n_m=\langle\hat a^\dagger\hat a\rangle$.
To benchmark the single-magnon character of the steady state, we evaluate the equal-time second-order magnon correlation function $g_m^{(2)}(0)=
{\langle \hat a^{\dagger2} \hat a^2 \rangle}/{\langle \hat a^\dagger \hat a\rangle^2}$~\cite{Glauber1963,Kimble1977,Scully1997,GardinerZoller2004,Carmichael1999}.
Here $g_m^{(2)}(0)<1$ signals magnon antibunching~\cite{PhysRevA.101.042331,Xu2021,Wang2022,PhysRevA.108.053702,PhysRevA.110.012459,PhysRevA.110.023725,Gupta2024InterferenceBlockade}, with the ideal single-magnon Fock state satisfying $P_1=1$, $n_m=1$, and $g_m^{(2)}(0)=0$.

For the single-qubit configuration, we tune the qubit resonantly to the number-selective transition $\ket{0,e}\leftrightarrow\ket{1,g}$, while the dispersive shift detunes the next transition $\ket{1,e}\leftrightarrow\ket{2,g}$.
As a result, the incoherently pumped qubit injects one magnon efficiently but strongly suppresses further excitation.
This behavior is shown in Fig.\,\ref{fig1}(a): at $\wp/g=2$, the steady state is dominated by the single-magnon population, whereas the vacuum and two-magnon components remain small.
\begin{figure}[htb]
\centerline{\includegraphics[width=9cm]{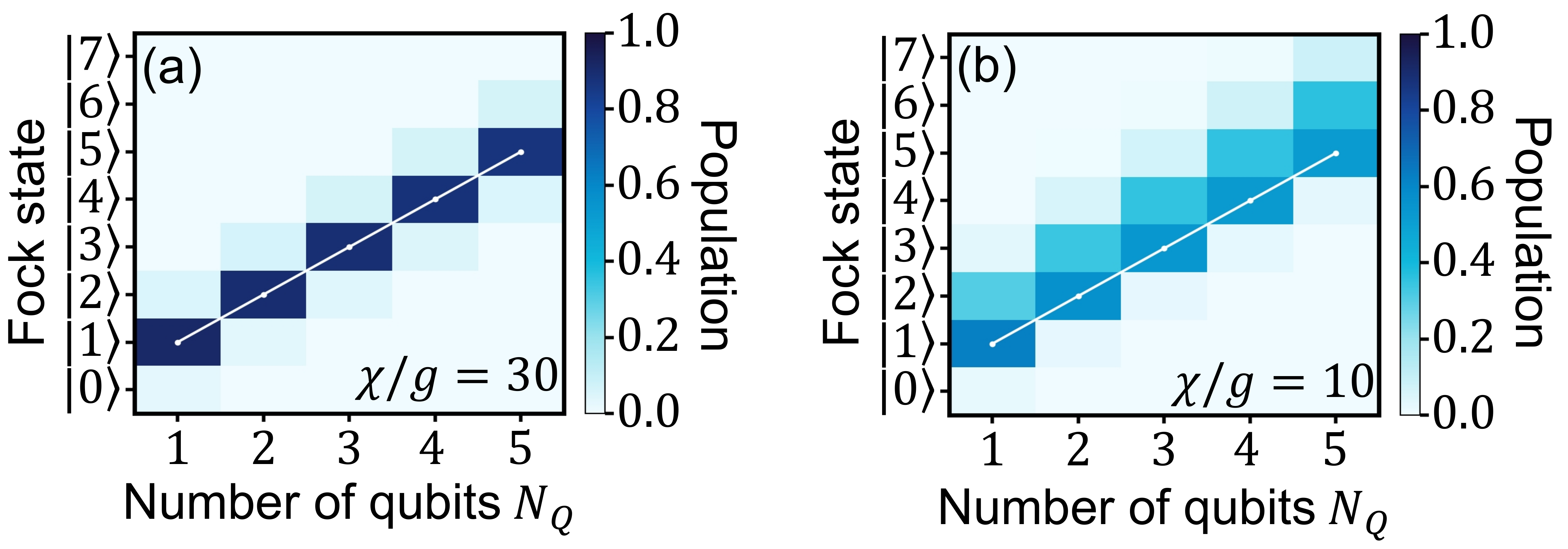}}
\caption{Cascaded stabilization near selected magnon Fock states with multiple incoherently pumped qubits.
The color maps show the numerical steady-state populations $P_n$ versus the number of qubits $N_Q$ and the magnon number $n$.
For each $N_Q$, the qubits are tuned by $\Delta_j=(N_Q-2j-2)\chi$ with $j=0,\ldots,N_Q-1$, corresponding to $\bar z_k=+1$.
The white line marks the target relation $n=N_Q$.
Common parameters are $\wp/g=2$, $\kappa/g=0.01$, and $\gamma/g=0.05$.
(a) $\chi/g=30$.
(b) $\chi/g=10$.
}
\label{fig2}
\end{figure}
The corresponding Wigner function, shown in Fig.\,\ref{fig1}(b), exhibits a negative region around the phase-space origin, demonstrating that the stabilized state is not merely antibunched but genuinely nonclassical in phase space with Wigner negativity $\mathcal{N}_{\rm neg}=0.134$.
The pump dependence in Fig.\,\ref{fig1}(c) further shows an intermediate regime where population is transferred from the vacuum to the single-magnon state, with $n_m$ approaching unity while $P_2$ remains suppressed; the decrease at stronger pumping reflects pump-induced broadening and self-quenching, a behavior analogous to that of strongly coupled light--matter systems under incoherent excitation~\cite{PhysRevLett.105.233601,PhysRevA.84.043816,Mu1992,Rice1994,McKeever2003,Astafiev2007,PhysRevA.82.053802,PhysRevA.103.013718}.
Consistently, Fig.\,\ref{fig1}(d) shows pronounced antibunching, $g_m^{(2)}(0)<1$, in the same intermediate pump regime where $P_1$ is maximized.
At stronger pumping, the finite dispersive selectivity permits residual off-resonant excitation of the $|1,e\rangle\leftrightarrow |2,g\rangle$ transition, increasing the two-magnon population and hence $g_m^{(2)}(0)$.
The value $P_1\simeq0.8$ in Fig.\,\ref{fig1} is not a fundamental limit, but reflects the moderate dispersive shift used there; larger $\chi$ suppresses the residual off-resonant leakage and drives $P_1$ closer to unity.
The agreement between the numerical steady state obtained with QuTiP~\cite{JOHANSSON20121760,JOHANSSON20131234} and the analytical steady-state populations calculated from Eq.\,(S59) in Sec.\,S1\,F of the SM~\cite{supplementary2025} validates the effective birth--death rate-equation picture.
The single-magnon steady state can be further optimized with an unpumped auxiliary qubit that selectively suppresses residual two-magnon leakage (see End Matter).

\prlrunin{Multimagnon steady states}
We next examine the multiqubit cascade introduced in Eq.\,\eqref{eq4}.
Figure~\ref{fig2} shows that the population maximum follows the target relation $n=N_Q$, confirming number-selective ladder stabilization analogous to engineered oscillator Fock-state trapping~\cite{Poyatos1996,Weidinger1999,Brattke2001,Sayrin2011,Kienzler2015}, rather than broad thermal-like excitation~\cite{Scully1997,Carmichael1999,GardinerZoller2004,Breuer2007,PhysRevA.84.043816}.

\begin{figure}[htb]
\centerline{\includegraphics[width=1.0\linewidth]{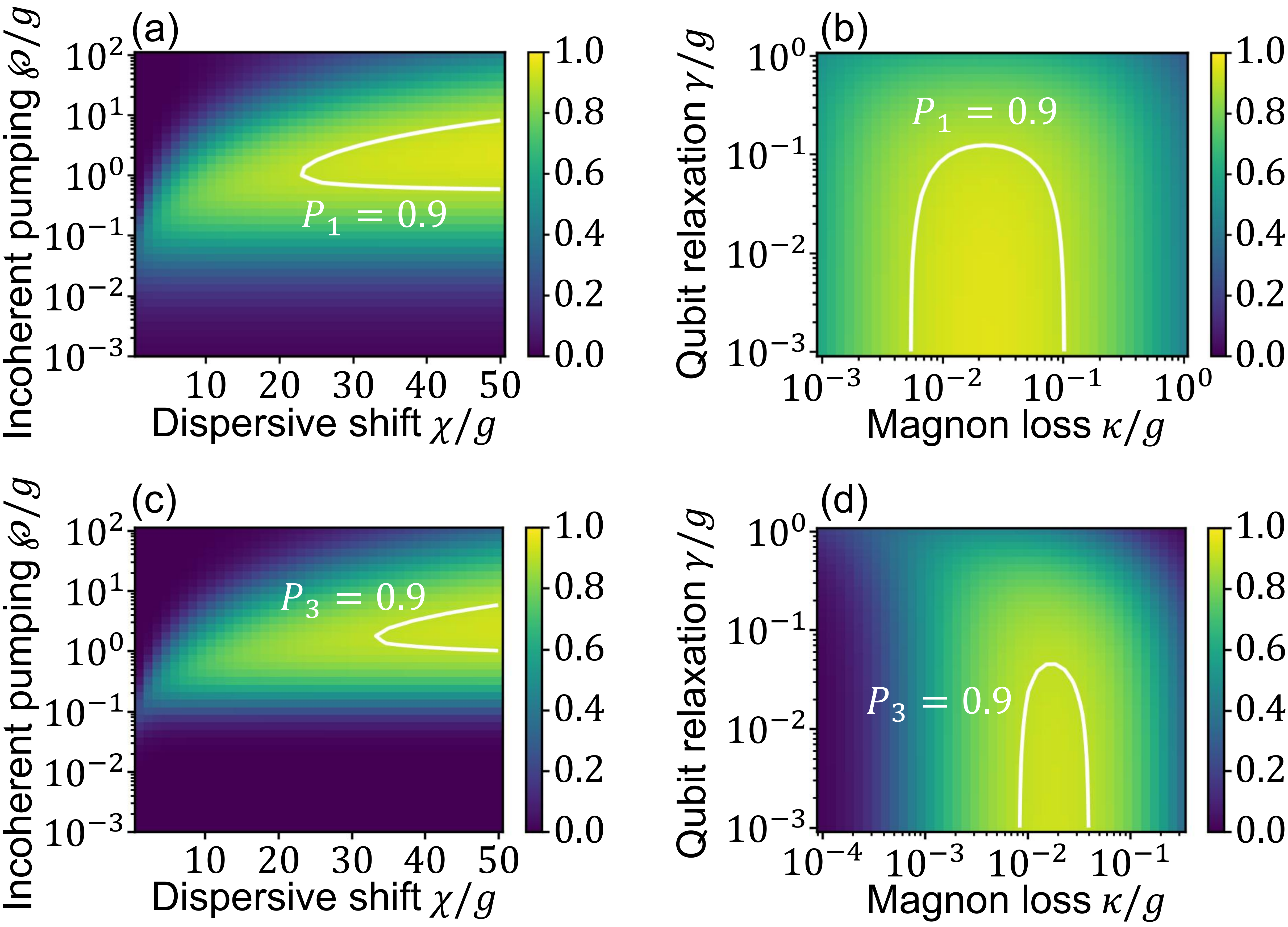}}
\caption{
Parameter dependence of the target Fock-state population.
The color maps show numerical steady-state populations of the selected magnon Fock state.
(a, b) Single-qubit case with target state $|1\rangle$, showing $P_1$.
(c, d) Three-qubit case with target state $|3\rangle$, showing $P_3$.
White contours mark $P_1=0.9$ in (a, b) and $P_3=0.9$ in (c, d).
In the multiqubit case, the qubits are tuned by $\Delta_j=(N_Q-2j-2)\chi$ for $j=0,\ldots,N_Q-1$.
(a, c) Dependence on $\chi/g$ and $\wp/g$, with $\kappa/g=0.01$ and $\gamma/g=0.05$.
(b, d) Dependence on $\kappa/g$ and $\gamma/g$, with $\chi/g=30$ and $\wp/g=2$.
}
\label{fig3}
\end{figure}

The comparison between Figs.\,\ref{fig2}(a) and \ref{fig2}(b) highlights the role of the dispersive shift $\chi$.
For larger $\chi$, the populations are sharply concentrated around the target states, reflecting better-resolved magnon-number transitions and strongly suppressed off-resonant leakage.
For smaller $\chi$, the target-state populations are reduced, with increased weight in neighboring Fock states because adjacent number-selective transitions are less spectrally resolved.
This comparison demonstrates that dispersive number selectivity controls the confinement of the cascaded steady state around the target Fock state.

Figure~\ref{fig3} maps $P_1$ and $P_3$ versus the pump, dispersive shift, magnon loss, and qubit relaxation, revealing broad high-population regions and an intermediate-pump optimum.
This nonmonotonic pump dependence can be understood from the effective birth--death model derived in Sec.\,S1 of the SM~\cite{supplementary2025}.
For the qubit-assisted transition $|n,e_j;v_j\rangle\leftrightarrow |n+1,g_j;v_j\rangle$, eliminating the transition coherence gives the exchange coefficient
\begin{equation}
R_{j,n}(v_j)
=
\frac{2g_j^2(n+1)\Gamma_{j,n}}
{\Gamma_{j,n}^2+\delta_j^2(n;v_j)} ,
\end{equation}
where $\Gamma_{j,n}$ is the total decay rate of the transition coherence.
In the reduced population equation, the exchange coefficient evaluated at the compensated spectator configuration, $R_{j,n}\equiv R_{j,n}(\bar v_j)$, enters the birth and death rates as
\begin{equation}
\lambda_n=\sum_{j=0}^{N_Q-1}q_{e,j}R_{j,n},\qquad
\mu_{n+1}=(n+1)\kappa+\sum_{j=0}^{N_Q-1}q_{g,j}R_{j,n},
\end{equation}
\begin{figure}[htb]
\centerline{\includegraphics[width=1.0\linewidth]{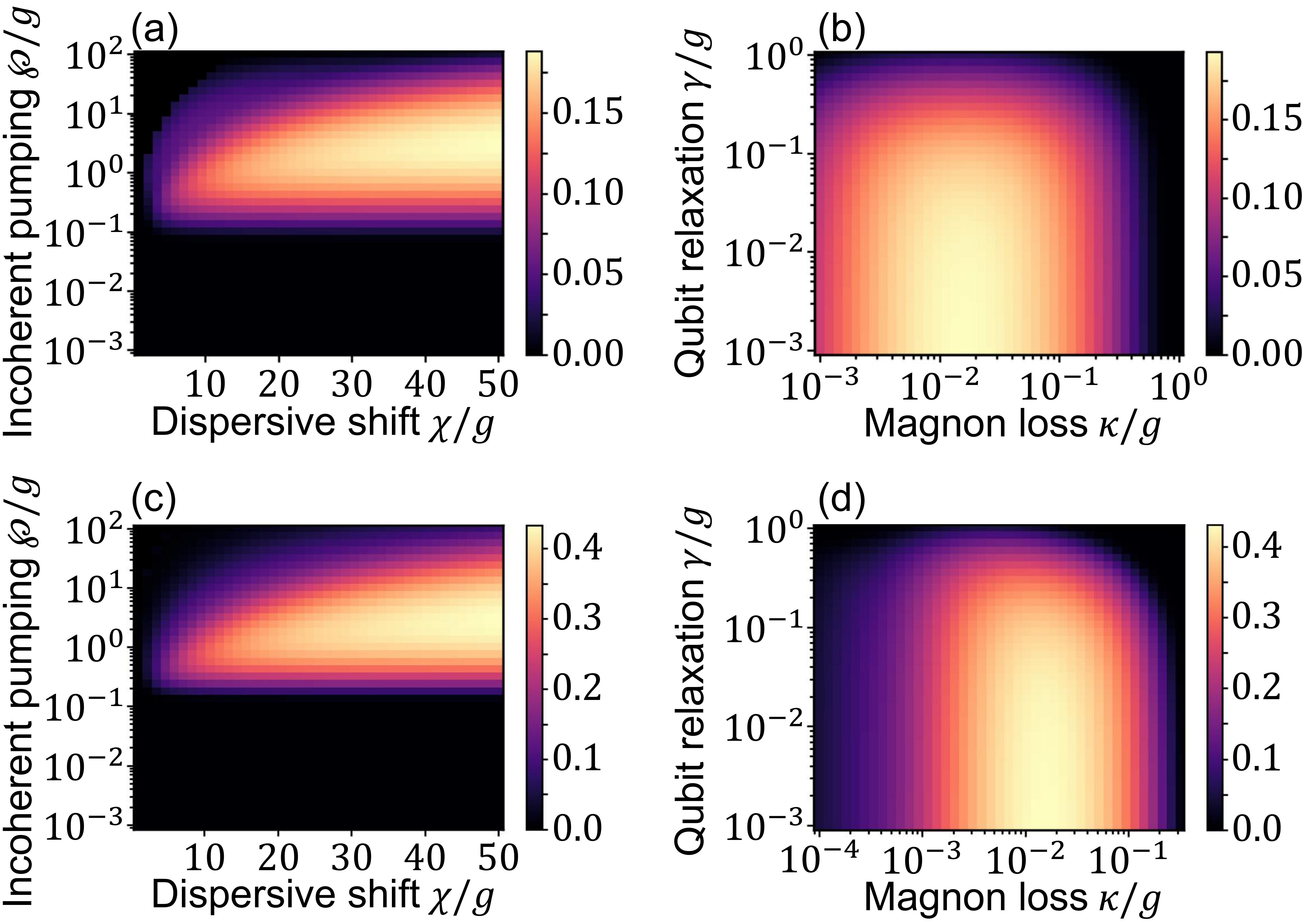}}
\caption{
Wigner negativity of the reduced magnon steady state.
Color maps show the numerical integrated negativity $\mathcal{N}_{\rm neg}$ for (a, b) $N_Q=1$, targeting $|1\rangle$, and (c, d) $N_Q=3$, targeting $|3\rangle$.
Qubit detunings are set by $\Delta_j=(N_Q-2j-2)\chi$ for $j=0,\ldots,N_Q-1$.
(a, c) Dependence on $\chi/g$ and $\wp/g$, with $\kappa/g=0.01$ and $\gamma/g=0.05$.
(b, d) Dependence on $\kappa/g$ and $\gamma/g$, with $\chi/g=30$ and $\wp/g=2$.
}
\label{fig4}
\end{figure}
where $q_{e,j}=\wp_j/(\wp_j+\gamma_j)$ and $q_{g,j}=\gamma_j/(\wp_j+\gamma_j)$ are the excited- and ground-state populations of an isolated qubit in steady state, as used in the reduced population closure.
Within this reduced description, the steady-state balance condition $\lambda_nP_n=\mu_{n+1}P_{n+1}$ gives $P_{n+1}/P_n=\lambda_n/\mu_{n+1}$.
Thus, at low to moderate pump rates, increasing $\wp_j$ increases $q_{e,j}/q_{g,j}$ and therefore increases the addressed birth rates $\lambda_n$ relative to the corresponding death rates, transferring population upward along the ladder toward the target state.

Increasing $\wp_j$ also increases $\Gamma_{j,n}$ through its contribution to the transition-coherence decay rate.
On the addressed transition $n=j$, where $\delta_j(j;\bar v_j)=0$, the exchange coefficient reduces to $R_{j,j}^{\rm res}={2g_j^2(j+1)}/{\Gamma_{j,j}}$.
Excessive pumping therefore reduces the resonant exchange coefficient and weakens the intended injection steps.
At the same time, for detuned transitions in the resolved regime, the off-resonant exchange coefficient scales as $R_{j,n}^{\rm off}\simeq{2g_j^2(n+1)\Gamma_{j,n}}/[{\delta_j^2(n;\bar v_j)}]$,
so increasing $\Gamma_{j,n}$ increases the residual birth rates feeding Fock states above the target.
The target population is therefore maximized at an intermediate pump strength, where the addressed birth rates are large enough to drive the cascade upward, while resonant self-quenching and upper-tail birth rates remain sufficiently small, as shown in Figs.\,\ref{fig3}(a) and \ref{fig3}(c).

Figures~\ref{fig3}(a) and \ref{fig3}(c) further show that increasing $\chi$ expands the high-target-population region.
This behavior is consistent with the analytical leakage estimates in Sec.\,S1\,F of the SM~\cite{supplementary2025}, where the first and second upper-tail populations scale as $\chi^{-2}$ and $\chi^{-4}$, respectively, in the resolved-transition regime.
These scalings identify the suppression of upper-tail leakage as the origin of the improved target-state confinement at larger $\chi$.

The dependence on magnon loss in Figs.\,\ref{fig3}(b) and \ref{fig3}(d) is also nonmonotonic. A large $\kappa$ depletes the target state faster than the cascade can replenish it, thereby reducing $P_{N_Q}$.
However, an overly small $\kappa$ is not optimal either: residual off-resonant excitation can populate states above the target, and the analytical rate solution gives upper-tail ratios such as $P_{N_Q+1}/P_{N_Q}\simeq\lambda_{N_Q}/\mu_{N_Q+1}$, which scale approximately as $1/\kappa$ when the return rate is dominated by magnon loss.
A finite magnon loss rate therefore plays a constructive role by removing population leaked into the upper tail, while the number-selective cascade replenishes the target state after loss events.
By contrast, qubit relaxation does not provide such a useful return channel: increasing $\gamma$ reduces the excited-to-ground population ratio $q_{e,j}/q_{g,j}$ and enhances the qubit-induced downward rates, so smaller $\gamma$ consistently improves the target-state population.
Figure~\ref{fig3} thus identifies the operating window for high-population magnon Fock-state stabilization: strong dispersive selectivity, intermediate incoherent qubit pumping, weak qubit relaxation, and a finite but not excessive magnon loss rate.

Figure~\ref{fig4} provides a complementary phase-space benchmark by showing the integrated Wigner negativity $\mathcal{N}_{\rm neg}$ over the same parameter space as in Fig.\,\ref{fig3}.
The close correspondence between the negativity and target-population maps shows that the same parameter region supports both high target-state populations and sizable Wigner negativity, confirming that the cascaded incoherent-pumping mechanism governs both target-state stabilization and phase-space nonclassicality.

The SM~\cite{supplementary2025} provides the full rate-equation derivation and pump-directionality analysis in Secs.\,S1 and S2, spectator-compensation tests in Sec.\,S3, additional single- and two-magnon analyses in Secs.\,S4 and S5, robustness and scaling tests in Secs.\,S6--S8, and experimental-feasibility discussion in Sec.\,S9.

\prlrunin{Conclusion}
Incoherently pumped qubits stabilize Wigner-negative magnon steady states through dispersive number selectivity.
A single pumped qubit stabilizes an antibunched steady state dominated by the single-magnon component, while a multiqubit cascade stabilizes steady states dominated by higher magnon Fock components by sequentially addressing the transitions $\ket{j}\rightarrow\ket{j+1}$. Together, the numerical and analytical results establish this mechanism as an autonomous
dissipative route to number-selective nonclassical magnon-state stabilization.

\prlrunin{Acknowledgments}
We acknowledge Ying Wu and Xin-You
L\"{u} for useful discussions. The present research is supported in
part by the National Natural Science Foundation of China (NSFC)
through Grant No.\,12275092 and by the National Key Research and
Development Program of China under Contract No.\,2021YFA1400700.

\prlrunin{Data availability}
The data supporting this study are available from the corresponding author upon reasonable request.

\bibliographystyle{apsrev4-2-1}
\bibliography{WF-sub-arXiv}

	\onecolumngrid
	\vspace{0.8cm}
	\begin{center}
		\rule{0.5\textwidth}{0.4pt}\\[6pt]
		{\large\textbf{End Matter}}\\[6pt]
		\rule{0.5\textwidth}{0.4pt}
	\end{center}
	\twocolumngrid

\prlrunin{Relation to pulsed state preparation}
The present scheme differs from coherent pulsed protocols for bosonic Fock-state preparation.
In circuit QED, photon Fock states were synthesized by repeatedly
exciting a qubit and resonantly transferring each excitation to a
microwave resonator~\cite{Hofheinz2008}.
More recently, a single-magnon Fock state and a vacuum--single-magnon superposition
were prepared using qubit rotations, Autler--Townes-assisted resonance tuning, and coherent qubit--magnon exchange, followed by Wigner tomography~\cite{PhysRevLett.130.193603}.
These protocols generate the target state at a prescribed time through calibrated coherent control, after which the prepared state undergoes dissipative decay without continuous stabilization.

The present work instead addresses autonomous steady-state stabilization.
Incoherent qubit pumping and dispersive magnon-number selectivity continuously replenish the target-state population depleted by magnon loss and establish a Wigner-negative steady state without timed qubit rotations or resonant swap pulses.
The same principle extends to higher target Fock states by assigning the successive transitions
$\ket{j}\rightarrow\ket{j+1}$ to different pumped qubits.

The two approaches are complementary: coherent pulsed control enables the preparation of Fock states and coherent superpositions at prescribed times, whereas the present scheme continuously maintains the target-state population and Wigner negativity.
Related autonomous, reservoir-engineered, and feedback-based approaches to oscillator-state stabilization have been explored in other bosonic platforms~\cite{Harrington2022,PhysRevLett.132.203602, Li2024,Sayrin2011,Kienzler2015}.
In the present magnon setting, incoherently pumped qubits and dispersive number selectivity provide an autonomous stabilization mechanism, together with a cascaded route to higher target Fock states.

\prlrunin{Auxiliary-qubit-enhanced magnon blockade}
In the single-qubit case discussed in the main text, the incoherently pumped qubit selectively assists the transition $|0\rangle\rightarrow |1\rangle$.
However, because the dispersive selectivity is finite, a residual population can still appear in the two-magnon state $|2\rangle$.
This unwanted population in $|2\rangle$ increases $g_m^{(2)}(0)$ and therefore weakens the magnon antibunching.
This observation suggests that, rather than further strengthening the forward injection into $\ket{1}$, the blockade may be enhanced by actively suppressing leakage into $\ket{2}$.

The idea is motivated by the steady-state balance condition of the effective birth--death model, $P_{n+1}/P_n=\lambda_n/\mu_{n+1}$.
Although this relation follows from the reduced birth--death model, it indicates that the relative population of adjacent Fock states can be controlled by engineering the effective upward and downward rates.
In the two-qubit cascade discussed in the main text, the second qubit is used to drive the population from $|1\rangle$ to $|2\rangle$.
Here qubit $1$ is used as an unpumped auxiliary qubit to assist the reverse $\lvert2\rangle\to\lvert1\rangle$ process.
Qubit $0$ remains incoherently pumped and supports the single-magnon injection process, while qubit $1$ is not pumped and is tuned to assist the reverse transition from $|2\rangle$ to $|1\rangle$.
Since qubit $1$ is not incoherently pumped, a sufficiently large relaxation rate $\gamma_1$ keeps it predominantly in the ground state, such that $q_{g,1}\gg q_{e,1}$.
The resonant exchange $\ket{2,g_1;v_1}\leftrightarrow\ket{1,e_1;v_1}$, followed by the rapid relaxation $\ket{1,e_1;v_1}\rightarrow\ket{1,g_1;v_1}$, therefore provides an effectively irreversible pathway that transfers population from the two-magnon state to the single-magnon state and suppresses $P_2$.

\begin{figure}[htb]
\centerline{\includegraphics[width=9cm]{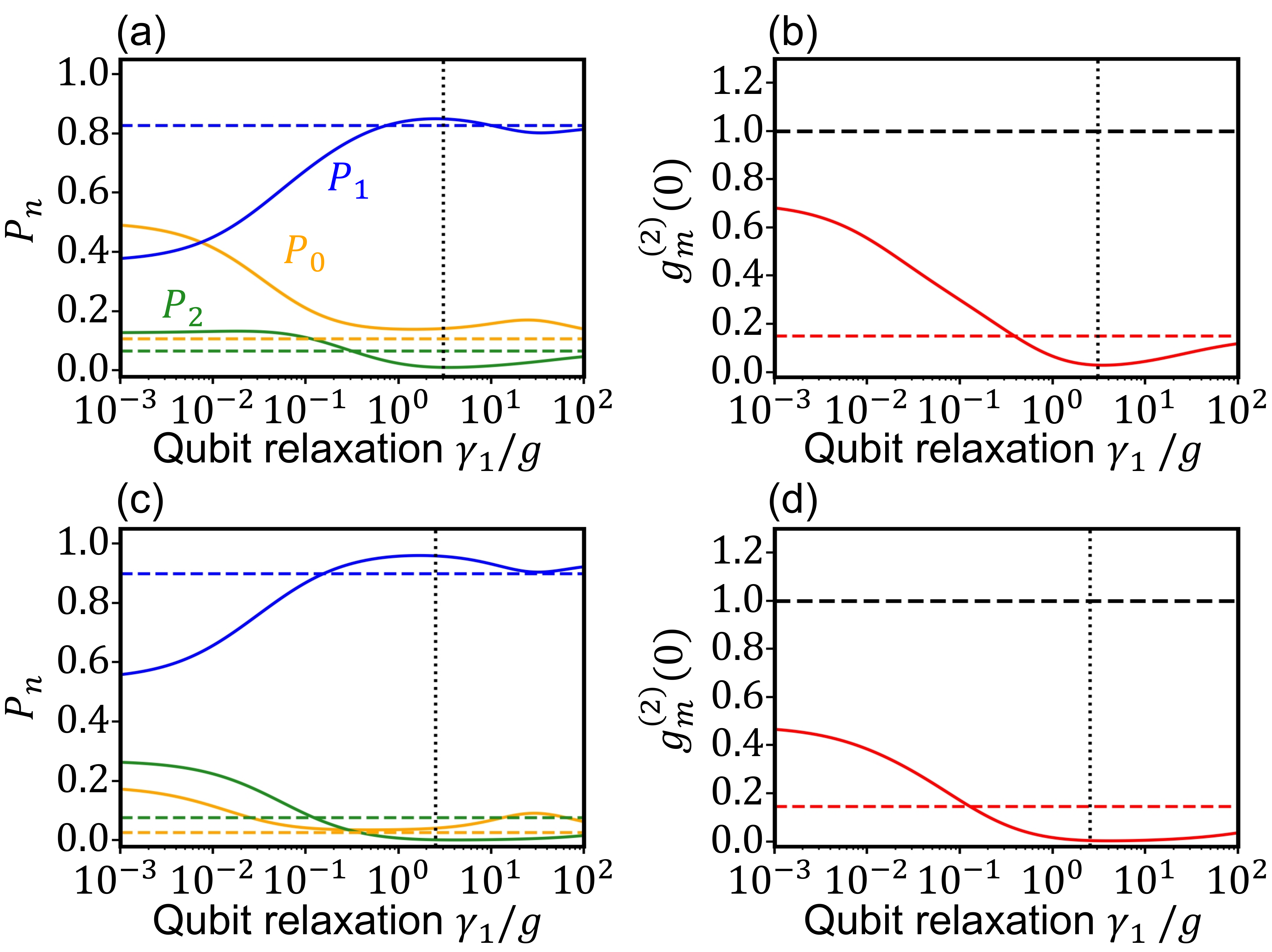}}
\caption{Auxiliary-qubit-enhanced magnon blockade.
Steady-state magnon populations $P_n$ [(a), (c)] and equal-time correlation $g_m^{(2)}(0)$ [(b), (d)] versus the auxiliary-qubit relaxation rate $\gamma_1/g$.
Solid curves show the two-qubit results, while colored dashed horizontal lines show the single-qubit case without qubit $1$.
In (a) and (c), blue, orange, and green curves denote $P_1$, $P_0$, and $P_2$, respectively.
The qubit detunings are chosen as $\Delta_0=-\chi_0-\chi_1$ and $\Delta_1=\chi_0-3\chi_1$.
For the two-qubit results, $\wp_1=0$ and $g_0=g_1=g$.
The parameters are $\chi_0/g=10$, $\chi_1/g=7$, $\wp_0/g=3$, $\kappa/g=0.1$,
and $\gamma_0/g=0.05$ in (a) and (b), and $\chi_0/g=30$, $\chi_1/g=7$,
$\wp_0/g=3$, $\kappa/g=0.01$, and $\gamma_0/g=0.05$ in (c) and (d).
For the single-qubit cases, only qubit $0$ is included, with $\Delta_0=-\chi_0$ and the same $\chi_0$, $\wp_0$, $\kappa$, and $\gamma_0$ as in the corresponding two-qubit cases.
The vertical dotted lines mark the values of $\gamma_1/g$ at which $P_1$ is maximized.
The black dashed lines in (b) and (d) mark $g_m^{(2)}(0)=1$.
}
\label{fig5}
\end{figure}

Figure~\ref{fig5} illustrates the auxiliary-qubit-enhanced blockade mechanism.
Panels (a) and (b) show the case with a moderate dispersive shift, $\chi_0/g=10$.
When $\gamma_1$ is very small, qubit $1$ has little relaxation-induced directionality.
Although it is not incoherently pumped, the number-selective exchange associated with qubit $1$ still opens a resonant path between the single- and two-magnon states.
As a result, part of the population injected into $|1\rangle$ can be transferred to the unwanted state $|2\rangle$, leading to a visible two-magnon population and a relatively large
$g_m^{(2)}(0)$.

Accordingly, increasing $\gamma_1$ from the weak-relaxation regime suppresses the effective population flow from $\ket{1}$ to $\ket{2}$ and favors the reverse process that removes population from $\ket{2}$.
Consequently, $P_2$ is strongly reduced over an intermediate range of $\gamma_1$, and this
reduction is accompanied by a pronounced decrease of $g_m^{(2)}(0)$.
Therefore, the auxiliary qubit enhances the magnon blockade not by strengthening the forward injection into $\ket{1}$, but by selectively suppressing the two-magnon population $P_2$.
The finite vacuum population in panels (a) and (b) is mainly due to the relatively large magnon loss rate, $\kappa/g=0.1$, which also drives the decay $|1\rangle\rightarrow |0\rangle$.

Panels (c) and (d) show the same mechanism for a larger dispersive shift, $\chi_0/g=30$, together with a smaller magnon loss rate, $\kappa/g=0.01$.
In this regime, for the single-qubit case the unwanted off-resonant excitation toward $|2\rangle$ is already weaker because the magnon-number-dependent transitions are better resolved.
The auxiliary qubit then further removes the remaining two-magnon leakage through the same relaxation-assisted reverse process.
As a result, $P_1$ remains close to unity, $P_2$ remains strongly suppressed, and $g_m^{(2)}(0)$ is reduced over a broader range of $\gamma_1$.

The comparison between the two parameter sets clarifies the role of the dispersive shift.
For $\chi_0/g=10$, the auxiliary qubit is needed to suppress appreciable leakage into $|2\rangle$, and the blockade improvement is visible only in a limited range of $\gamma_1$.
For $\chi_0/g=30$, the stronger number selectivity suppresses the undesired transition more effectively from the outset, so that the auxiliary-qubit filtering yields a steady state with a larger single-magnon population, reduced two-magnon leakage, and a more robust antibunching window.
When $\gamma_1$ becomes too large, the resonant auxiliary exchange is overdamped, with
$R_{1,1}^{\mathrm{res}}\propto\Gamma_{1,1}^{-1}$, while relaxation-induced broadening also reduces the number selectivity.
The blockade enhancement therefore gradually weakens.
These results suggest that the same auxiliary-qubit strategy may be extended to higher target Fock states by selectively suppressing neighboring unwanted Fock components.

\providecommand{\D}{\mathcal D}
\providecommand{\Tr}{\operatorname{Tr}}
\providecommand{\ImPart}{\operatorname{Im}}
\providecommand{\Hc}{\mathrm{H.c.}}
\providecommand{\ii}{\mathrm{i}}

%
\clearpage
\onecolumngrid

\setcounter{page}{1}
\setcounter{section}{0}
\setcounter{subsection}{0}
\setcounter{subsubsection}{0}
\setcounter{equation}{0}
\setcounter{figure}{0}
\setcounter{table}{0}

\renewcommand{\thesection}{S\arabic{section}}
\renewcommand{\theequation}{S\arabic{equation}}
\renewcommand{\thefigure}{S\arabic{figure}}
\renewcommand{\thetable}{S\arabic{table}}

\providecommand{\theHsection}{}
\providecommand{\theHsubsection}{}
\providecommand{\theHsubsubsection}{}
\providecommand{\theHequation}{}
\providecommand{\theHfigure}{}
\providecommand{\theHtable}{}
\renewcommand{\theHsection}{supp.section.\arabic{section}}
\renewcommand{\theHsubsection}{supp.subsection.\arabic{section}.\arabic{subsection}}
\renewcommand{\theHsubsubsection}{supp.subsubsection.\arabic{section}.\arabic{subsection}.\arabic{subsubsection}}
\renewcommand{\theHequation}{supp.equation.\arabic{equation}}
\renewcommand{\theHfigure}{supp.figure.\arabic{figure}}
\renewcommand{\theHtable}{supp.table.\arabic{table}}

\begingroup
\renewcommand{\thefootnote}{\fnsymbol{footnote}}
\begin{center}
{\Large\bfseries Supplemental Material for\\[0.35em]
``Wigner-Negative Magnon Steady States from Incoherent Qubit Pumping''\par}
\end{center}
\endgroup

\vspace{1.0em}

\setcounter{tocdepth}{2}
\tableofcontents
\clearpage

This Supplemental Material provides the detailed analytical derivations, numerical checks, and experimental-feasibility considerations supporting the results presented in the main text.
In Sec.\,\ref{sec:analytical_rate_model}, we derive the analytical rate model for the magnon-number populations starting from the full Lindblad master equation.
The derivation identifies the number-selective qubit-assisted transitions, includes the spectator-qubit-induced dispersive shifts, eliminates the transition coherences, and reduces the driven-dissipative dynamics to a finite birth--death equation with a closed-form steady-state solution.
In Sec.\,\ref{sec:irreversibility}, we clarify how the incoherent qubit pump, rather than the Jaynes--Cummings exchange alone, provides the directional bias required for magnon injection.
In Sec.\,\ref{sec:spectator_compensation}, we test the compensation of spectator-qubit dispersive shifts and justify the reference configuration $\bar z_k=+1$ used in the main-text simulations.
In Sec.\,\ref{sec:single_magnon_additional_scans}, we analyze the single-magnon limit and resolve how vacuum depletion, upper-tail leakage, detuning, dispersive selectivity, pump strength, and magnon loss determine the achievable single-magnon population.
In Sec.\,\ref{sec:two_magnon_even_fock}, we analyze the two-qubit cascade for stabilizing the two-magnon Fock state and examine the corresponding target population and Wigner negativity.
In Sec.\,\ref{sec:representative_states_inhomogeneity}, we present representative steady-state Fock distributions and Wigner functions and assess the robustness of the cascaded mechanism against qubit-parameter inhomogeneity.
In Sec.\,\ref{sec:dephasing_robustness}, we study the influence of magnon and qubit pure dephasing and compare the full master-equation results with the analytical rate-equation predictions.
In Sec.\,\ref{sec:higher_fock_extrapolation}, we use the reduced rate equation to extrapolate the cascaded stabilization mechanism to higher magnon Fock states beyond the range accessible to full master-equation simulations.
Finally, in Sec.\,\ref{sec:experimental_feasibility}, we discuss how the effective exchange--dispersive magnon--qubit Hamiltonian can be obtained in a realistic cavity-mediated superconducting-qubit--magnon platform and relate the model parameters to experimentally motivated coupling, dissipation, pumping, and sideband-control scales.

\section{Analytical rate model for magnon-number populations}
\label{sec:analytical_rate_model}
The main text proposes a cascaded injection scheme in which incoherently pumped qubits inject magnons through the sequence $\ket0\to\ket1\to\cdots\to\ket{N_Q}$~\cite{Poyatos1996,Kraus2008,Verstraete2009_DissipativeQE,Diehl2008a,Harrington2022}. 
The purpose of this section is to show how that driven-dissipative dynamics reduces to an analytically solvable birth--death equation for the magnon populations~\cite{GardinerZoller2004,Carmichael1999,WallsMilburn2008,Breuer2007,Alicki2007}.
Starting from the full master equation, we identify the number-selective qubit-assisted transitions, include the spectator-qubit-induced dispersive shifts, eliminate the transition coherences, and obtain a finite birth--death equation for the magnon Fock-state populations.
The resulting closed-form steady-state distribution explains how the incoherent qubit pump, number-selective resonance, and magnon loss combine to stabilize the target nonclassical magnon state.

\subsection{Model and notation}
\label{sec:model}

We first fix the model and notation used throughout the analytical derivation. 
We set $\hbar=1$. 
The magnon annihilation and creation operators are $\hat a$ and $\hat a^\dagger$, and $\hat n=\hat a^\dagger\hat a$ is the magnon-number operator. 
The total number of qubits is denoted by $N_Q$, and the qubits are labeled by $j=0,1,\ldots,N_Q-1$. 
For qubit $j$ of effective transition frequency $\omega_j$, the raising, lowering, and Pauli-$z$ operators are defined as
\begin{equation}
\hat\sigma_j^+=\ket{e_j}\bra{g_j},\qquad
\hat\sigma_j^-=\ket{g_j}\bra{e_j},\qquad
\hat\sigma_{z,j}=\ket{e_j}\bra{e_j}-\ket{g_j}\bra{g_j}.
\label{eq:pauli_def}
\end{equation}
In the frame rotating at the effective magnon frequency $\omega_{\rm m}$, the Hamiltonian is
\begin{equation}
\hat H
=
\sum_{j=0}^{N_Q-1}\Delta_j\hat\sigma_j^+\hat\sigma_j^-
+
\sum_{j=0}^{N_Q-1}\chi_j\hat n\hat\sigma_{z,j}
+
\sum_{j=0}^{N_Q-1}g_j
\left(\hat a\hat\sigma_j^++\hat a^\dagger\hat\sigma_j^-\right).
\label{eq:H}
\end{equation}
Here $\Delta_j=\omega_j-\omega_{\rm m}$ is the detuning of the qubit effective transition frequency from the effective magnon frequency, $\chi_j$ is the magnon-number-dependent dispersive shift, and $g_j$ is the Jaynes--Cummings exchange coupling between qubit $j$ and the magnon mode~\cite{Jaynes1963,PhysRev.140.A1051,Blais2004CircuitQED,RevModPhys.93.025005,Tabuchi2015,LachanceQuirion2017,LachanceQuirion2020,ZareRameshti2022,Yuan2022}.
In addition to the magnon loss, qubit relaxation, and incoherent qubit pump considered in the main text, realistic hybrid systems are also subject to fluctuations that randomize the relative phases of quantum states without directly changing their populations, a process known as pure dephasing~\cite{Breuer2007,RevModPhys.93.025005,Kjaergaard2020,ZareRameshti2022,Yuan2022}. 
For the qubits, such pure dephasing can originate, for example, from fluctuations of their transition frequencies induced by low-frequency flux or charge noise. 
For the magnon mode, fluctuations of the local magnetic field, magnetic anisotropy, or resonance frequency can lead to an analogous loss of phase coherence.
Including these pure-dephasing processes, the dynamics of the density operator $\hat\rho$ is governed by
\begin{align}
\dot{\hat\rho}
={}&-i[\hat H,\hat\rho]
+\kappa\D[\hat a]\hat\rho
+
\sum_{j=0}^{N_Q-1}\gamma_j\D[\hat\sigma_j^-]\hat\rho
+
\sum_{j=0}^{N_Q-1}\wp_j\D[\hat\sigma_j^+]\hat\rho
\nonumber\\
&+
\kappa_\phi\D[\hat n]\hat\rho
+
\sum_{j=0}^{N_Q-1}\gamma_{\phi,j}\D[\hat\sigma_{z,j}]\hat\rho,
\label{eq:ME}
\end{align}
where the Lindblad dissipator is defined by $\mathcal D[\hat o]\hat\rho
=
\hat o\hat\rho\hat o^\dagger
-\frac{1}{2}\hat o^\dagger\hat o\hat\rho
-\frac{1}{2}\hat\rho\hat o^\dagger\hat o$~\cite{Gorini1976,Lindblad1976,Breuer2007,GardinerZoller2004}. 
The parameters $\kappa$, $\gamma_j$, and $\wp_j$ denote the magnon loss rate, qubit relaxation rate, and incoherent qubit pump rate, respectively.  The parameters $\kappa_\phi$ and $\gamma_{\phi,j}$ are the pure-dephasing rates of the magnon mode and qubit $j$. 
With this convention, $\gamma_{\phi,j}$ is the Lindblad prefactor multiplying $\mathcal D[\hat\sigma_{z,j}]$, so that the active-qubit coherence decays at the rate $2\gamma_{\phi,j}$.
Unless otherwise stated, we set the pure-dephasing rates to zero, $\kappa_\phi=0$ and $\gamma_{\phi,j}=0$, in the numerical simulations and analytical estimates.

Throughout this derivation, an overdot denotes a time derivative, $\dot X\equiv dX/dt$. 
We use the notation $\left.\dot X\right|_{\alpha}$ to denote the contribution of the physical process labeled by $\alpha$ to the equation of motion of $X$. 
In particular, $\alpha=H$ denotes the Hamiltonian contribution, whereas $\alpha=\gamma_j,\wp_j,\kappa,\gamma_{\phi,j},\kappa_\phi$ denote the contributions from qubit relaxation, incoherent qubit pumping, magnon loss, qubit pure dephasing, and magnon pure dephasing, respectively.
Following the convention for the magnon Fock states adopted in the main text, we denote by $\hat\rho_m=\Tr_{\mathrm q}\hat\rho$ the reduced density operator of the magnon mode, where $\Tr_{\mathrm q}$ denotes the partial trace over all qubit degrees of
freedom. 
The population of the magnon Fock state $\ket n$ is then defined as $P_n=\mel{n}{\hat\rho_m}{n}$.

For the cascaded injection scheme introduced in the main text, the
$N_Q$ number-selective injection steps form the sequence
\begin{equation}
\ket{0}\rightarrow\ket{1}\rightarrow\cdots\rightarrow\ket{N_Q},
\end{equation}
which dynamically concentrates the magnon population near the target Fock state $\ket{N_Q}$. 
To obtain a finite set of analytical population equations while retaining the population transferred beyond $\ket{N_Q}$, we truncate the magnon Hilbert space as
\begin{equation}
\mathcal H_m^{(M)}
=
\operatorname{span}
\left\{
\ket{0},\ket{1},\ldots,\ket{M}
\right\},
\qquad
M=N_Q+2.
\label{eq:sm_cutoff}
\end{equation}
The truncated space therefore contains $M+1=N_Q+3$ magnon Fock states, ranging from the vacuum state $\ket{0}$ to the highest retained state $\ket{N_Q+2}$. 
In addition to the states involved in the intended injection sequence, the two states $\ket{N_Q+1}$ and $\ket{N_Q+2}$ are retained to describe the leading population leakage beyond the target Fock state.

\subsection{Number-selective transition detuning}
\label{sec:detuning}

In this section, we evaluate the transition energy of each qubit-assisted magnon transition, including the dispersive shifts induced by the spectator qubits~\cite{Schuster2007PhotonNumber,Gambetta2006,Boissonneault2009,LachanceQuirion2017,LachanceQuirion2020,PhysRevLett.125.117701}. 
We then determine the compensated detuning $\Delta_j$ for which qubit $j$ is resonant with the intended transition $\ket j\leftrightarrow\ket{j+1}$, while the other adjacent transitions remain off resonant.

We use the product basis 
\begin{equation}
\ket{n;s_0,s_1,\ldots,s_{N_Q-1}}
=
\ket{n}\bigotimes_{k=0}^{N_Q-1}\ket{s_k},
\qquad
s_k\in\{g_k,e_k\}.
\label{eq:product_basis}
\end{equation}
Here $\ket{n}$ is the magnon Fock state with $n$ magnons, and $s_k$ denotes the internal state of qubit $k$. 
When the transition generated by qubit $j$ is considered, the states of all other qubits are collected into the spectator configuration
\begin{equation}
v_j=(s_0,\ldots,s_{j-1},s_{j+1},\ldots,s_{N_Q-1}),
\qquad
s_k\in\{g_k,e_k\}\quad
(k=0,\ldots,N_Q-1,\;k\neq j).
\label{eq:vj_def}
\end{equation}
Thus $v_j$ denotes one assignment of all spectator qubits, and there are $2^{N_Q-1}$ such assignments for a fixed active qubit $j$. 
For each spectator qubit $k\neq j$, the state $s_k$ contributes to the diagonal energy in two ways: through the dispersive term $\chi_k\hat n\hat\sigma_{z,k}$ and, when the spectator is excited, through the qubit energy term $\Delta_k\hat\sigma_k^+\hat\sigma_k^-$. 
We therefore introduce $z_k(v_j)$ to record the $\hat\sigma_{z,k}$ eigenvalue of $s_k$, and $\eta_k(v_j)$ to record whether $s_k$ is the excited state
\begin{equation}
z_k(v_j)=\mel{s_k}{\hat\sigma_{z,k}}{s_k}=
\begin{cases}
+1, & s_k=e_k,\\
-1, & s_k=g_k,
\end{cases}
\qquad
\eta_k(v_j)=\frac{1+z_k(v_j)}{2}.
\label{eq:z_eta_def}
\end{equation}
Thus $z_k(v_j)$ enters the spectator-induced dispersive shift, while $\eta_k(v_j)$ enters the spectator-qubit excitation energy.  Both are fixed once the spectator configuration $v_j$ is specified.

For a fixed $j$, $n$, and $v_j$, the exchange term $g_j(\hat a\hat\sigma_j^++\hat a^\dagger\hat\sigma_j^-)$ couples exactly the two basis states that differ by one magnon and one qubit excitation
\begin{equation}
\ket{A_{j,n;v_j}}=\ket{n,e_j;v_j},
\qquad
\ket{B_{j,n;v_j}}=\ket{n+1,g_j;v_j}.
\label{eq:A_B_def}
\end{equation}
These two states define the qubit-assisted transition between the adjacent magnon-number states $\ket n$ and $\ket{n+1}$.  To obtain the corresponding energy difference, we use the diagonal part of Eq.\,\eqref{eq:H}
\begin{equation}
\hat H_0=
\sum_{r=0}^{N_Q-1}\Delta_r\hat\sigma_r^+\hat\sigma_r^-
+
\sum_{r=0}^{N_Q-1}\chi_r\hat n\hat\sigma_{z,r}.
\label{eq:H0}
\end{equation}
Taking the expectation value of $\hat H_0$ in the two states defined in Eq.\,\eqref{eq:A_B_def} gives
\begin{align}
E_A
&\equiv
\mel{A_{j,n;v_j}}{\hat H_0}{A_{j,n;v_j}}
=
\Delta_j+n\chi_j
+
\sum_{\substack{k=0\\ k\neq j}}^{N_Q-1}\Delta_k\eta_k(v_j)
+n\sum_{\substack{k=0\\ k\neq j}}^{N_Q-1}\chi_kz_k(v_j),
\label{eq:EA}\\
E_B
&\equiv
\mel{B_{j,n;v_j}}{\hat H_0}{B_{j,n;v_j}}
=
-(n+1)\chi_j
+
\sum_{\substack{k=0\\ k\neq j}}^{N_Q-1}\Delta_k\eta_k(v_j)
+(n+1)\sum_{\substack{k=0\\ k\neq j}}^{N_Q-1}\chi_kz_k(v_j).
\label{eq:EB}
\end{align}
The spectator excitation energies proportional to $\Delta_k$ cancel in the transition energy because the exchange of qubit $j$ does not change $v_j$. 
We define the transition detuning by
\begin{equation}
\delta_j(n;v_j)=E_B-E_A.
\label{eq:delta_def}
\end{equation}
Using Eqs.\,\eqref{eq:EA} and \eqref{eq:EB}, one obtains
\begin{equation}
\delta_j(n;v_j)
=
-\Delta_j-(2n+1)\chi_j
+
\sum_{\substack{k=0\\ k\neq j}}^{N_Q-1}\chi_kz_k(v_j).
\label{eq:delta_general}
\end{equation}
The transition $\ket{n,e_j;v_j}\leftrightarrow\ket{n+1,g_j;v_j}$ is resonant when $\delta_j(n;v_j)=0$.

The cascaded injection scheme uses qubit $j$ to address the transition $\ket j\leftrightarrow\ket{j+1}$. 
Since the dispersive shift depends on the spectator configuration, the resonance condition must be specified with respect to a reference configuration $\bar v_j$. 
This reference configuration is the configuration around which the analytical rate model is built.  We let $\bar z_k=z_k(\bar v_j)$. 
Imposing $\delta_j(j;\bar v_j)=0$ gives the compensated detuning
\begin{equation}
\Delta_j
=-(2j+1)\chi_j+
\sum_{\substack{k=0\\ k\neq j}}^{N_Q-1}\chi_k\bar z_k.
\label{eq:compensated_detuning}
\end{equation}
Substituting Eq.\,\eqref{eq:compensated_detuning} back into Eq.\,\eqref{eq:delta_general} gives the residual detuning
\begin{equation}
\delta_j(n;v_j)
=2(j-n)\chi_j+
\sum_{\substack{k=0\\ k\neq j}}^{N_Q-1}\chi_k\left[z_k(v_j)-\bar z_k\right].
\label{eq:delta_residual}
\end{equation}
For the reference spectator configuration, Eq.\,\eqref{eq:delta_residual} reduces to
\begin{equation}
\delta_j(n;\bar v_j)=2(j-n)\chi_j.
\label{eq:delta_reference}
\end{equation}
Thus the transition selected by qubit $j$ is resonant at $n=j$, whereas other transitions with $n\neq j$ remain detuned by multiples of $2\chi_j$.

In the strongly pumped regime considered in the main text, $\wp_k\gg\gamma_k$, the spectator qubits are predominantly excited. 
Therefore, taking $\bar z_k=+1$ for all spectator indices $k=0,\ldots,N_Q-1$ with $k\neq j$ and identical dispersive shifts $\chi_k=\chi$ gives
\begin{equation}
\Delta_j=-(2j+1)\chi+(N_Q-1)\chi=(N_Q-2j-2)\chi.
\label{eq:identical_detuning}
\end{equation}
This is the equally spaced compensated detuning used for the cascaded injection ladder.

\subsection{Single-transition coherence and effective exchange rate}
\label{sec:local_coherence}

Having fixed the number-selective detuning, we next derive the population current generated by one qubit-assisted adjacent transition.
This step connects the coherent Jaynes--Cummings exchange term in the full master equation to the effective transition rates used in the reduced population model~\cite{Carmichael1999,GardinerZoller2004,WallsMilburn2008,Breuer2007}.
For fixed qubit $j$, magnon number $n$, and spectator configuration $v_j$, we keep the transition coherence between $|n,e_j;v_j\rangle$ and $|n+1,g_j;v_j\rangle$.
Coherences involving a different qubit, a different magnon-number transition, or a different spectator configuration are neglected when deriving this single-transition current.
The same single-transition reduction is then applied to all qubits and all retained adjacent magnon-number transitions, and the resulting qubit-induced population currents are assembled into the finite birth--death equation.

For fixed qubit $j$, magnon number $n$, and spectator configuration $v_j$, we abbreviate
\begin{equation}
\ket A\equiv\ket{n,e_j;v_j},
\qquad
\ket B\equiv\ket{n+1,g_j;v_j},
\qquad
\Omega_{j,n}=g_j\sqrt{n+1}.
\label{eq:AB_Omega}
\end{equation}
For this transition, we define the coherence and the two diagonal populations as
\begin{equation}
C_{j,n}(v_j)=\mel{A}{\hat\rho}{B},
\qquad
P_A=\mel{A}{\hat\rho}{A},
\qquad
P_B=\mel{B}{\hat\rho}{B}.
\label{eq:C_PA_PB}
\end{equation}
The Hamiltonian of Eq.\,\eqref{eq:H} projected onto the subspace spanned by $\ket A$ and $\ket B$ is
\begin{align}
\hat H^{\rm proj}_{j,n}(v_j)
&=
\hat H^{\rm diag}_{j,n}(v_j)+\hat H^{\rm ex}_{j,n},
\label{eq:Hloc_decomp}\\
\hat H^{\rm diag}_{j,n}(v_j)
&=
E_A\ket A\bra A+E_B\ket B\bra B,
\qquad
\hat H^{\rm ex}_{j,n}
=
\Omega_{j,n}\left(\ket B\bra A+\ket A\bra B\right).
\label{eq:Hdiag_Hex}
\end{align}
The diagonal term $\hat H^{\rm diag}_{j,n}$ gives the phase evolution of the transition coherence $C_{j,n}(v_j)$, whereas the exchange term transfers population between $\ket A$ and $\ket B$.
The Hamiltonian part of the population current is therefore
\begin{align}
\left.\dot P_A\right|_{H^{\rm proj}_{j,n}(v_j)}
&=
-i\mel{A}{[\hat H^{\rm proj}_{j,n}(v_j),\hat\rho]}{A}
=
-2\Omega_{j,n}\ImPart [C_{j,n}(v_j)],
\label{eq:PA_loc}\\
\left.\dot P_B\right|_{H^{\rm proj}_{j,n}(v_j)}
&=
-i\mel{B}{[\hat H^{\rm proj}_{j,n}(v_j),\hat\rho]}{B}
=
+2\Omega_{j,n}\ImPart [C_{j,n}(v_j)].
\label{eq:PB_loc}
\end{align}
Thus a positive $2\Omega_{j,n}\ImPart[C_{j,n}]$ transfers population from
$\ket A=\ket{n,e_j;v_j}$ to $\ket B=\ket{n+1,g_j;v_j}$, corresponding to the transition $\ket n\to\ket{n+1}$ of the magnon mode.
The equation for the transition coherence is obtained by projecting the master equation onto the matrix element $\mel{A}{\hat\rho}{B}$.  The Hamiltonian contribution is
\begin{equation}
\left.\dot C_{j,n}\right|_{H^{\rm proj}_{j,n}(v_j)}
=
-i\mel{A}{[\hat H^{\rm proj}_{j,n}(v_j),\hat\rho]}{B}
=
i\delta_j(n;v_j)C_{j,n}(v_j)
-i\Omega_{j,n}(P_B-P_A).
\label{eq:C_loc_hamiltonian}
\end{equation}
The term proportional to $\delta_j(n;v_j)$ describes the phase evolution due to the energy difference $E_B-E_A$, while the term proportional to $\Omega_{j,n}$ is generated by the exchange coupling between $\ket A$ and $\ket B$.

In the projected two-state description, the dissipators enter through the decay rate of the transition coherence. 
Physically, any process that distinguishes $\ket A$ from $\ket B$, or removes amplitude from either state, broadens the transition and therefore changes the effective exchange coefficient. 
The notation $\left.\dot C_{j,n}\right|_{\alpha}$ denotes the contribution to $\dot C_{j,n}$ obtained by projecting the dissipator labeled by $\alpha$ onto the matrix element $\mel{A}{\dot{\hat\rho}}{B}$. 
The relevant projected dissipative contributions are
\begin{align}
\left.\dot C_{j,n}\right|_{\kappa}
&=
\mel{A}{\kappa\D[\hat a]\hat\rho}{B}
=
-\frac{(2n+1)\kappa}{2}C_{j,n}(v_j)
+\kappa\sqrt{(n+1)(n+2)}\,C_{j,n+1}(v_j),
\label{eq:C_kappa_local}\\
\left.\dot C_{j,n}\right|_{\kappa_\phi}
&=
\mel{A}{\kappa_\phi\D[\hat n]\hat\rho}{B}
=
-\frac{\kappa_\phi}{2}C_{j,n}(v_j),
\label{eq:C_magnon_deph_projection}\\
\left.\dot C_{j,n}\right|_{\gamma_j}
&=
\mel{A}{\gamma_j\D[\hat\sigma_j^-]\hat\rho}{B}
=
-\frac{\gamma_j}{2}C_{j,n}(v_j),
\label{eq:C_qubit_relax_projection}\\
\left.\dot C_{j,n}\right|_{\wp_j}
&=
\mel{A}{\wp_j\D[\hat\sigma_j^+]\hat\rho}{B}
=
-\frac{\wp_j}{2}C_{j,n}(v_j),
\label{eq:C_qubit_pump_projection}\\
\left.\dot C_{j,n}\right|_{\gamma_{\phi,j}}
&=
\mel{A}{\gamma_{\phi,j}\D[\hat\sigma_{z,j}]\hat\rho}{B}
=
-2\gamma_{\phi,j}C_{j,n}(v_j).
\label{eq:C_qubit_deph_projection}
\end{align}
The last term in Eq.\,\eqref{eq:C_kappa_local} contains the coherence of the adjacent higher transition,
\begin{equation}
C_{j,n+1}(v_j)
=
\mel{n+1,e_j;v_j}{\hat\rho}{n+2,g_j;v_j}.
\label{eq:C_neighbor_def}
\end{equation}
When deriving the reduced equation for $C_{j,n}(v_j)$, we omit this coupling to $C_{j,n+1}(v_j)$.
Dissipators acting on spectator qubits are not included in Eqs.\,\eqref{eq:C_kappa_local}-\eqref{eq:C_qubit_deph_projection}. 
Relaxation and pump of the spectator qubits change the spectator configuration $v_j$, and therefore couple $C_{j,n}(v_j)$ to coherences with different spectator configurations.
By contrast, for pure dephasing of a spectator qubit $k\neq j$, the spectator state is the same on the bra and ket sides of $C_{j,n}(v_j)$, and hence $\mel{A}{\D[\hat\sigma_{z,k}]\hat\rho}{B}=0$.
With these terms omitted, the equation for $C_{j,n}(v_j)$ becomes a single damped, driven linear equation.

Combining the Hamiltonian contribution with the retained dissipative contributions to $C_{j,n}(v_j)$ gives
\begin{equation}
\dot C_{j,n}(v_j)
=
-\left[\Gamma_{j,n}-i\delta_j(n;v_j)\right]C_{j,n}(v_j)
-i\Omega_{j,n}(P_B-P_A),
\label{eq:C_local}
\end{equation}
where the decay rate of this transition coherence is
\begin{equation}
\Gamma_{j,n}
=
\frac{\gamma_j+\wp_j}{2}
+
\frac{(2n+1)\kappa}{2}
+2\gamma_{\phi,j}
+
\frac{\kappa_\phi}{2}.
\label{eq:Gamma}
\end{equation}
The terms in $\Gamma_{j,n}$ are the contributions to the decay of the transition coherence from active-qubit $j$ relaxation and pump, magnon loss, active-qubit pure dephasing, and magnon pure dephasing.

For the stationary solution, setting $\dot C_{j,n}\simeq0$ gives the transition coherence. 
The same expression can be used dynamically when the coherence relaxes faster than the populations~\cite{GardinerZoller2004,Carmichael1999,Breuer2007}
\begin{equation}
C_{j,n}(v_j)
\simeq
\frac{-i\Omega_{j,n}(P_B-P_A)}{\Gamma_{j,n}-i\delta_j(n;v_j)}.
\label{eq:C_stationary}
\end{equation}

The imaginary part of the coherence is the only part that enters the population current in Eqs.\,\eqref{eq:PA_loc} and \eqref{eq:PB_loc}. 
Taking the imaginary part of Eq.\,\eqref{eq:C_stationary} yields
\begin{equation}
\ImPart [C_{j,n}(v_j)]
=
\frac{\Omega_{j,n}\Gamma_{j,n}}{\Gamma_{j,n}^2+\delta_j(n;v_j)^2}(P_A-P_B).
\label{eq:ImC}
\end{equation}
Substituting Eq.\,\eqref{eq:ImC} into Eqs.\,\eqref{eq:PA_loc} and \eqref{eq:PB_loc}, the qubit-induced population current associated with the transition $n\leftrightarrow n+1$ is
\begin{equation}
J_{j,n}(v_j)
=2\Omega_{j,n}\ImPart [C_{j,n}(v_j)]
=R_{j,n}(v_j)(P_A-P_B).
\label{eq:J_local}
\end{equation}
This equation defines the effective exchange coefficient for this qubit-assisted transition
\begin{equation}
R_{j,n}(v_j)
=
\frac{2g_j^2(n+1)\Gamma_{j,n}}{\Gamma_{j,n}^2+\delta_j(n;v_j)^2}.
\label{eq:R_local}
\end{equation}
The detuning dependence in Eq.\,\eqref{eq:R_local} gives rise to number selectivity: for fixed $j$ and $v_j$, only transitions with $|\delta_j(n;v_j)|\lesssim \Gamma_{j,n}$ have appreciable exchange rates, whereas far-detuned transitions are suppressed.  For a resonant transition, $\delta_j(n;v_j)=0$, the exchange coefficient becomes
\begin{equation}
R_{j,n}^{\rm res}=\frac{2g_j^2(n+1)}{\Gamma_{j,n}}.
\label{eq:R_res}
\end{equation}
For a far-detuned transition, $|\delta_j(n;v_j)|\gg\Gamma_{j,n}$, Eq.\,\eqref{eq:R_local} reduces to
\begin{equation}
R_{j,n}^{\rm off}
\simeq
\frac{2g_j^2(n+1)\Gamma_{j,n}}{\delta_j(n;v_j)^2}.
\label{eq:R_off}
\end{equation}
These two limits show the dual role of the coherence decay rate $\Gamma_{j,n}$.  The dissipative contributions to the transition coherence enter the effective exchange coefficient through $\Gamma_{j,n}$: increasing $\Gamma_{j,n}$ suppresses resonant exchange in Eq.\,\eqref{eq:R_res}, but enhances off-resonant leakage in Eq.\,\eqref{eq:R_off}.
Pure dephasing is one such contribution, and therefore affects the population dynamics through the exchange coefficient even though it does not directly transfer population.

\subsection{Population closure and effective transition rates}
\label{sec:closure}

The coefficient $R_{j,n}$ is positive and sets the magnitude of the exchange for the addressed transition.
The direction of the population current is determined by the sign of $P_A-P_B$: if $P_A>P_B$, population flows from $\ket A=\ket{n,e_j;v_j}$ to $\ket B=\ket{n+1,g_j;v_j}$, increasing the magnon number, whereas if $P_A<P_B$, the current reverses.
Thus the exchange coefficient itself does not impose a preferred direction in magnon number; it only converts a population imbalance between the two states into a transition current.
A sustained upward bias appears only after the populations of the two states are tied to the nonequilibrium qubit populations maintained by incoherent pump and relaxation~\cite{Poyatos1996,Verstraete2009_DissipativeQE,Harrington2022,PhysRevLett.109.183602,PhysRevLett.110.120501,Magnard2018PRL,Sunada2022PRApplied}.
We now use this pump-maintained qubit population imbalance to close $P_A$ and $P_B$ in terms of the reduced magnon populations.

The reduced magnon population is $P_n=\mel{n}{\hat\rho_m}{n}$. 
After eliminating the transition coherence, we close the remaining population equations by neglecting correlations between the magnon population and the qubit states.
In the strong-pump regime considered in the main text, the reference spectator configuration $\bar v_j$ is the one in which all spectator qubits are in their excited states.
The population weight of this configuration $\bar v_j$ is therefore
\begin{equation}
\Pi_j^{(\bar v_j)}
\equiv
\prod_{\substack{k=0\\k\neq j}}^{N_Q-1}
q_{e,k}
\simeq
1,
\label{eq:sm_reference_configuration_weight}
\end{equation}
where $q_{e,k}$ is the excited-state population of spectator qubit $k$. 
The approximation in Eq.\,\eqref{eq:sm_reference_configuration_weight} follows from
$\wp_k\gg\gamma_k$, for which $q_{e,k}\simeq1$.
The populations of the two states involved in the selected transition are then factorized as
\begin{align}
P_A
&=
\mel{n,e_j;\bar v_j}{\hat\rho}{n,e_j;\bar v_j}
\simeq
P_n q_{e,j}\Pi_j^{(\bar v_j)}
\simeq
P_n q_{e,j},
\label{eq:sm_PA_population_closure}
\\
P_B
&=
\mel{n+1,g_j;\bar v_j}{\hat\rho}{n+1,g_j;\bar v_j}
\simeq
P_{n+1}q_{g,j}\Pi_j^{(\bar v_j)}
\simeq
P_{n+1}q_{g,j}.
\label{eq:sm_PB_population_closure}
\end{align}
Here $\bar v_j$ is the reference spectator configuration used in the detuning compensation, and $q_{e,j}$ and $q_{g,j}$ are the excited- and ground-state populations of the active qubit $j$. 
The first approximations in Eqs.\,\eqref{eq:sm_PA_population_closure} and \eqref{eq:sm_PB_population_closure} neglect correlations among the magnon population, the active-qubit state, and the spectator-qubit states. 
The second approximations use $\Pi_j^{(\bar v_j)}\simeq1$.
For an isolated incoherently pumped qubit, these steady-state populations are
\begin{equation}
q_{e,j}=\frac{\wp_j}{\wp_j+\gamma_j},
\qquad
q_{g,j}=\frac{\gamma_j}{\wp_j+\gamma_j}.
\label{eq:qeg}
\end{equation}
Equations~\eqref{eq:sm_PA_population_closure} and \eqref{eq:sm_PB_population_closure} constitute the population closure used in the reduced model. 
The accuracy of this reduction is assessed by comparison with the numerical solution of the full master equation.

In the following reduced rate equations, we evaluate the exchange coefficient at the reference spectator configuration $\bar v_j$ used to compensate the transition detuning. We therefore define
\begin{equation}
R_{j,n}\equiv R_{j,n}(\bar v_j)
=
\frac{2g_j^2(n+1)\Gamma_{j,n}}
{\Gamma_{j,n}^2+\delta_j(n;\bar v_j)^2}.
\label{eq:R_reference}
\end{equation}
Using Eqs.\,\eqref{eq:J_local}, \eqref{eq:sm_PA_population_closure}, and \eqref{eq:sm_PB_population_closure}, the population current induced by qubit $j$ associated with the transition $n\leftrightarrow n+1$ becomes
\begin{equation}
J_{j,n}
\simeq
R_{j,n}
\left(q_{e,j}P_n-q_{g,j}P_{n+1}\right).
\label{eq:J_jn}
\end{equation}
In the strong-pump regime, $q_{e,j}/q_{g,j}=\wp_j/\gamma_j\gg 1$.
Therefore, the net current $J_{j,n}$ can remain positive even when the upper magnon state is more populated than the lower one, $P_{n+1}>P_n$.
The condition for an upward current is explicitly
\begin{equation}
J_{j,n}>0
\quad \Longleftrightarrow \quad
\frac{P_{n+1}}{P_n}<\frac{q_{e,j}}{q_{g,j}} .
\end{equation}
Thus the incoherent pump controls the range over which the qubit-induced exchange drives population from $\ket{n}$ to $\ket{n+1}$: increasing the pump-to-relaxation ratio increases $q_{e,j}/q_{g,j}$ and strengthens the upward bias on the resonantly addressed transition.
These qubit-induced population currents are then summed over all qubits and combined with magnon loss to obtain the finite birth--death equation.

\subsection{Finite birth--death equation}
\label{sec:birth_death}

We now combine the qubit-induced currents with magnon loss. 
This gives a closed nearest-neighbor rate equation for the magnon-number Fock-state populations~\cite{GardinerZoller2004,Carmichael1999,WallsMilburn2008,Breuer2007,Alicki2007}. 
The current $J_{j,n}$ in Eq.\,\eqref{eq:J_jn} contains an upward part $R_{j,n}q_{e,j}P_n$ and a downward absorption part $R_{j,n}q_{g,j}P_{n+1}$.
The magnon loss term gives the familiar downward process $\ket{n}\to\ket{n-1}$. 
After tracing out the qubits and projecting $\kappa\mathcal{D}[\hat a]\hat\rho_m$ onto the magnon Fock state $|n\rangle$, one obtains
\begin{equation}
\left.\dot P_n\right|_{\kappa}
=
\kappa\mel{n}{\D[\hat a]\hat\rho_m}{n}
=
\kappa\left[(n+1)P_{n+1}-nP_n\right].
\label{eq:magnon_loss_population}
\end{equation}
In the finite magnon-number basis retained in the reduced rate model, $|0\rangle,\ldots,|M\rangle$, the population outside the retained space is set to zero, so that $P_{M+1}=0$. 
This boundary convention removes the loss-induced inflow from the unretained state $|M+1\rangle$ into $|M\rangle$, while the physical decay from each retained state $|n\rangle$ to $|n-1\rangle$ remains included with rate $n\kappa$. 
The two retained states above the target capture the leading leakage when further excitation to higher magnon numbers is suppressed by the dispersive detuning.
We therefore combine the qubit-induced population currents with the magnon-loss current to define the finite birth--death rates below.

For each retained adjacent magnon-number transition, the qubit-induced population current is obtained by summing the contributions from all qubits. The detuning dependence of each contribution is already contained in the exchange coefficient $R_{j,n}$ defined in Eq.\,\eqref{eq:R_reference}. The upward rate from $|n\rangle$ to $|n+1\rangle$ is therefore
\begin{equation}
\lambda_n
=
\sum_{j=0}^{N_Q-1} q_{e,j} R_{j,n},
\qquad
n=0,1,\ldots,M-1 .
\label{eq:lambda_def}
\end{equation}
The downward rate from $|n\rangle$ to $|n-1\rangle$ is
\begin{equation}
\mu_n
=
n\kappa+
\sum_{j=0}^{N_Q-1} q_{g,j} R_{j,n-1},
\qquad
n=1,2,\ldots,M .
\label{eq:mu_def}
\end{equation}
The term $n\kappa$ is the magnon-loss rate, while the second term in $\mu_n$ describes qubit-induced absorption from $|n\rangle$ to $|n-1\rangle$. The boundary rates are
\begin{equation}
\mu_0=0,
\qquad
\lambda_M=0 .
\label{eq:boundary_rates}
\end{equation}
Here $\lambda_n$ denotes the birth rate for the transition $n\to n+1$, and $\mu_n$ denotes the death rate for the transition $n\to n-1$.

The population equation follows by projecting the master equation onto the magnon state $\ket{n}$
\begin{equation}
\dot P_n=\Tr\left[\ket{n}\bra{n}\,\dot{\hat\rho}_m\right],
\label{eq:pn_projection_start}
\end{equation}
and collecting the population currents through the two neighboring transitions. 
The resulting birth--death equation is
\begin{equation}
\dot P_n
=
\lambda_{n-1}P_{n-1}
+\mu_{n+1}P_{n+1}
-(\lambda_n+\mu_n)P_n,
\label{eq:birth_death}
\end{equation}
for $n=0,1,\ldots,M$.  The boundary convention is $\lambda_{-1}P_{-1}=0$, $\mu_0P_0=0$, $\lambda_M=0$, and $P_{M+1}=0$.  Pure dephasing does not appear as a separate birth or death process because $\D[\hat n]$ and $\D[\hat\sigma_{z,j}]$ do not directly change diagonal populations in the number basis.  It affects Eq.\,\eqref{eq:birth_death} through $\Gamma_{j,n}$ and hence through $R_{j,n}$, $\lambda_n$, and $\mu_n$.

\subsection{Closed-form steady state and target population}
\label{sec:solution}

The population birth--death equation obtained in the previous section reduces the master equation to a finite set of nearest-neighbor rate equations for the magnon Fock-state populations. 
These equations can be solved recursively, yielding a closed-form steady-state solution. 
This solution separates the target-state population into three physically distinct contributions: the desired population in the target Fock state, the lower-number tail from states below the target, and the upper leakage tail resulting from residual off-resonant excitation above the target.

Define the net population current across the adjacent transition $n\leftrightarrow n+1$ with the positive direction chosen from $|n\rangle$ to $|n+1\rangle$
\begin{equation}
\mathcal J_n
=
\lambda_n P_n-\mu_{n+1}P_{n+1}.
\label{eq:current_def}
\end{equation}
Here $\lambda_n P_n$ is the upward contribution associated with $|n\rangle\to |n+1\rangle$, whereas $\mu_{n+1}P_{n+1}$ is the downward contribution associated with $|n+1\rangle\to |n\rangle$. Therefore $\mathcal J_n>0$ denotes a net population current toward larger magnon number, while $\mathcal J_n<0$ denotes a net current toward smaller magnon number.

With this sign convention, the population birth--death equation~\eqref{eq:birth_death} can be written in the continuity form
\begin{equation}
\dot P_n
=
\mathcal J_{n-1}-\mathcal J_n .
\label{eq:continuity_form}
\end{equation}
The first term is the net inflow into $|n\rangle$ through the transition $(n-1)\leftrightarrow n$, and the second term is the net outflow from $|n\rangle$ through the transition $n\leftrightarrow n+1$. At steady state, $\dot P_n=0$ implies $\mathcal J_{n-1}=\mathcal J_n$, so the net current is independent of $n$. For the finite set of retained states, the boundary conditions $\mu_0=0$ and $\lambda_M=0$ imply that there is no population current entering from below $|0\rangle$ or leaving above $|M\rangle$. The steady-state net population current therefore vanishes, giving
\begin{equation}
\lambda_n P_n
=
\mu_{n+1}P_{n+1},
\qquad
n=0,1,\ldots,M-1 .
\label{eq:zero_current_balance}
\end{equation}
This gives the recursion relation
\begin{equation}
P_{n+1}=\frac{\lambda_n}{\mu_{n+1}}P_n.
\label{eq:recursion}
\end{equation}
Repeated use of Eq.\,\eqref{eq:recursion} gives
\begin{equation}
P_n=P_0\prod_{\ell=0}^{n-1}\frac{\lambda_\ell}{\mu_{\ell+1}},
\qquad
n=1,2,\ldots,M.
\label{eq:pn_p0}
\end{equation}
Using the normalization condition $\sum_{n=0}^{M}P_n=1$, the prefactor $P_0$ is fixed as
\begin{equation}
P_0=
\left[
1+
\sum_{m=1}^{M}
\prod_{\ell=0}^{m-1}\frac{\lambda_\ell}{\mu_{\ell+1}}
\right]^{-1}.
\label{eq:p0}
\end{equation}
Thus, the steady-state populations of the reduced model are
\begin{equation}
P_n
=
\frac{
\displaystyle
\prod_{\ell=0}^{n-1}\lambda_\ell/\mu_{\ell+1}
}{
\displaystyle
1+
\sum_{m=1}^{M}
\prod_{\ell=0}^{m-1}\lambda_\ell/\mu_{\ell+1}
},
\qquad
n=0,1,\ldots,M.
\label{eq:pn_final}
\end{equation}
Equation~\eqref{eq:pn_final} is exact for the reduced nearest-neighbor birth--death equations; any approximations enter only through the rates $\lambda_n$ and $\mu_n$.
Here and below, an empty product is understood as unity.

For the cascaded injection scheme, we set $M=N_Q+2$, so that the retained magnon-number basis contains the target state $|N_Q\rangle$ and the two states $|N_Q+1\rangle$ and $|N_Q+2\rangle$ above it. 
Dividing the normalization condition by $P_{N_Q}$ gives
\begin{equation}
P_{N_Q}^{-1}
=
1+
(\sum_{m=0}^{N_Q-1}
\prod_{\ell=m}^{N_Q-1}\frac{\mu_{\ell+1}}{\lambda_\ell}
)+
\frac{\lambda_{N_Q}}{\mu_{N_Q+1}}
+
\frac{\lambda_{N_Q}\lambda_{N_Q+1}}{\mu_{N_Q+1}\mu_{N_Q+2}} .
\label{eq:p_target}
\end{equation}
Each term in Eq.\,\eqref{eq:p_target} has a direct interpretation as a population ratio relative to the target state. 
The first term is the ratio of the target population to itself, $P_{N_Q}/P_{N_Q}=1$. 
The term with index $m$ in the sum is the ratio $P_m/P_{N_Q}$ for a state below the target. This ratio is obtained by multiplying the inverse rate ratios $\mu_{\ell+1}/\lambda_\ell$ along the path from $|N_Q\rangle$ down to $|m\rangle$. 
The sum therefore gives the total lower-number population relative to $P_{N_Q}$. 
It is small when the addressed upward rates $\lambda_\ell$ dominate over the corresponding downward rates $\mu_{\ell+1}$ for $0\leq \ell\leq N_Q-1$.
The last two terms are the relative populations of the two retained states above the target,
\begin{equation}
\frac{P_{N_Q+1}}{P_{N_Q}}=\frac{\lambda_{N_Q}}{\mu_{N_Q+1}},
\qquad
\frac{P_{N_Q+2}}{P_{N_Q}}
=
\frac{\lambda_{N_Q}\lambda_{N_Q+1}}{\mu_{N_Q+1}\mu_{N_Q+2}} .
\label{eq:tail_ratios}
\end{equation}
They quantify leakage above $|N_Q\rangle$. The first ratio is controlled by the competition between the upward rate $\lambda_{N_Q}$ from $|N_Q\rangle$ to $|N_Q+1\rangle$ and the downward rate $\mu_{N_Q+1}$. 
The second ratio contains one additional upward step, from $|N_Q+1\rangle$ to $|N_Q+2\rangle$, and is further suppressed when $\lambda_{N_Q+1}\ll\mu_{N_Q+2}$.

High population of the target state therefore requires two conditions: population must be driven upward along the addressed part of the ladder, while excitation above the target must be suppressed. 
In terms of the effective rates,
\begin{equation}
\lambda_\ell\gg\mu_{\ell+1}\quad(0\leq\ell\leq N_Q-1),
\qquad
\lambda_{N_Q}\ll\mu_{N_Q+1},
\qquad
\lambda_{N_Q+1}\ll\mu_{N_Q+2}.
\label{eq:target_condition}
\end{equation}
The first set of inequalities suppresses the lower-number population relative to $P_{N_Q}$, whereas the last two suppress the populations of $|N_Q+1\rangle$ and $|N_Q+2\rangle$ relative to the target state.

We now use the Lorentzian form of the exchange coefficient to estimate the rates that control the confinement of the steady-state population near the target state. 
This estimates the upper leakage retained by the finite cutoff $M=N_Q+2$, since the two retained states above the target, $|N_Q+1\rangle$ and $|N_Q+2\rangle$, are populated only through residual off-resonant excitation.
For the reference spectator configuration, Eq.\,\eqref{eq:R_reference} gives
\begin{equation}
\delta_j(n;\bar v_j)=2(j-n)\chi_j .
\end{equation}
Thus qubit $j=n$ is resonant with the addressed transition $|n\rangle\leftrightarrow |n+1\rangle$, whereas the same transition is detuned from all qubits with $j\ne n$. 
For $0\le n\le N_Q-1$, we define $\lambda_n^{\rm res}$ as the resonant estimate of the addressed birth rate, obtained by retaining the resonant term $j=n$ in $\lambda_n=\sum_j q_{e,j}R_{j,n}$. 
Since $\delta_n(n;\bar v_n)=0$, the Lorentzian denominator reduces to $\Gamma_{n,n}^2$, giving
\begin{equation}
\lambda_n^{\rm res}
\simeq
q_{e,n}R_{n,n}(\bar v_n)
=
q_{e,n}\frac{2g_n^2(n+1)}{\Gamma_{n,n}},
\qquad
n=0,1,\ldots,N_Q-1 .
\label{eq:lambda_res}
\end{equation}
The first transition above the target, $|N_Q\rangle\leftrightarrow |N_Q+1\rangle$, has no resonant qubit in the cascaded injection scheme. 
Its birth rate is therefore generated by the off-resonant Lorentzian tails of all qubits. Substituting $\delta_j(N_Q;\bar v_j)=2(j-N_Q)\chi_j$ into Eq.\,\eqref{eq:R_reference} gives
\begin{equation}
\lambda_{N_Q}
\simeq
\sum_{j=0}^{N_Q-1}
q_{e,j}
\frac{2g_j^2(N_Q+1)\Gamma_{j,N_Q}}
{\Gamma_{j,N_Q}^2+4(N_Q-j)^2\chi_j^2}.
\label{eq:first_leakage_rate}
\end{equation}
Similarly, the second retained leakage transition, $|N_Q+1\rangle\leftrightarrow |N_Q+2\rangle$, is also off resonant for every qubit. 
Using $\delta_j(N_Q+1;\bar v_j)=2(j-N_Q-1)\chi_j$, one obtains
\begin{equation}
\lambda_{N_Q+1}
\simeq
\sum_{j=0}^{N_Q-1}
q_{e,j}
\frac{2g_j^2(N_Q+2)\Gamma_{j,N_Q+1}}
{\Gamma_{j,N_Q+1}^2+4(N_Q+1-j)^2\chi_j^2}.
\label{eq:second_leakage_rate}
\end{equation}
Equations~\eqref{eq:first_leakage_rate} and \eqref{eq:second_leakage_rate} show that the population leakage above $|N_Q\rangle$ is controlled by the off-resonant tails of the same Lorentzian exchange coefficients. 
When $|\chi_j|\gg \Gamma_{j,n}$ for the unaddressed transitions, these tails are strongly suppressed, so the retained upper states provide a controlled estimate of the leakage beyond the target.

Combining these estimates with Eq.\,\eqref{eq:tail_ratios} gives the corresponding upper-tail populations. In the resolved-transition limit $|\chi_j|\gg \Gamma_{j,n}$ for the unaddressed transitions, the first leakage ratio becomes
\begin{equation}
\frac{P_{N_Q+1}}{P_{N_Q}}
\simeq
\frac{1}{\mu_{N_Q+1}}
\sum_{j=0}^{N_Q-1}
q_{e,j}
\frac{
g_j^2 (N_Q+1)\Gamma_{j,N_Q}
}{
2 (N_Q-j)^2\chi_j^2
}.
\label{eq:first_tail_large_chi}
\end{equation}
Similarly, it also gives
\begin{equation}
\frac{P_{N_Q+2}}{P_{N_Q}}
\simeq
\frac{1}{\mu_{N_Q+1}\mu_{N_Q+2}}
\left[
\sum_{j=0}^{N_Q-1}
q_{e,j}
\frac{
g_j^2 (N_Q+1)\Gamma_{j,N_Q}
}{
2 (N_Q-j)^2\chi_j^2
}
\right]
\left[
\sum_{j=0}^{N_Q-1}
q_{e,j}
\frac{
g_j^2 (N_Q+2)\Gamma_{j,N_Q+1}
}{
2 (N_Q+1-j)^2\chi_j^2
}
\right].
\label{eq:second_tail_large_chi}
\end{equation}
Thus the population of $|N_Q+1\rangle$ is suppressed by the off-resonant Lorentzian factor proportional to $\chi_j^{-2}$, while the population of $|N_Q+2\rangle$ contains two successive off-resonant steps and is suppressed at the next order. In the strong-pump regime, $q_{g,j}\ll q_{e,j}$, the return rates $\mu_{N_Q+1}$ and $\mu_{N_Q+2}$ are often dominated by magnon loss, giving the transparent estimates
\begin{equation}
\frac{P_{N_Q+1}}{P_{N_Q}}
\simeq
\frac{1}{\kappa}
\sum_{j=0}^{N_Q-1}
q_{e,j}
\frac{
g_j^2\Gamma_{j,N_Q}
}{
2 (N_Q-j)^2\chi_j^2
},
\end{equation}
and
\begin{equation}
\frac{P_{N_Q+2}}{P_{N_Q}}
\simeq
\frac{1}{\kappa^2}
\left[
\sum_{j=0}^{N_Q-1}
q_{e,j}
\frac{
g_j^2\Gamma_{j,N_Q}
}{
2 (N_Q-j)^2\chi_j^2
}
\right]
\left[
\sum_{j=0}^{N_Q-1}
q_{e,j}
\frac{
g_j^2\Gamma_{j,N_Q+1}
}{
2 (N_Q+1-j)^2\chi_j^2
}
\right].
\end{equation}
These estimates make explicit that increasing the dispersive shift reduces the first upper-tail population as $\chi_j^{-2}$ and the second upper-tail population as $\chi_j^{-4}$, up to the weak $n$ dependence of $\Gamma_{j,n}$ and the return rates.
These estimates also show that reducing the magnon loss rate $\kappa$ is not always beneficial for target-state preparation. When the return rates $\mu_{N_Q+1}$ and $\mu_{N_Q+2}$ are dominated by magnon loss, the upper-tail ratios scale inversely with $\kappa$ and $\kappa^2$, respectively. Therefore, too small a value of $\kappa$ can allow population to accumulate in the states above the target, reducing the occupation of $|N_Q\rangle$. The magnon loss rate must therefore be small enough not to overwhelm the addressed upward cascade, but large enough to remove residual population leaked above the target.

As a concrete specialization, for $N_Q=4$ and $M=6$, Eq.\,\eqref{eq:p_target} gives
\begin{align}
P_4^{-1}
={}&
1
+\frac{\mu_4}{\lambda_3}
+\frac{\mu_3\mu_4}{\lambda_2\lambda_3}
+\frac{\mu_2\mu_3\mu_4}{\lambda_1\lambda_2\lambda_3}
+\frac{\mu_1\mu_2\mu_3\mu_4}{\lambda_0\lambda_1\lambda_2\lambda_3}
\nonumber\\
&+
\frac{\lambda_4}{\mu_5}
+
\frac{\lambda_4\lambda_5}{\mu_5\mu_6}.
\label{eq:N4_example}
\end{align}

\section{Role of the incoherent qubit pump}
\label{sec:irreversibility}

The main text emphasizes that the qubit pump is incoherent. 
The rate construction above provides the corresponding population-level interpretation of this point. 
As follows from the coherence equation in Eq.\,\eqref{eq:C_loc_hamiltonian}, eliminating the transition coherence gives a current of the form $J_{j,n}=R_{j,n}(P_A-P_B)$, with $R_{j,n}>0$. 
Thus the Jaynes--Cummings exchange term itself does not select a preferred direction in magnon number. 
It transfers population according to the instantaneous population difference between $|A\rangle=\ket{n,e_j;v_j}$ and $|B\rangle=\ket{n+1,g_j;v_j}$, and the exchange current vanishes when $P_A=P_B$.

The directionality described in the main text is introduced by the incoherent pump~\cite{Poyatos1996,Verstraete2009_DissipativeQE,Harrington2022,PhysRevLett.109.183602,PhysRevLett.110.120501,Magnard2018PRL,Sunada2022PRApplied}. 
After an emission event $|n,e_j;v_j\rangle\to |n+1,g_j;v_j\rangle$, the pump repopulates the excited state of qubit $j$ without changing the magnon number. 
In the population closure this gives $P_A\simeq q_{e,j}P_n$ and $P_B\simeq q_{g,j}P_{n+1}$, so that the current becomes
\begin{equation}
J_{j,n}
\simeq
R_{j,n}\left(q_{e,j}P_n-q_{g,j}P_{n+1}\right).
\end{equation}
In the strong-pump regime, the pump maintains $q_{e,j}>q_{g,j}$ and therefore biases the resonantly addressed transition toward $|n\rangle\to |n+1\rangle$. 
This is the rate-equation counterpart of the main-text statement that the incoherent pump restores the active qubit after magnon emission, while the number-dependent detuning prevents the same qubit from resonantly driving the next higher transition. 
Hence the directed magnon injection arises from the incoherent pump together with number-selective resonance, rather than from the Jaynes--Cummings exchange coefficient alone.

\section{Sensitivity to spectator-qubit compensation}
\label{sec:spectator_compensation}

\begin{figure}[htb]
\centerline{\includegraphics[width=9cm]{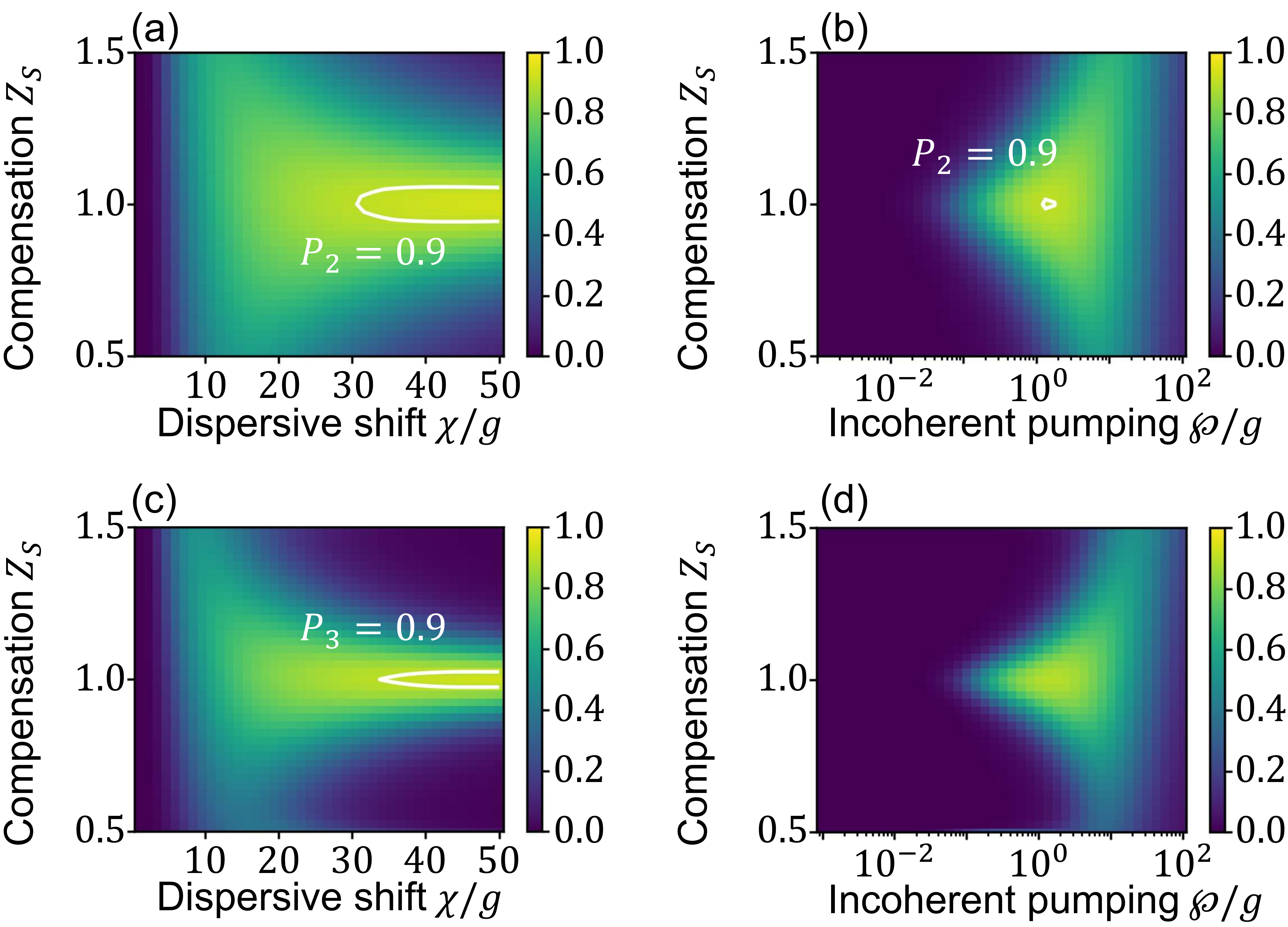}}
\caption{
Effect of spectator-qubit compensation on cascaded Fock-state preparation. 
Color maps show the numerical target-state population $P_{N_Q}$ versus the compensation parameter $Z_s$ and either $\chi/g$ or $\wp/g$. 
The qubit detunings are chosen as $\Delta_j=-(2j+1)\chi+(N_Q-1)Z_s\chi$ for $j=0,\ldots,N_Q-1$.
(a, b) Two-qubit cascade targeting $|2\rangle$, showing $P_2$.
(c, d) Three-qubit cascade targeting $|3\rangle$, showing $P_3$.
(a, c) Dependence on $Z_s$ and $\chi/g$ at $\wp/g=2$.
(b, d) Dependence on $Z_s$ and $\wp/g$ at $\chi/g=30$.
White contours mark $P_{N_Q}=0.9$. 
Other parameters are $\kappa/g=0.01$ and $\gamma/g=0.05$.
}
\label{figS1}
\end{figure}

In the multiqubit cascaded scheme, the qubit assigned to the transition $|j\rangle\rightarrow |j+1\rangle$ is not isolated from the other qubits.
Although the spectator qubits $k\neq j$ do not participate in this exchange process, their dispersive couplings shift the corresponding transition frequency through the terms $\chi_k \hat n\hat\sigma_{z,k}$~\cite{Schuster2007PhotonNumber,Gambetta2006,Boissonneault2009,LachanceQuirion2017,LachanceQuirion2020,PhysRevLett.125.117701}.
Therefore, the qubit detuning used to address the $j$-th step depends on a reference spectator configuration.
In the main text we choose this reference configuration as the fully excited one, namely $\bar z_k=+1$ for all spectator qubits.
This choice is physically motivated by the operating regime of the scheme: the qubits are continuously incoherently pumped, with $\wp_k\gg \gamma_k$, so that a spectator qubit spends most of the time in its excited state and hence has $\langle\hat\sigma_{z,k}\rangle$ close to $+1$.

To provide a numerical check of this assumption, we introduce a common spectator-compensation parameter $Z_s$ in the detuning condition,
\begin{equation}
\Delta_j
=
-(2j+1)\chi_j
+
\sum_{k\neq j} Z_s\chi_k .
\end{equation}
We then vary $Z_s$ and calculate the steady-state target population $P_{N_Q}$. 
In this test, $Z_s$ is used only as a calibration parameter for the spectator-induced dispersive shift. 
A larger value of $P_{N_Q}$ indicates a better compensation of the resonant injection steps in the cascade. 
Therefore, if the maximum target population occurs near $Z_s=+1$, the scan provides a direct numerical justification for the fully excited spectator-qubit compensation adopted in the main text.

Figure~\ref{figS1} shows a representative numerical test of the spectator-qubit compensation for the two-qubit and three-qubit cascades. 
In both cases, the high-target-population regions are localized around $Z_s\simeq 1$. 
In particular, the contours with $P_2=0.9$ and $P_3=0.9$ appear only near this value, whereas a clear deviation of $Z_s$ from unity rapidly reduces the target-state population. 
This indicates that, for the strongly pumped parameters considered here, the detuning compensation based on fully excited spectator qubits provides the most favorable calibration of the number-selective injection steps. 
The same trend is observed when either the dispersive shift $\chi$ or the incoherent pump $\wp$ is varied. 
Therefore, these representative scans support the choice $\bar z_k=+1$ used in the main-text simulations and in the analytical rate-equation model. 
When $Z_s$ is chosen away from unity, the imposed detunings compensate an incorrect spectator-induced dispersive shift, so the addressed transitions become less resonant and the cascade loses target-state selectivity.

\section{Single-qubit cascade for stabilizing the single-magnon Fock state}
\label{sec:single_magnon_additional_scans}

For a single incoherently pumped qubit, $N_Q=1$, we truncate the reduced birth--death equation at $M=N_Q+2=3$, retaining the magnon Fock states $n=0,1,2,3$. 
The only qubit has index $j=0$. 
Since there is no spectator qubit, the compensated detuning is simply $\Delta_0=-\chi_0$, and Eq.\,\eqref{eq:delta_reference} gives
\begin{equation}
\delta_0(n;\bar v_0)=-2n\chi_0 .
\label{eq:single_qubit_delta}
\end{equation}
Thus the transition $\ket{0}\leftrightarrow\ket{1}$ is resonant, whereas the higher transitions $\ket{1}\leftrightarrow\ket{2}$ and $\ket{2}\leftrightarrow\ket{3}$ are detuned by $-2\chi_0$ and $-4\chi_0$, respectively.

Using Eqs.\,\eqref{eq:lambda_def} and \eqref{eq:mu_def}, the birth and death rates reduce to
\begin{equation}
\lambda_n=q_{e,0}R_{0,n},
\qquad
\mu_{n+1}=(n+1)\kappa+q_{g,0}R_{0,n},
\qquad n=0,1,2,
\label{eq:single_lambda_mu}
\end{equation}
where
\begin{equation}
q_{e,0}=\frac{\wp_0}{\wp_0+\gamma_0},
\qquad
q_{g,0}=\frac{\gamma_0}{\wp_0+\gamma_0},
\end{equation}
and
\begin{equation}
R_{0,n}
=
\frac{2g_0^2(n+1)\Gamma_{0,n}}
{\Gamma_{0,n}^2+4n^2\chi_0^2} .
\label{eq:single_R}
\end{equation}
The steady-state recursion relation $P_{n+1}/P_n=\lambda_n/\mu_{n+1}$ therefore gives
\begin{equation}
\frac{P_{n+1}}{P_n}
=
\frac{2g_0^2\wp_0\Gamma_{0,n}}
{\kappa(\wp_0+\gamma_0)\left(\Gamma_{0,n}^2+4n^2\chi_0^2\right)
+2\gamma_0g_0^2\Gamma_{0,n}},
\qquad n=0,1,2 .
\label{eq:single_qubit_ratio}
\end{equation}
Defining $\zeta_{n+1}=P_{n+1}/P_n$ for $n=0,1,2$, the normalized populations are
\begin{align}
P_0&=\left(1+\zeta_1+\zeta_1\zeta_2+\zeta_1\zeta_2\zeta_3\right)^{-1},
\nonumber\\
P_1&=\zeta_1P_0,
\qquad
P_2=\zeta_1\zeta_2P_0,
\qquad
P_3=\zeta_1\zeta_2\zeta_3P_0 .
\label{eq:single_qubit_populations}
\end{align}
Equivalently, the target-state population can be written as
\begin{equation}
P_1^{-1}
=
1+\frac{1}{\zeta_1}+\zeta_2+\zeta_2\zeta_3 .
\label{eq:single_target_inverse}
\end{equation}
This form makes the requirements for a large $P_1$ explicit: $\zeta_1\gg1$ suppresses the vacuum contribution relative to $P_1$, whereas $\zeta_2\ll1$ and $\zeta_2\zeta_3\ll1$ suppress the population in $\ket{2}$ and $\ket{3}$.
\begin{figure}[htb]
\centerline{\includegraphics[width=9cm]{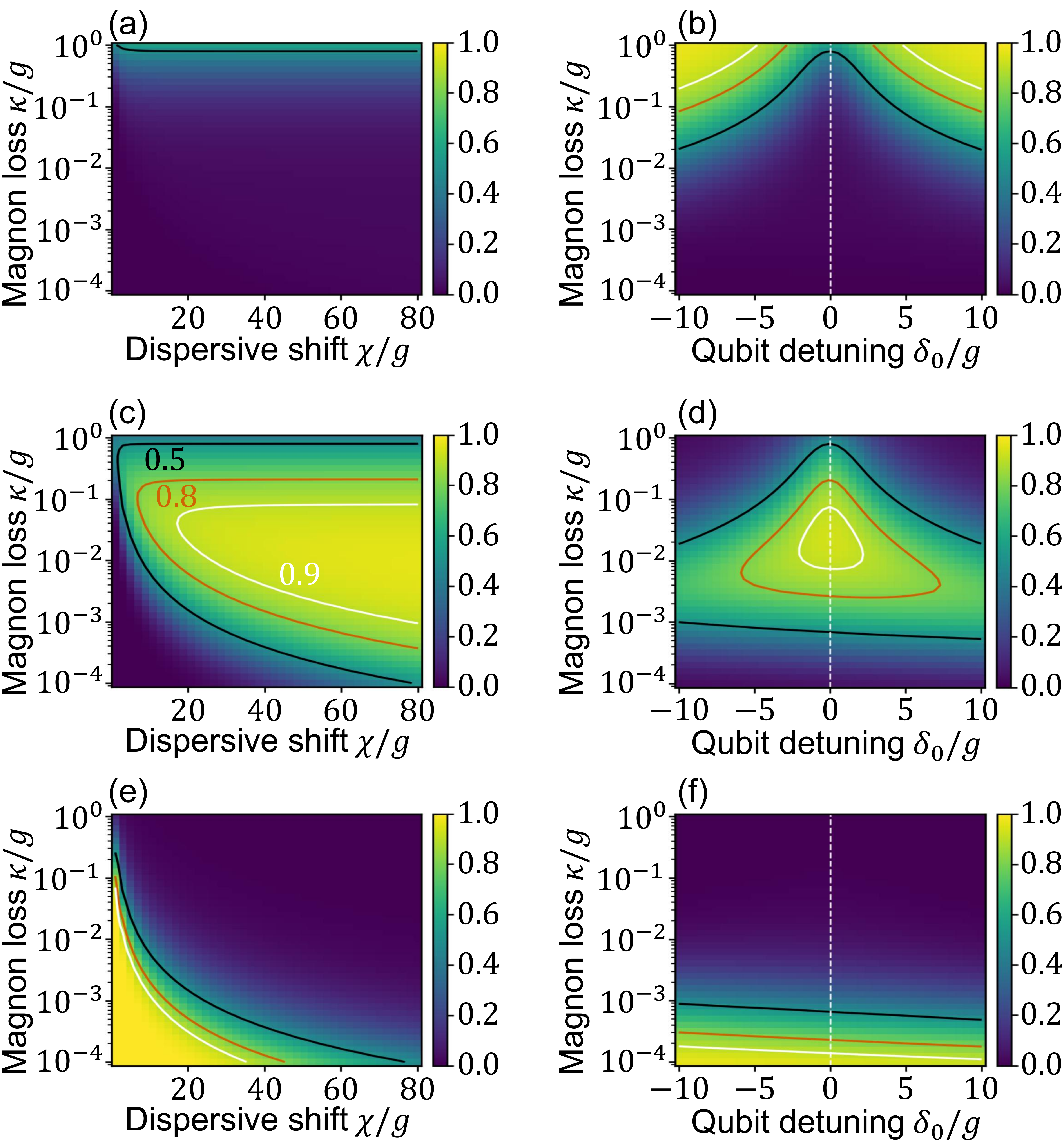}}
\caption{
Additional parameter scans for the single-magnon steady state generated by one incoherently pumped qubit. 
Color maps show the numerical steady-state populations: (a, b) $P_0$, (c, d) $P_1$, and (e, f) $P_{\geq2}=1-P_0-P_1$.
(a, c, e) Dependence on $\chi/g$ and $\kappa/g$ at the resonant detuning $\Delta_0=-\chi$.
(b, d, f) Dependence on the detuning offset $\delta_0/g$ and $\kappa/g$ at $\chi/g=30$, where $\delta_0=\Delta_0-\Delta_0^{\rm res}$ and $\Delta_0^{\rm res}=-\chi$; the vertical dashed line marks $\delta_0=0$.
Fixed parameters are $\wp/g=2$ and $\gamma/g=0.05$. 
White, red, and black contours mark population levels $0.9$, $0.8$, and $0.5$, respectively.
}
\label{figS2}
\end{figure}

The first ratio corresponds to the resonant transition $\ket{0}\leftrightarrow\ket{1}$. Setting $n=0$ in Eq.\,\eqref{eq:single_qubit_ratio} gives
\begin{equation}
\frac{P_1}{P_0}
=
\frac{2g_0^2\wp_0}
{\kappa(\wp_0+\gamma_0)\Gamma_{0,0}+2\gamma_0g_0^2} .
\label{eq:single_resonant_ratio}
\end{equation}
In the strongly pumped regime, $\gamma_0\ll\wp_0$, and when the downward process in which a ground-state qubit reabsorbs the magnon is negligible, i.e., $2\gamma_0 g_0^2\ll \kappa(\wp_0+\gamma_0)\Gamma_{0,0}$, this reduces to
\begin{equation}
\frac{P_1}{P_0}
\simeq
\frac{2g_0^2}{\kappa\Gamma_{0,0}}
\simeq
\frac{4g_0^2}{\kappa\wp_0},
\label{eq:single_resonant_ratio_asymptotic}
\end{equation}
where the last expression assumes that the pump dominates the decay rate of transition coherence, $\Gamma_{0,0}\simeq\wp_0/2$.

For the higher transitions, $n\ge1$, the qubit is off resonant. In the resolved-transition regime $4n^2\chi_0^2\gg\Gamma_{0,n}^2$, Eq.\,\eqref{eq:single_qubit_ratio} gives
\begin{equation}
\frac{P_{n+1}}{P_n}
\simeq
\frac{g_0^2\wp_0\Gamma_{0,n}}
{2\kappa(\wp_0+\gamma_0)n^2\chi_0^2},
\qquad n=1,2 .
\label{eq:single_off_resonant_ratio}
\end{equation}
For $\gamma_0\ll\wp_0$ and $\Gamma_{0,n}\simeq\wp_0/2$, this becomes
\begin{equation}
\frac{P_{n+1}}{P_n}
\simeq
\frac{g_0^2\wp_0}{4\kappa n^2\chi_0^2},
\qquad n=1,2 .
\label{eq:single_off_resonant_ratio_pump_dominated}
\end{equation}
In particular,
\begin{equation}
\frac{P_2}{P_1}
\simeq
\frac{g_0^2\wp_0}{4\kappa\chi_0^2},
\qquad
\frac{P_3}{P_2}
\simeq
\frac{g_0^2\wp_0}{16\kappa\chi_0^2} .
\label{eq:single_tail_ratios}
\end{equation}
Therefore the leading upper-tail population above the single-magnon target scales as $P_2/P_1\propto\chi_0^{-2}$, while the next tail component scales as $P_3/P_1\propto\chi_0^{-4}$. 
This is the single-qubit limit of the cascaded mechanism: the resonant injection opens the first step $\ket{0}\to\ket{1}$, whereas the following steps are suppressed by dispersive detuning~\cite{Hofheinz2008,Hofheinz2009,Deleglise2008,Schuster2007PhotonNumber}.

Equations~\eqref{eq:single_resonant_ratio_asymptotic} and \eqref{eq:single_tail_ratios} also clarify the parameter window for maximizing the single-magnon population. 
The resonant ratio $P_1/P_0$ is enhanced by a sufficiently strong qubit-induced injection rate, but it decreases when the magnon loss $\kappa$ is too large. 
By contrast, the unwanted upper-tail ratios $P_2/P_1$ and $P_3/P_2$ are suppressed by increasing the dispersive shift $\chi_0$, because the transitions above $\ket{1}$ are off resonant and scale as $\chi_0^{-2}$. 
The pump rate $\wp_0$ should be intermediate: a weak pump gives an insufficient excited-qubit population and hence weak injection into $\ket{1}$, whereas an excessively strong pump broadens the qubit--magnon transition, reduces the resonant exchange rate, and increases off-resonant leakage into $\ket{2}$ and higher Fock states. 
The magnon loss rate $\kappa$ also has an optimal finite range. 
If $\kappa$ is too small, population that leaks into $\ket{2}$ and $\ket{3}$ is not efficiently removed; if $\kappa$ is too large, the stabilized $\ket{1}$ population is depleted. 
Finally, a small qubit relaxation rate $\gamma_0$ is favorable because it maintains the pump-induced qubit inversion and suppresses qubit-induced reabsorption of the magnon.
These analytical trends are consistent with the main-text parameter scans and motivate the additional single-qubit scans below, where the roles of vacuum depletion, target-state accumulation, and upper-tail population are resolved separately.

In these scans, we separately plot the vacuum component $P_0$, the target component $P_1$, and the upper-tail population $P_{\geq2}=1-P_0-P_1$.
This decomposition makes it possible to distinguish whether the reduction of $P_1$ is caused mainly by incomplete depletion of the vacuum or by residual population in higher Fock states.

The left column shows the resonant case, $\Delta_0=-\chi$, as a function of $\chi/g$ and $\kappa/g$. 
For weak dispersive selectivity, population can leak into higher Fock states, as seen from the large $P_{\geq2}$ region in Fig.\,\ref{figS2}(e). 
Increasing $\chi/g$ suppresses this upper-tail population and therefore allows the target population $P_1$ in Fig.\,\ref{figS2}(c) to approach unity. 
However, the magnon loss rate cannot be chosen arbitrarily. 
If $\kappa$ is too small, the off-resonantly generated population in $\ket{2}$ and higher states is not efficiently removed. 
If $\kappa$ is too large, the single-magnon state is depleted by magnon decay and the system is driven back toward the vacuum, as indicated by the increase of $P_0$ in Fig.\,\ref{figS2}(a). Thus the high-$P_1$ region appears at large enough $\chi/g$ and at an intermediate range of $\kappa/g$.

The right column examines the sensitivity to the residual qubit detuning $\delta_0=\Delta_0-\Delta_0^{\rm res}$ at fixed $\chi/g=30$. 
The single-magnon population is maximized near $\delta_0=0$, where the intended transition $\ket{0}\leftrightarrow\ket{1}$ is resonant. 
When $|\delta_0|$ increases, the qubit-induced injection into $\ket{1}$ becomes less efficient, so the population is no longer concentrated in the target state. 
This reduction of $P_1$ is accompanied mainly by an increase of the vacuum component $P_0$, especially for larger $\kappa$, because loss then returns the system to $\ket{0}$ faster than the detuned qubit can repump it. 
At very small $\kappa$, the upper-tail population $P_{\geq2}$ remains visible, reflecting the fact that weak loss cannot efficiently remove population that has reached $\ket{2}$ and higher Fock states through residual off-resonant processes.

Overall, Fig.\,\ref{figS2} shows that the single-magnon steady state is limited by two different mechanisms on opposite sides of the optimal window. 
Large loss or detuning leaves too much population in the vacuum, whereas weak loss or insufficient dispersive selectivity allows population to accumulate above the target state. 
The near-unity $P_1$ region therefore results from a balance between resonant injection into $\ket{1}$, dispersive suppression of the unwanted transitions above $\ket{1}$, and a finite magnon loss rate that removes the upper-tail population without emptying the target state.

\section{Two-qubit cascade for stabilizing the two-magnon Fock state}
\label{sec:two_magnon_even_fock}

We next specialize the reduced rate equation to two incoherently pumped qubits. 
For $N_Q=2$, the analytical rate-equation description is truncated at $M=N_Q+2=4$, so that the retained magnon populations are $P_0$, $P_1$, $P_2$, $P_3$, and $P_4$. 
The qubits have indices $j=0$ and $j=1$, with qubit $0$ assigned to $\ket{0}\leftrightarrow\ket{1}$ and qubit $1$ assigned to $\ket{1}\leftrightarrow\ket{2}$. 
At the reference spectator configuration, Eq.\,\eqref{eq:delta_reference} gives
\begin{equation}
\delta_0(n;\bar v_0)=-2n\chi_0,
\qquad
\delta_1(n;\bar v_1)=2(1-n)\chi_1 .
\label{eq:two_qubit_detunings}
\end{equation}
For identical dispersive shifts and fully excited spectator compensation, this corresponds to the main-text detunings $\Delta_0=0$ and $\Delta_1=-2\chi$.

For compactness, we define the nearest-neighbor population ratios
\begin{equation}
\zeta_m\equiv\frac{P_m}{P_{m-1}}
=
\frac{\lambda_{m-1}}{\mu_m},
\qquad m=1,2,3,4 .
\label{eq:two_eta_def}
\end{equation}
Using Eqs.\,\eqref{eq:lambda_def} and \eqref{eq:mu_def}, these ratios are
\begin{equation}
\zeta_m
=
\frac{q_{e,0}R_{0,m-1}+q_{e,1}R_{1,m-1}}
{m\kappa+q_{g,0}R_{0,m-1}+q_{g,1}R_{1,m-1}},
\qquad m=1,2,3,4 .
\label{eq:two_qubit_exact_ratios}
\end{equation}
The normalized populations are then
\begin{align}
P_0&=
\left(1+\zeta_1+\zeta_1\zeta_2+\zeta_1\zeta_2\zeta_3+\zeta_1\zeta_2\zeta_3\zeta_4\right)^{-1},
\nonumber\\
P_1&=\zeta_1P_0,
\qquad
P_2=\zeta_1\zeta_2P_0,
\qquad
P_3=\zeta_1\zeta_2\zeta_3P_0,
\qquad
P_4=\zeta_1\zeta_2\zeta_3\zeta_4P_0 .
\label{eq:two_qubit_populations}
\end{align}
Equivalently, the two-magnon target population satisfies
\begin{equation}
P_2^{-1}
=
1+\frac{1}{\zeta_2}+\frac{1}{\zeta_1\zeta_2}+\zeta_3+\zeta_3\zeta_4 .
\label{eq:two_target_inverse}
\end{equation}
Thus a large $P_2$ requires $\zeta_1\gg1$ and $\zeta_2\gg1$ along the addressed part of the cascade, together with $\zeta_3\ll1$ and $\zeta_3\zeta_4\ll1$ for the two retained states above the target.

The first ratio, $\zeta_1=P_1/P_0$, corresponds to the transition $\ket{0}\leftrightarrow\ket{1}$. 
Qubit $0$ is resonant with this transition, whereas qubit $1$ is detuned by $2\chi_1$. 
In the resolved-transition regime, the off-resonant contribution from qubit $1$ is small, and
\begin{equation}
\frac{P_1}{P_0}
\simeq
\frac{q_{e,0}R_{0,0}}
{\kappa+q_{g,0}R_{0,0}}
=
\frac{2q_{e,0}g_0^2/\Gamma_{0,0}}
{\kappa+2q_{g,0}g_0^2/\Gamma_{0,0}} .
\label{eq:two_first_ratio}
\end{equation}
If the downward process in which qubit $0$ reabsorbs the magnon on the addressed $\ket{0}\leftrightarrow\ket{1}$ transition is negligible compared with magnon loss, i.e., $2q_{g,0}g_0^2/\Gamma_{0,0}\ll\kappa$, this further reduces to
\begin{equation}
\frac{P_1}{P_0}
\simeq
\frac{2q_{e,0}g_0^2}{\kappa\Gamma_{0,0}} .
\label{eq:two_first_ratio_simplified}
\end{equation}

The second ratio, $\zeta_2=P_2/P_1$, corresponds to the transition $\ket{1}\leftrightarrow\ket{2}$. Qubit $1$ is resonant with this transition, while qubit $0$ is detuned by $-2\chi_0$. Keeping the dominant resonant contribution gives
\begin{equation}
\frac{P_2}{P_1}
\simeq
\frac{q_{e,1}R_{1,1}}
{2\kappa+q_{g,1}R_{1,1}}
=
\frac{4q_{e,1}g_1^2/\Gamma_{1,1}}
{2\kappa+4q_{g,1}g_1^2/\Gamma_{1,1}} .
\label{eq:two_second_ratio}
\end{equation}
\begin{figure}[htb]
\centerline{\includegraphics[width=9cm]{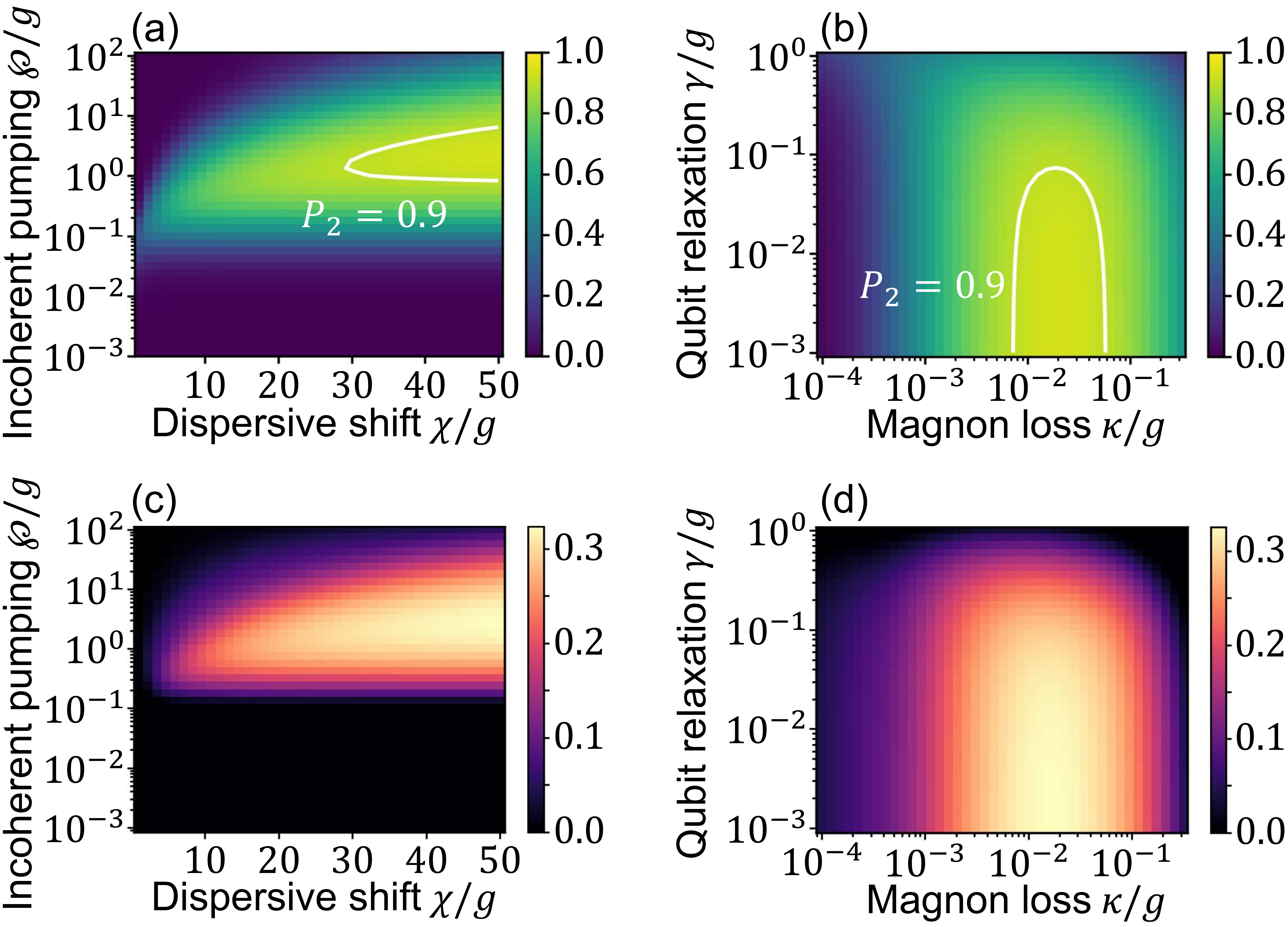}}
\caption{Two-magnon target state in the two-qubit cascade. 
Color maps show numerical steady-state results for (a, b) the target population $P_2=\langle2|\rho_m|2\rangle$ and (c, d) the integrated Wigner negativity $\mathcal{N}_{\rm neg}$. 
The two qubits are tuned by $\Delta_j=(N_Q-2j-2)\chi$ for $j=0,\ldots,N_Q-1$ with $N_Q=2$.
(a, c) Dependence on $\chi/g$ and $\wp/g$, with $\kappa/g=0.01$ and $\gamma/g=0.05$.
(b, d) Dependence on $\kappa/g$ and $\gamma/g$, with $\chi/g=30$ and $\wp/g=2$.
White contours in (a, b) mark $P_2=0.9$.
}
\label{figS3}
\end{figure}
If the downward process in which qubit $1$ reabsorbs the magnon on the addressed $\ket{1}\leftrightarrow\ket{2}$ transition is negligible compared with magnon loss, i.e., $4q_{g,1}g_1^2/\Gamma_{1,1}\ll2\kappa$, then
\begin{equation}
\frac{P_2}{P_1}
\simeq
\frac{2q_{e,1}g_1^2}{\kappa\Gamma_{1,1}} .
\label{eq:two_second_ratio_simplified}
\end{equation}
In the strongly pumped limit, where $q_{e,j}\simeq1$ and the coherence decay rate is dominated by the incoherent pump, $\Gamma_{j,j}\simeq\wp_j/2$, Eqs.\,\eqref{eq:two_first_ratio_simplified} and \eqref{eq:two_second_ratio_simplified} become
\begin{equation}
\frac{P_1}{P_0}
\simeq
\frac{4g_0^2}{\kappa\wp_0},
\qquad
\frac{P_2}{P_1}
\simeq
\frac{4g_1^2}{\kappa\wp_1} .
\label{eq:two_resonant_ratios_pump_dominated}
\end{equation}
The two addressed steps are therefore not directly suppressed by the dispersive detuning.

We finally estimate the leading population above the target state $\ket{2}$. 
The ratio $\zeta_3=P_3/P_2$ corresponds to the unaddressed transition $\ket{2}\leftrightarrow\ket{3}$. Its detunings are $\delta_1(2;\bar v_1)=-2\chi_1$ and $\delta_0(2;\bar v_0)=-4\chi_0$. 
Hence, for $|\chi_j|\gg\Gamma_{j,n}$, it becomes
\begin{equation}
\frac{P_3}{P_2}
\simeq
\frac{q_{e,1}g_1^2\Gamma_{1,2}}{2\kappa\chi_1^2}
+
\frac{q_{e,0}g_0^2\Gamma_{0,2}}{8\kappa\chi_0^2},
\label{eq:two_first_leakage_ratio}
\end{equation}
where the denominator has been approximated by the magnon-loss contribution $\mu_3\simeq3\kappa$. 
The next ratio $\zeta_4=P_4/P_3$ corresponds to $\ket{3}\leftrightarrow\ket{4}$, with detunings $\delta_1(3;\bar v_1)=-4\chi_1$ and $\delta_0(3;\bar v_0)=-6\chi_0$. With $\mu_4\simeq4\kappa$, one obtains
\begin{equation}
\frac{P_4}{P_3}
\simeq
\frac{q_{e,1}g_1^2\Gamma_{1,3}}{8\kappa\chi_1^2}
+
\frac{q_{e,0}g_0^2\Gamma_{0,3}}{18\kappa\chi_0^2} .
\label{eq:two_second_leakage_ratio}
\end{equation}
If the pump dominates the coherence decay rates and $q_{e,j}\simeq1$, these become
\begin{equation}
\frac{P_3}{P_2}
\simeq
\frac{g_1^2\wp_1}{4\kappa\chi_1^2}
+
\frac{g_0^2\wp_0}{16\kappa\chi_0^2},
\qquad
\frac{P_4}{P_3}
\simeq
\frac{g_1^2\wp_1}{16\kappa\chi_1^2}
+
\frac{g_0^2\wp_0}{36\kappa\chi_0^2} .
\label{eq:two_leakage_ratios_pump_dominated}
\end{equation}
Thus the leading leakage above the two-magnon target scales as $P_3/P_2\propto\chi^{-2}$, while the next retained tail scales as $P_4/P_2\propto\chi^{-4}$ for comparable dispersive shifts. This is the two-qubit cascaded limit: the first two number-selective transitions are resonantly opened, while the first unaddressed transition above $\ket{2}$ is dispersively suppressed.

Figure~\ref{figS3} provides a numerical benchmark of the two-qubit analytical picture discussed above. 
In the reduced rate-equation description, a large two-magnon population requires the two addressed ratios $\zeta_1=P_1/P_0$ and $\zeta_2=P_2/P_1$ to be larger than unity, while the upper-tail ratios $\zeta_3=P_3/P_2$ and $\zeta_4=P_4/P_3$ must remain small. 
The maps of $P_2$ in Figs.\,\ref{figS3}(a,b) show precisely this balance. 
In Fig.\,\ref{figS3}(a), increasing the dispersive shift $\chi$ enlarges the high-$P_2$ region because the first unaddressed transition above the target, $\ket{2}\rightarrow\ket{3}$, becomes more strongly detuned. 
This is consistent with the analytical estimates $P_3/P_2\propto\chi^{-2}$ and $P_4/P_2\propto\chi^{-4}$. 
At the same time, the incoherent pump $\wp$ cannot be chosen arbitrarily. 
For weak pumping, the qubits are not sufficiently inverted and the two resonant injection steps are inefficient. 
For overly strong pumping, the transition coherence decay rates increase, which weakens the resonant exchange and enhances residual off-resonant excitation above $\ket{2}$. 
The high-$P_2$ region therefore appears at large enough $\chi/g$ and at an intermediate pump strength.

Figure~\ref{figS3}(b) shows the complementary dependence on magnon loss and qubit relaxation. 
A small qubit relaxation rate $\gamma$ is favorable because it preserves the pump-induced inversion of both qubits and suppresses qubit-induced reabsorption on the addressed transitions. 
The dependence on $\kappa$ is nonmonotonic. 
If $\kappa$ is too large, magnon decay removes population from the target state $\ket{2}$ before the cascade can replenish it. 
If $\kappa$ is too small, population that has leaked into $\ket{3}$ and higher Fock states through off-resonant transitions is not efficiently returned downward. 
The optimal region is therefore a finite window of magnon loss, where leakage above the target is removed without strongly depleting the stabilized two-magnon component.

The Wigner-negativity maps in Figs.\,\ref{figS3}(c,d) follow the same operating window as the population maps. 
This correspondence is expected because the reduced magnon state becomes close to an even Fock state only when the population is concentrated near $\ket{2}$ and the upper tail remains small. 
Unlike the odd-parity single-magnon state, the Wigner function of an even-parity Fock state is not negative at the phase-space origin; nevertheless, it has negative annular regions, which are captured by the integrated Wigner negativity $\mathcal{N}_{\rm neg}$~\cite{Wigner1932,Cahill1969,Kenfack2004,Lutterbach1997,Deleglise2008}. 
The simultaneous enhancement of $P_2$ and $\mathcal{N}_{\rm neg}$ therefore supports the interpretation that the two-qubit cascade stabilizes a nonclassical even-Fock magnon state, rather than merely producing a broadened incoherent mixture.

\section{Steady states and qubit-parameter inhomogeneity}
\label{sec:representative_states_inhomogeneity}

\begin{figure}[htb]
\centerline{\includegraphics[width=9cm]{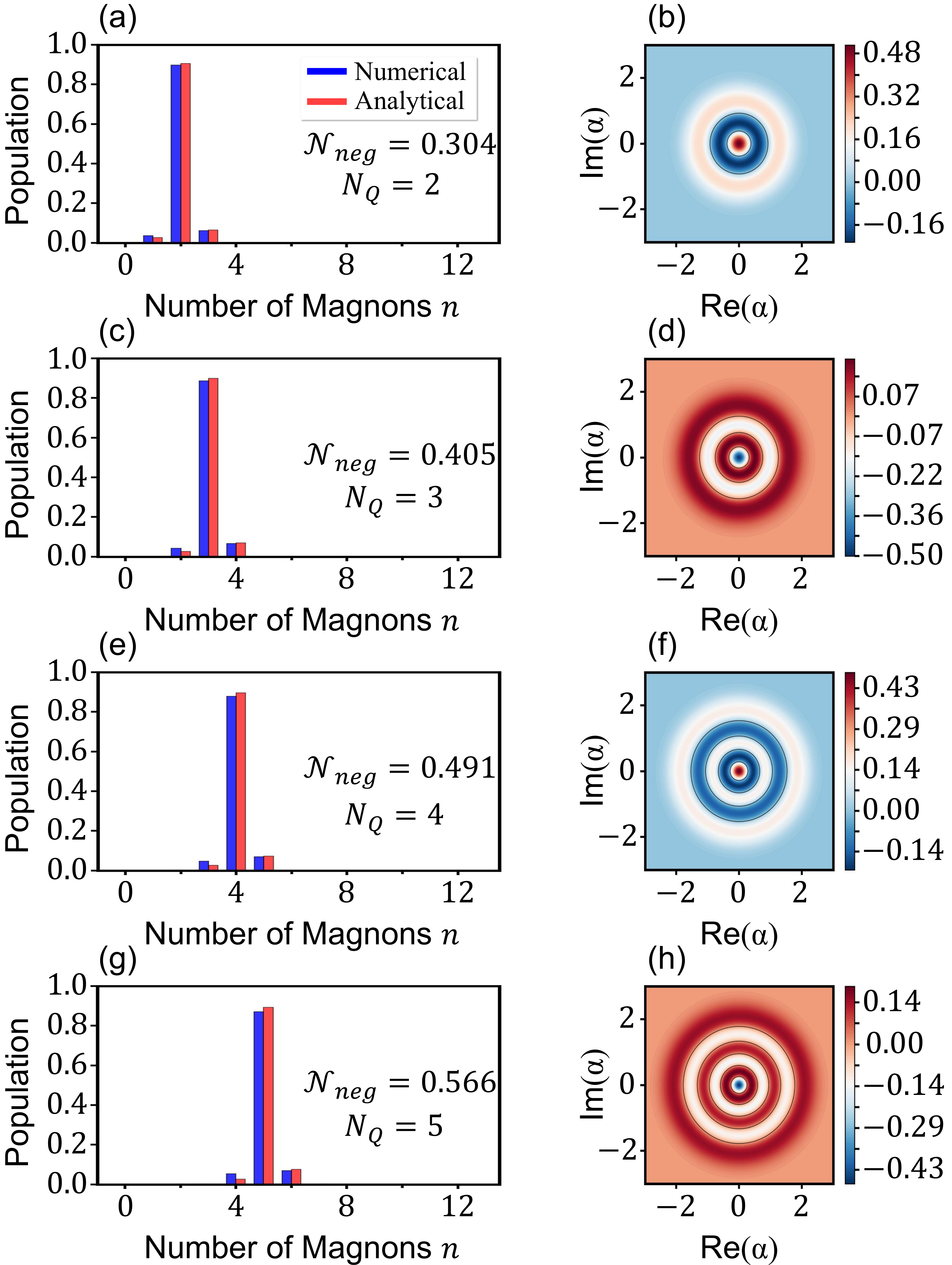}}
\caption{
Representative higher-magnon Fock-state steady states generated by the multiqubit cascade. 
Rows correspond to $N_Q=2,3,4,5$, targeting $|N_Q\rangle$.
(a, c, e, g) Magnon populations $P_n$, with blue and red bars denoting numerical and analytical results, respectively.
(b, d, f, h) Numerical Wigner functions $W(\alpha)$ of the corresponding reduced magnon states. 
The integrated negativities are $\mathcal{N}_{\rm neg}=0.304$, $0.405$, $0.491$, and $0.566$ for $N_Q=2$, $3$, $4$, and $5$, respectively.
Common parameters are $\chi/g=30$, $\kappa/g=0.01$, $\gamma/g=0.05$, and $\wp/g=2$; the qubits are tuned by $\Delta_j=(N_Q-2j-2)\chi$ for $j=0,\ldots,N_Q-1$.
}
\label{figS4}
\end{figure}
\begin{figure}[htb]
\centerline{\includegraphics[width=9cm]{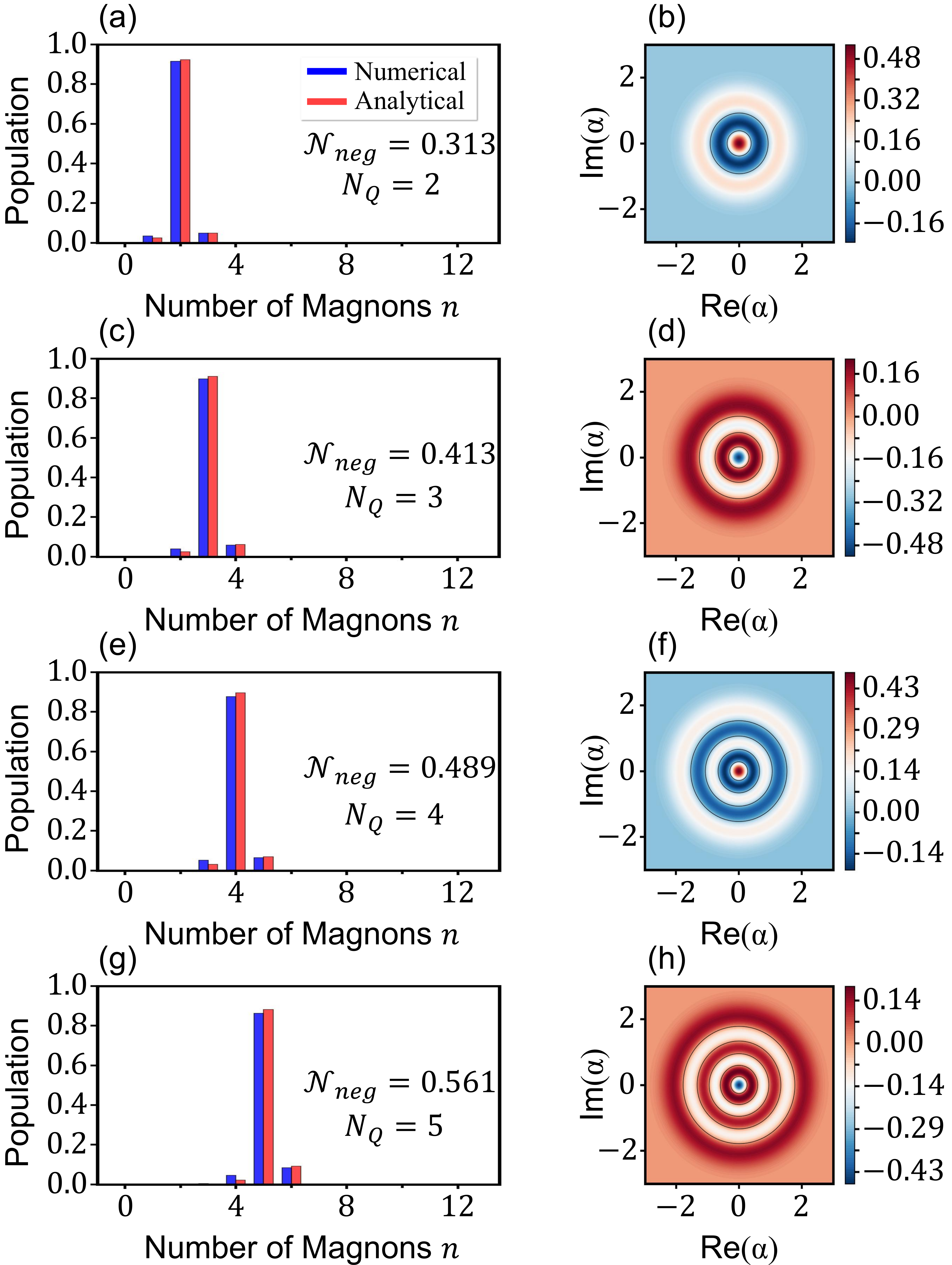}}
\caption{
Robustness against qubit-parameter inhomogeneity. Rows correspond to $N_Q=2,3,4,5$, targeting $|N_Q\rangle$. 
(a, c, e, g) Magnon populations $P_n$, with blue and red bars denoting numerical and analytical results, respectively. 
(b, d, f, h) Numerical Wigner functions $W(\alpha)$ of the corresponding reduced magnon states. 
The integrated negativities are $\mathcal{N}_{\rm neg}=0.313$, $0.413$, $0.489$, and $0.561$ for $N_Q=2$, $3$, $4$, and $5$, respectively. 
All rates are normalized by $g_0$; common parameters are $\kappa/g_0=0.01$. 
The detunings are chosen using the $\bar z_k=1$ compensation, qubit detunings $\Delta_j=-(2j+1)\chi_j+\sum_{k\neq j}\chi_k$. 
For each $N_Q$, the entries of the following vectors are ordered by the qubit index $j=0,\ldots,N_Q-1$, so that the $j$th entry gives the parameter of qubit $j$: for $N_Q=2$, $(g_j/g_0)=(1, 0.87)$, $(\chi_j/g_0)=(30.0, 32.5)$, $(\gamma_j/g_0)=(0.05, 0.043)$, and $(\wp_j/g_0)=(2.0, 2.18)$; for $N_Q=3$, $(g_j/g_0)=(1, 1.12, 0.91)$, $(\chi_j/g_0)=(30.0, 26.8, 34.2)$, $(\gamma_j/g_0)=(0.05, 0.058, 0.046)$, and $(\wp_j/g_0)=(2.0, 1.76, 2.24)$; for $N_Q=4$, $(g_j/g_0)=(1, 0.90, 1.10, 0.95)$, $(\chi_j/g_0)=(30.0, 27.0, 33.0, 29.0)$, $(\gamma_j/g_0)=(0.05, 0.06, 0.045, 0.055)$, and $(\wp_j/g_0)=(2.0, 1.8, 2.2, 1.9)$; for $N_Q=5$, $(g_j/g_0)=(1, 0.94, 1.08, 0.89, 1.15)$, $(\chi_j/g_0)=(30.0, 28.4, 33.1, 26.7, 31.6)$, $(\gamma_j/g_0)=(0.05, 0.057, 0.044, 0.061, 0.047)$, and $(\wp_j/g_0)=(2.0, 1.85, 2.16, 1.72, 2.28)$.
}
\label{figS5}
\end{figure}

In this section, we provide two complementary checks of the multiqubit cascaded-pumping mechanism.
First, we present representative steady-state magnon distributions and Wigner functions to visualize the Fock-space and phase-space structure of the stabilized states~\cite{Wigner1932,Cahill1969,Kenfack2004,Lutterbach1997,Deleglise2008}.
Second, we examine the sensitivity of the mechanism to qubit-parameter nonuniformity, which is an important consideration for realistic hybrid magnonic platforms.

Figure~\ref{figS4} displays the steady-state Fock-state populations and the corresponding Wigner functions for representative higher-Fock target states generated by the multiqubit cascade. 
The population distributions are strongly localized around the intended target states $\ket{N_Q}$, indicating that the cascade produces selective population accumulation rather than an uncontrolled spread over the magnon ladder. 
The analytical birth--death model also reproduces the dominant numerical population peaks, supporting the rate-equation interpretation of the stabilization mechanism. 
The associated Wigner functions exhibit negative regions in phase space, providing a direct visualization of the nonclassicality quantified by the integrated Wigner negativity $\mathcal{N}_{\rm neg}$~\cite{Wigner1932,Kenfack2004,Veitch2013,Chabaud2021WitnessingWN}. 
These examples therefore connect the population-based diagnosis, $P_n$, with the phase-space diagnosis, $\mathcal{N}_{\rm neg}$, and illustrate that the operating windows identified above correspond to Wigner-negative magnon steady states.

Figure~\ref{figS5} further examines the effect of qubit-parameter inhomogeneity using representative multiqubit examples. 
In the main simulations, identical or nearly identical qubit parameters are used to expose the underlying mechanism most clearly. 
The essential requirement, however, is not strict parameter uniformity, but rather that each qubit remains predominantly resonant with its assigned number-selective transition after the appropriate detuning compensation. 
Variations in $g_j$, $\chi_j$, $\wp_j$, or $\gamma_j$ mainly change the effective rate and resonance condition of each addressed step, provided that the transitions remain spectrally resolved and the upward population flow along the cascade is maintained.
The results in Fig.\,\ref{figS5} show that the target-state localization persists for the representative inhomogeneous parameter sets considered here. 
For $N_Q=2,3,4,5$, the magnon population remains concentrated near the intended target state $\ket{N_Q}$, and the analytical birth--death model continues to reproduce the main population distribution. 
The corresponding Wigner functions also retain clear negative regions, with integrated negativities comparable to the homogeneous-parameter cases. 
The target state would be degraded if the parameter variations became large enough to shift an addressed transition away from resonance, weaken a required injection step, or enhance off-resonant excitation above the target. 
Thus, Fig.\,\ref{figS5} indicates that the mechanism does not require perfectly identical qubits for these representative cases, although the tolerance remains finite and is set by the same resolved-transition and upward-bias conditions discussed above.

\begin{figure}[htb]
\centerline{\includegraphics[width=9cm]{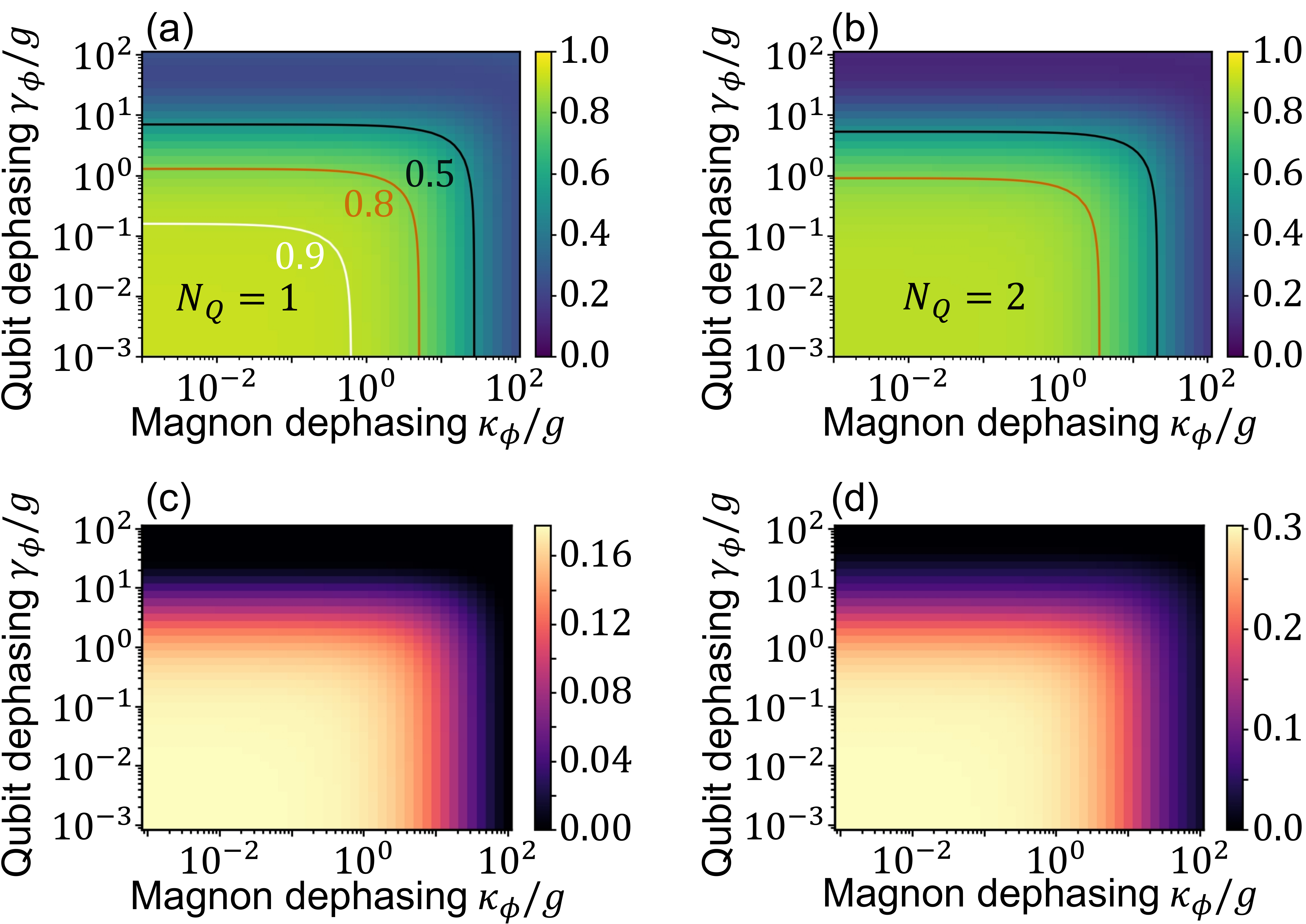}}
\caption{
Robustness against pure dephasing. 
Color maps show numerical steady-state results versus the magnon and qubit pure-dephasing rates, $\kappa_\phi/g$ and $\gamma_\phi/g$.
(a, c) Single-qubit case targeting $|1\rangle$: (a) target population $P_1$ and (c) integrated Wigner negativity $\mathcal{N}_{\rm neg}$.
(b, d) Two-qubit cascade targeting $|2\rangle$: (b) target population $P_2$ and (d) $\mathcal{N}_{\rm neg}$.
Qubit detunings are $\Delta_j=(N_Q-2j-2)\chi$ for $j=0,\ldots,N_Q-1$, with $N_Q=1$ in (a,c) and $N_Q=2$ in (b,d).
Common parameters are $\chi/g=30$, $\kappa/g=0.01$, $\gamma/g=0.05$, and $\wp/g=2$.
In (a, b), black, red, and white contours mark target-population levels $0.5$, $0.8$, and $0.9$, respectively. 
}
\label{figS6}
\end{figure}

\begin{figure}[htb]
\centerline{\includegraphics[width=9cm]{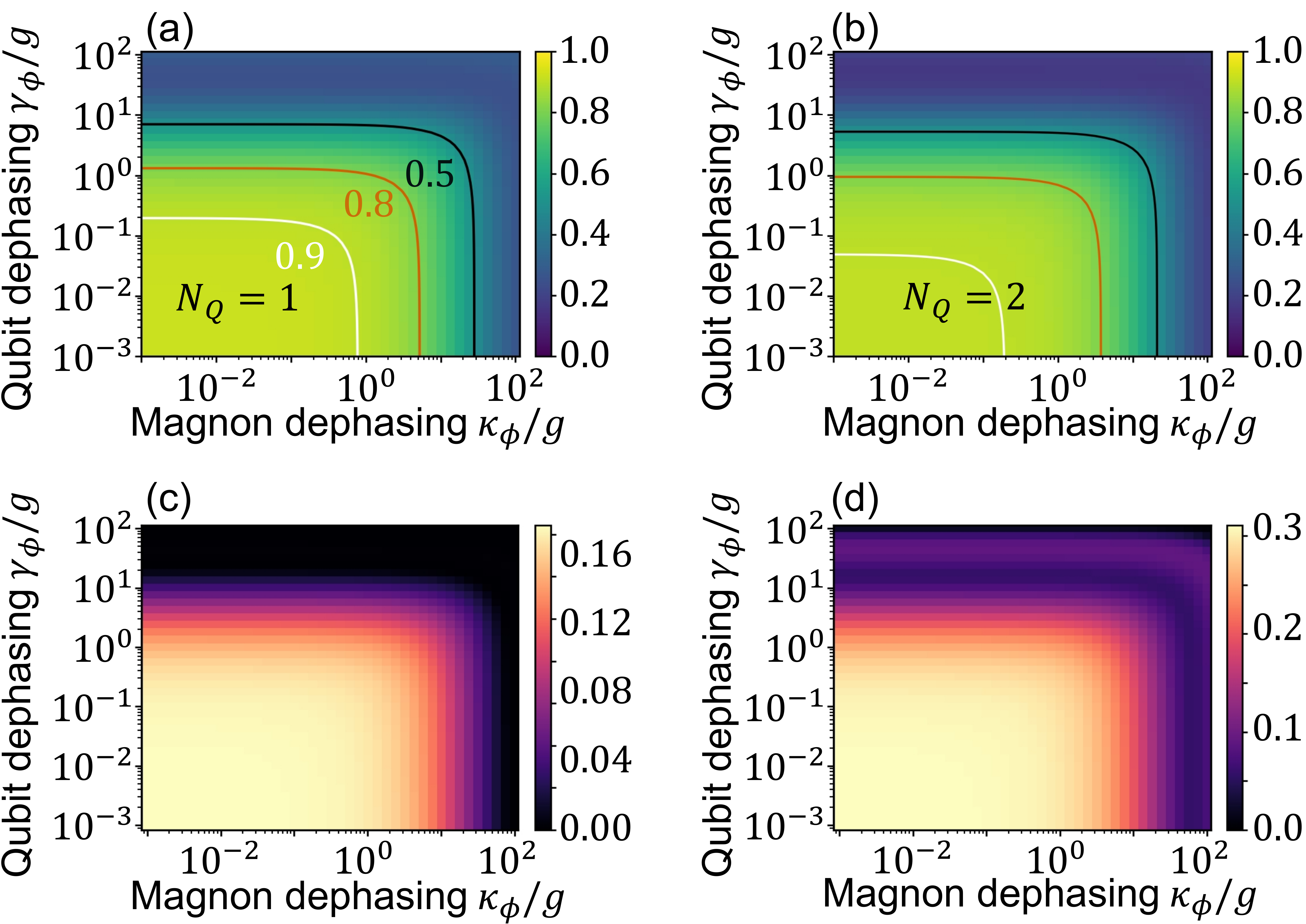}}
\caption{Analytical verification of the dephasing robustness shown in Fig.\,\ref{figS6}. 
Color maps are obtained from the steady-state birth--death rate equation using the same parameters as in Fig.\,\ref{figS6}.
(a, c) Single-qubit case targeting $|1\rangle$: (a) target population $P_1$ and (c) integrated Wigner negativity $\mathcal{N}_{\rm neg}$.
(b, d) Two-qubit cascade targeting $|2\rangle$: (b) target population $P_2$ and (d) $\mathcal{N}_{\rm neg}$.
In (a,b), black, red, and white contours mark target-population levels $0.5$, $0.8$, and $0.9$, respectively. 
}
\label{figS7}
\end{figure}

\section{Robustness against pure dephasing and analytical validation}
\label{sec:dephasing_robustness}

We now examine the effect of pure dephasing on the stabilized magnon Fock states~\cite{Breuer2007,RevModPhys.93.025005,Kjaergaard2020,ZareRameshti2022,Yuan2022}. 
In this section we take a homogeneous qubit pure-dephasing rate, $\gamma_\phi=\gamma_{\phi,j}$, for all qubits.
Figure~\ref{figS6} shows the target-state population and the integrated Wigner negativity as functions of the magnon and qubit pure-dephasing rates, $\kappa_\phi/g$ and $\gamma_\phi/g$. 
The upper panels show the stabilized Fock-state populations for the single-qubit and two-qubit cases, while the lower panels show the corresponding Wigner negativities. 
Both diagnostics remain sizable in the low-dephasing region, indicating that the cascaded stabilization mechanism is not restricted to the ideal zero-dephasing limit. 
However, the maps also show that qubit pure dephasing is particularly detrimental: increasing $\gamma_\phi$ broadens the qubit-assisted transitions, weakens the number-selective exchange, and rapidly suppresses both the target-state population and the Wigner negativity. Magnon pure dephasing is tolerated over a finite range, but sufficiently large $\kappa_\phi$ also degrades the stabilized state by reducing the phase-space nonclassicality and eventually lowering the target population. 
The similar boundaries of the high-$P_{N_Q}$ and high-$\mathcal{N}_{\rm neg}$ regions show that the loss of Wigner negativity follows the same degradation of Fock-state localization caused by dephasing.

The degradation at large dephasing rates can be understood from the effective rate-equation picture. 
Pure dephasing increases the decay rate of the coherence associated with each qubit-assisted transition and therefore enters the effective exchange rate [see Eq.\,\eqref{eq:R_local}]
\begin{equation}
R_{j,n}(v_j)
=
\frac{2g_j^2(n+1)\Gamma_{j,n}}
{\Gamma_{j,n}^2+\delta_j^2(n;v_j)} ,
\end{equation}
where $\Gamma_{j,n}$ includes relaxation, incoherent pumping, magnon damping, and pure-dephasing contributions. 
On the addressed resonant transition, $\delta_j(n; v_j)=0$, this rate becomes
\begin{equation}
R_{j,n}^{\rm res}
=
\frac{2g_j^2(n+1)}{\Gamma_{j,n}} ,
\end{equation}
so increasing $\gamma_{\phi,j}$ reduces the resonant injection rate and weakens the upward cascade. By contrast, for an undesired off-resonant transition in the resolved regime, $|\delta_j(n)|\gg\Gamma_{j,n}$, one has
\begin{equation}
R_{j,n}^{\rm off}
\simeq
\frac{2g_j^2(n+1)\Gamma_{j,n}}{\delta_j^2(n)} .
\end{equation}
Thus a larger coherence decay rate increases the residual off-resonant exchange into undesired Fock states. 
Pure dephasing therefore degrades the stabilized state in two complementary ways: it suppresses the intended resonant injection steps and enhances leakage through off-resonant transitions. 
These effects explain the simultaneous decrease of the target-state population and the integrated Wigner negativity at large dephasing rates.

\begin{figure}[htb]
\centerline{\includegraphics[width=9cm]{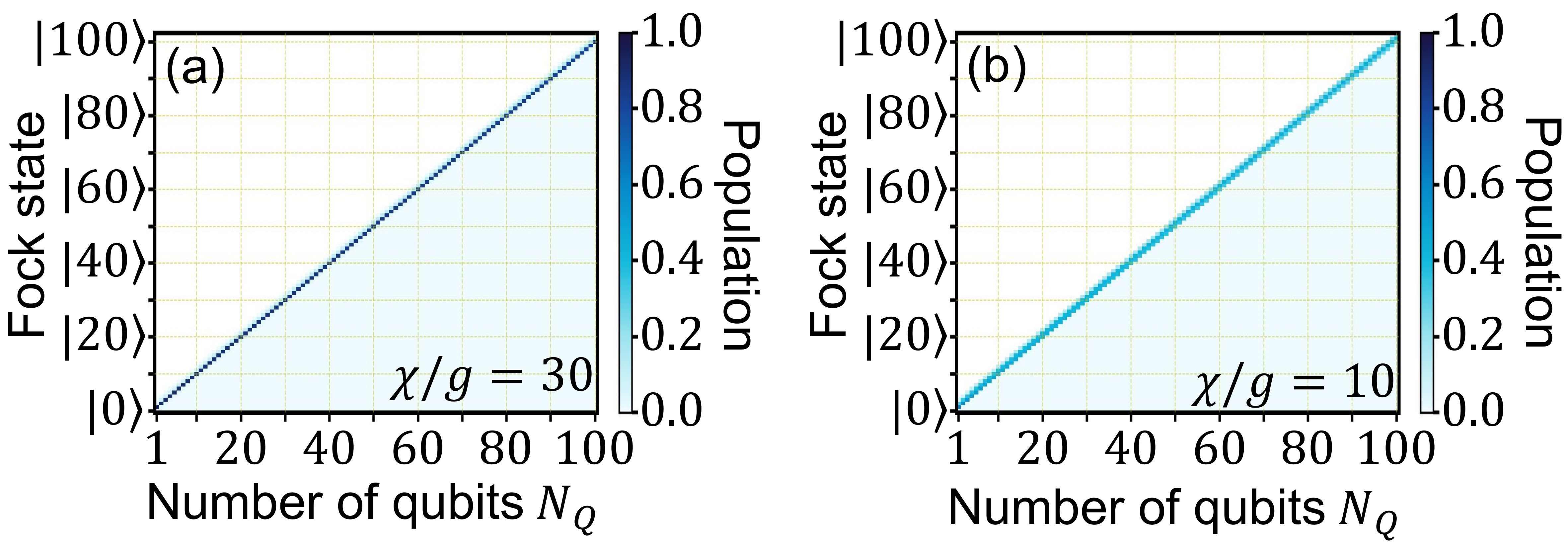}}
\caption{Analytical rate-equation extrapolation to the stabilization of higher magnon Fock states.
Color maps show the steady-state magnon populations $P_n$ versus the number of qubits $N_Q$ and the magnon number $n$, with $N_Q=1,\ldots,100$. For each $N_Q$, the qubits are tuned by the ladder condition $\Delta_j=(N_Q-2j-2)\chi$ for $j=0,\ldots,N_Q-1$, so that the target relation is $n=N_Q$.
(a) $\chi/g=30$.
(b) $\chi/g=10$.
Common parameters are $\wp/g=2$, $\kappa/g=0.01$, and $\gamma/g=0.05$. 
Dashed yellow lines mark intervals of ten in both $N_Q$ and $n$.
}
\label{figS8}
\end{figure}

Figure~\ref{figS7} provides an analytical counterpart to the full master-equation results in Fig.\,\ref{figS6}. 
The purpose is to check whether the effective birth--death rate equation can reproduce the dephasing dependence observed in the numerical simulations. 
Using the same parameters as in Fig.\,\ref{figS6}, we first solve the steady-state rate equation to obtain the Fock-state populations $P_n$. 
Since the rate equation contains only diagonal magnon populations, the Wigner function used for the analytical estimate is reconstructed as
\begin{equation}
W_{\rm bd}(\alpha)
=
\sum_{n=0}^{M} P_n W_n(\alpha),
\end{equation}
where
\begin{equation}
W_n(\alpha)
=
\frac{2}{\pi}(-1)^n
e^{-2|\alpha|^2}
L_n(4|\alpha|^2)
\end{equation}
is the Wigner function of the Fock state $\ket{n}$, where $L_n(x)$ is the Laguerre polynomial~\cite{Wigner1932,Cahill1969,Kenfack2004}. 
The integrated Wigner negativity $\mathcal{N}_{\rm neg}$ is then evaluated from $W_{\rm bd}(\alpha)$ using the same definition as in the main text. 
This Fock-diagonal construction neglects off-diagonal magnon coherences, but it is sufficient for testing whether the birth--death model captures the population-driven degradation of Wigner negativity.

The analytical maps in Fig.\,\ref{figS7} reproduce the main features of the full master-equation results in Fig.\,\ref{figS6}. 
In both the single-qubit and two-qubit cases, the target-state population remains large in the low-dephasing region and decreases when either the magnon or qubit dephasing rate becomes too large. 
The analytical Wigner-negativity maps also follow the same overall boundaries as the numerical results, showing that the reduction of $\mathcal{N}_{\rm neg}$ is largely governed by the redistribution of Fock-state populations. 
Small quantitative differences are expected because the birth--death treatment neglects off-diagonal magnon coherences and higher-order correlations, but the agreement in the operating-window structure supports the rate-equation interpretation of the dephasing response.

\section{Rate-equation extrapolation to higher magnon Fock states}
\label{sec:higher_fock_extrapolation}

In the main text and in the preceding sections, the cascaded incoherent-pumping mechanism was examined with full master-equation simulations for low target Fock states. 
Here we use the analytical birth--death rate equation to extrapolate the same mechanism to higher target magnon Fock states~\cite{GardinerZoller2004,Carmichael1999,WallsMilburn2008,Hofheinz2008,Hofheinz2009}.
This extrapolation is a rate-equation scaling test, not a full many-qubit master-equation prediction.

In this section, the extrapolation is used as an idealized rate-equation benchmark of the intrinsic number selectivity of the cascaded mechanism. 
The spectator configuration entering the compensated detunings is fixed to the reference configuration $\bar z_k=+1$, and its configuration weight is absorbed into the effective calibrated rates rather than treated as a product of independent spectator excitation probabilities; the plotted distributions are normalized in the full truncated basis with $M=N_{\rm Q}+2$, including the two retained upper-tail states above the target.

For a target state $\ket{N_Q}$, the scheme uses $N_Q$ incoherently pumped qubits to form a sequence of number-selective injection steps,
$\ket{0}\rightarrow \ket{1}\rightarrow\cdots\rightarrow \ket{N_Q}$.
Within the effective rate-equation description, each addressed step is characterized by a resonant upward transition rate, whereas population above the target state is controlled by off-resonant excitation rates and magnon loss. 
The predicted higher-Fock-state performance is therefore governed by the same ingredients identified in the full numerical simulations: efficient resonant injection along the addressed part of the ladder, sufficiently large dispersive separation $\chi$, and a finite loss channel that removes population in Fock states above the target without strongly depleting the target state itself.

Figure~\ref{figS8} shows the analytical birth--death prediction for the magnon Fock-state populations at larger target numbers. 
In contrast to the full master-equation calculations, which are limited to relatively small $N_Q$ by the rapidly increasing Hilbert-space dimension, the reduced rate equation allows us to extrapolate the cascaded mechanism up to $N_Q=100$. 
The population maximum follows the diagonal relation $n=N_Q$, indicating that adding one more incoherently pumped qubit adds one more number-selective injection step and shifts the selected magnon Fock state upward by one quantum.

The comparison between Figs.\,\ref{figS8}(a) and \ref{figS8}(b) highlights the role of dispersive selectivity in this extrapolation. 
For $\chi/g=30$, the population remains sharply concentrated along the target diagonal over the full range shown. 
For the smaller value $\chi/g=10$, the population is still centered near $n=N_Q$, but the distribution is visibly broader because the off-resonant transitions above and below the addressed ladder steps are less strongly suppressed. 
This behavior is consistent with the analytical leakage estimates derived above, where the upper-tail ratios scale inversely with powers of $\chi$.

These results should be interpreted as a rate-equation extrapolation rather than as a full many-qubit master-equation simulation. 
The calculation assumes that the number-selective transitions remain spectrally resolved, that diagonal Fock-state populations dominate the reduced magnon state, and that the effective transition rates continue to provide an upward bias along the addressed part of the ladder. Within these assumptions, Fig.\,\ref{figS8} suggests that the cascaded incoherent-pumping mechanism is not intrinsically limited to the first few magnon Fock states, although quantitative performance at very large $N_Q$ would depend on device-specific effects such as parameter crowding, detuning calibration, off-resonant leakage, and accumulated loss across many injection steps.

\section{Experimental feasibility}
\label{sec:experimental_feasibility}

This section explains how the effective Hamiltonian used in the main text can be obtained in a realistic superconducting-qubit--magnon device, and why the parameter regime used in the numerical simulations is experimentally motivated. 
In a selected rotating frame and setting $\hbar=1$, the effective qubit--magnon Hamiltonian reads
\begin{equation}
\label{eq:sm_model_H}
\hat H
=
\sum_{j=0}^{N_Q-1}
\Delta_j \hat\sigma_j^+\hat\sigma_j^-
+
\sum_{j=0}^{N_Q-1}
\chi_j \hat n\hat\sigma_{z,j}
+
\sum_{j=0}^{N_Q-1}
g_j
(\hat a^\dagger\hat\sigma_j^-+
\hat a\hat\sigma_j^+) .
\end{equation}
Here $\hat a$ and $\hat a^\dagger$ are the annihilation and creation operators of the Kittel-mode magnon, respectively, and $\hat n=\hat a^\dagger\hat a$ is the magnon number operator. 
The operators $\hat\sigma_j^+$ and $\hat\sigma_j^-$ are the raising and lowering operators of qubit $j$, respectively, $\hat\sigma_{z,j}$ is the corresponding Pauli-$z$ operator, and $\Delta_j=\omega_j-\omega_m$ is the effective detuning between the effective frequency $\omega_j$ of qubit $j$ and the effective magnon frequency $\omega_m$.
The central point is that Eq.\,\eqref{eq:sm_model_H} should be interpreted as a driven effective exchange--dispersive Hamiltonian, not as a bare resonant qubit--magnon Hamiltonian. 
The dispersive term $\chi_j\hat n\hat\sigma_{z,j}$ is produced by virtual exchange through a static off-resonant qubit--magnon coupling $G_j^{(0)}$, whereas the exchange coefficient $g_j$ is activated by an auxiliary drive or parametric modulation. 
This separation of a static dispersive background from a controllable sideband exchange channel is standard in circuit quantum electrodynamics (QED) and in hybrid qubit--magnon platforms~\cite{Blais2004CircuitQED,Wallraff2004,Schuster2007PhotonNumber,Gambetta2006,Boissonneault2009,RevModPhys.93.025005,Gu2017MicrowavePhotonics,Xiang2013HybridQuantumCircuits,Kurizki2015HybridSystems,Tabuchi2015,LachanceQuirion2017,LachanceQuirion2020,PhysRevLett.125.117701,PhysRevLett.130.193603,LachanceQuirion2019,ZareRameshti2022,Yuan2022}.

\subsection{Cavity--magnon--qubit Hamiltonian}
\label{subsec:microscopic_hamiltonian}

A direct magnetic-dipole coupling between an individual superconducting qubit and a Kittel-mode magnon can be small. 
In experimentally demonstrated devices, the microwave cavity is therefore used as a virtual bus that couples strongly both to the magnon mode and to the superconducting qubit~\cite{Soykal2010,Huebl2013,PhysRevLett.113.083603,Zhang2014,Goryachev2014,Bhoi2014,Tabuchi2015,TABUCHI2016729,LachanceQuirion2019,Li2020,Harder2021,Wang2020,ZareRameshti2022}.
Motivated by this cavity-mediated architecture, we describe the device using a minimal model consisting of a single microwave cavity mode, a Kittel magnon mode, and $N_Q$ superconducting qubits. 
Since the scheme operates in the low- to intermediate-excitation regime, the intrinsic Kerr nonlinearity of the Kittel mode is neglected, and the magnon mode is treated as an effective harmonic oscillator~\cite{PhysRev.73.155,Cherepanov1993,Serga2010,Chumak2015,ZareRameshti2022,Yuan2022}.
In the laboratory frame, the corresponding Hamiltonian can be written as
\begin{align}
\label{eq:sm_lab_H}
\hat H_{\rm lab}
&=
\omega_c\hat c^\dagger\hat c
+
\omega_{b,m}\hat a^\dagger\hat a
+
\sum_{j=0}^{N_Q-1}
\frac{\omega_{q,j}}{2}\hat\sigma_{z,j}
\notag\\
&\quad
+
g_m
\left(
\hat c^\dagger\hat a+\hat c\hat a^\dagger
\right)
+
\sum_{j=0}^{N_Q-1}
g_{c j}
\left(
\hat c^\dagger\hat\sigma_j^-
+
\hat c\hat\sigma_j^+
\right)
+
\sum_{j<k}
J_{jk}^{({\rm dir})}
\left(
\hat\sigma_j^+\hat\sigma_k^-
+
\hat\sigma_j^-\hat\sigma_k^+
\right).
\end{align}
Here $\hat c$ ($\hat c^\dagger$) is the annihilation (creation) operator of the microwave-cavity photon, and $\omega_c$, $\omega_{b,m}$, and $\omega_{q,j}$ are the bare frequencies of the cavity, Kittel-mode magnon, and qubit $j$, respectively. 
The parameters $g_m$ and $g_{c j}$ denote the cavity--magnon coupling and the cavity--qubit-$j$ coupling, respectively. 
The coefficient $J_{jk}^{({\rm dir})}$ accounts for possible direct qubit--qubit exchange due to residual capacitance, packaging modes, or microwave crosstalk.

We define the cavity detunings
\begin{equation}
\label{eq:sm_cavity_detunings}
\delta_m=\omega_{b,m}-\omega_c,
\qquad
\delta_j^c=\omega_{q,j}-\omega_c .
\end{equation}
The superscript $c$ in $\delta_j^c$ labels a qubit--cavity detuning; it is not an exponent. We work in the cavity-dispersive regime
\begin{equation}
\label{eq:sm_cavity_dispersive_condition}
|\delta_m|\gg g_m,
\qquad
|\delta_j^c|\gg g_{c j},
\qquad
j=0,\ldots,N_Q-1 .
\end{equation}
In this regime the cavity is only virtually populated, and it can be eliminated perturbatively~\cite{SchriefferWolff1966,Bravyi2011SW,Blais2004CircuitQED,Gambetta2006,Boissonneault2009,RevModPhys.93.025005}.

\subsection{Static Hamiltonian after eliminating the cavity}
\label{subsec:static_after_cavity}

Applying a Schrieffer--Wolff transformation to Eq.\,\eqref{eq:sm_lab_H} and retaining terms up to second order in $g_m/\delta_m$ and $g_{c j}/\delta_j^c$ gives the following static Hamiltonian in the zero-cavity-photon manifold
\begin{align}
\label{eq:sm_static_after_cavity}
\hat H_{\rm st}
&=
\widetilde\omega_m\hat a^\dagger\hat a
+
\sum_{j=0}^{N_Q-1}
\frac{\widetilde\omega_{q,j}}{2}\hat\sigma_{z,j}
\notag\\
&\quad
+
\sum_{j=0}^{N_Q-1}
G_j^{(0)}
\left(
\hat a^\dagger\hat\sigma_j^-
+
\hat a\hat\sigma_j^+
\right)
\notag\\
&\quad
+
\sum_{j<k}
J_{jk}^{(c)}
\left(
\hat\sigma_j^+\hat\sigma_k^-
+
\hat\sigma_j^-\hat\sigma_k^+
\right)
+
\sum_{j<k}
J_{jk}^{({\rm dir})}
\left(
\hat\sigma_j^+\hat\sigma_k^-
+
\hat\sigma_j^-\hat\sigma_k^+
\right).
\end{align}
All symbols in Eq.\,\eqref{eq:sm_static_after_cavity} are defined here. 
The symbols $\widetilde\omega_m$ and $\widetilde\omega_{q,j}$ denote the magnon and qubit frequencies dressed by virtual cavity photons, respectively. To leading order, they are
\begin{equation}
\label{eq:sm_dressed_frequencies}
\widetilde\omega_m
\simeq
\omega_{b,m}+\frac{g_m^2}{\delta_m},
\qquad
\widetilde\omega_{q,j}
\simeq
\omega_{q,j}+\frac{g_{c j}^2}{\delta_j^c},
\end{equation}
up to convention-dependent Lamb-shift signs associated with the chosen detuning definition.
The static cavity-mediated qubit--magnon transverse coupling is~\cite{Soykal2010,Huebl2013,PhysRevLett.113.083603,Zhang2014,Goryachev2014,Tabuchi2015,LachanceQuirion2017,LachanceQuirion2020,PhysRevLett.125.117701}
\begin{equation}
\label{eq:sm_G0}
G_j^{(0)}
\simeq
G_j^{({\rm dir})}
+
\frac{g_m g_{c j}}{2}
\left(
\frac{1}{\delta_m}
+
\frac{1}{\delta_j^c}
\right).
\end{equation}
Here $G_j^{({\rm dir})}$ denotes any small direct magnetic-dipole qubit--magnon coupling; in most cavity-mediated designs it can be neglected compared with the second term. 
For this reason, the direct qubit--magnon coupling is not explicitly included in the Hamiltonian of Eq.\,\eqref{eq:sm_lab_H}.
The superscript $(0)$ means the static or zeroth Fourier component of the qubit--magnon coupling at the parking point. 
It does not mean that the coupling is absent. 
The cavity-mediated qubit--qubit exchange is
\begin{equation}
\label{eq:sm_Jc}
J_{jk}^{(c)}
\simeq
\frac{g_{c j}g_{c k}}{2}
\left(
\frac{1}{\delta_j^c}
+
\frac{1}{\delta_k^c}
\right).
\end{equation}
The superscript $(c)$ labels the cavity-mediated part of the qubit--qubit exchange. Equation~\eqref{eq:sm_static_after_cavity} is the complete static Hamiltonian, to this perturbative order, needed for the feasibility discussion: it contains the dressed free magnon and qubit frequencies, the static qubit--magnon channel $G_j^{(0)}$, the cavity-mediated static qubit--qubit channel $J_{jk}^{(c)}$, and any direct qubit--qubit exchange $J_{jk}^{({\rm dir})}$.

The same equation also shows how unwanted static qubit--qubit exchange is suppressed. Define the dressed qubit--qubit detuning
\begin{equation}
\label{eq:sm_qubit_qubit_detuning}
\Delta_{jk}^q
=
\widetilde\omega_{q,j}-\widetilde\omega_{q,k},
\qquad j\neq k ,
\end{equation}
where the superscript $q$ indicates that this is a qubit--qubit detuning. 
Define the total static qubit--qubit exchange immediately after cavity elimination as
\begin{equation}
\label{eq:sm_Jqq_static}
J_{jk}^{({\rm st})}
=
J_{jk}^{(c)}+J_{jk}^{({\rm dir})}.
\end{equation}
In the interaction picture of the dressed qubit frequencies, the term $\hat\sigma_j^+\hat\sigma_k^-$ rotates as $e^{\ii\Delta_{jk}^q t}$. Therefore, if the qubits are parked at distinct frequencies satisfying
\begin{equation}
\label{eq:sm_static_qubit_suppression}
|\Delta_{jk}^q|
\gg
|J_{jk}^{({\rm st})}|,
\qquad j\neq k ,
\end{equation}
the static exchange in Eq.\,\eqref{eq:sm_static_after_cavity} is far off resonant~\cite{Blais2004CircuitQED,Gambetta2006,Boissonneault2009,RevModPhys.93.025005,McKay2016}. 
It then produces only a perturbative frequency shift of order $|J_{jk}^{({\rm st})}|^2/\Delta_{jk}^q$ and a leakage probability of order $|J_{jk}^{({\rm st})}/\Delta_{jk}^q|^2$. These small shifts are calibration parameters and can be absorbed into the experimentally chosen qubit parking frequencies and sideband-drive frequencies. 
Thus the cavity can enhance the qubit--magnon hardware coupling while the induced static qubit--qubit exchange remains suppressed by qubit--qubit detuning.

\subsection{Dispersive magnon--qubit interaction from the static coupling}
\label{subsec:static_dispersive}

The next step is not to use $G_j^{(0)}$ as a resonant exchange coupling. 
Instead, each qubit is parked far from the magnon mode~\cite{Schuster2007PhotonNumber,Gambetta2006,Boissonneault2009,LachanceQuirion2017,LachanceQuirion2020,PhysRevLett.125.117701}. 
We define the dressed bare qubit--magnon detuning
\begin{equation}
\label{eq:sm_bare_qm_detuning}
\Delta_j^{\rm b}
=
\widetilde\omega_{q,j}-\widetilde\omega_m .
\end{equation}
The superscript ${\rm b}$ means ``bare'' with respect to the later sideband drive: $\Delta_j^{\rm b}$ is the static dressed qubit--magnon detuning before the parametric activation of the exchange channel. The dispersive condition is
\begin{equation}
\label{eq:sm_qm_dispersive_condition}
|\Delta_j^{\rm b}|
\gg
|G_j^{(0)}|,
\qquad
j=0,\ldots,N_Q-1 .
\end{equation}
Under this condition, the static qubit--magnon exchange term in Eq.\,\eqref{eq:sm_static_after_cavity} can be removed by a second Schrieffer--Wolff transformation with generator
\begin{equation}
\label{eq:sm_SW_qm}
\hat S_{qm}
=
\sum_{j=0}^{N_Q-1}
\lambda_j
\left(
\hat a^\dagger\hat\sigma_j^-
-
\hat a\hat\sigma_j^+
\right),
\qquad
\lambda_j=\frac{G_j^{(0)}}{\Delta_j^{\rm b}} .
\end{equation}

Here $\lambda_j$ is the small dispersive expansion parameter for qubit $j$. 
To second order in $\lambda_j$, and neglecting higher-order corrections generated by off-resonant qubit--qubit exchange, the static qubit--magnon exchange is transformed into a dispersive interaction. 
The resulting static Hamiltonian can be written as
\begin{align}
\label{eq:sm_static_dispersive_H}
\hat H_{\rm disp}
&=
\bar\omega_m\hat n
+
\sum_{j=0}^{N_Q-1}
\frac{\bar\omega_{q,j}}{2}\hat\sigma_{z,j}
+
\sum_{j=0}^{N_Q-1}
\chi_j\hat n\hat\sigma_{z,j}
\notag\\
&\quad
+
\sum_{j<k}
J_{jk}^{({\rm res})}
\left(
\hat\sigma_j^+\hat\sigma_k^-
+
\hat\sigma_j^-\hat\sigma_k^+
\right)
+
\hat H_{\rm small}.
\end{align}
Keeping terms up to second order in $G_j^{(0)}/\Delta_j^{\rm b}$, the coefficients in Eq.\,\eqref{eq:sm_static_dispersive_H} are
\begin{align}
\label{eq:sm_disp_coefficients}
\bar\omega_m
&\simeq
\widetilde\omega_m,
\notag\\
\bar\omega_{q,j}
&\simeq
\widetilde\omega_{q,j}
+
\chi_j,
\notag\\
\chi_j
&\simeq
\frac{\left(G_j^{(0)}\right)^2}{\Delta_j^{\rm b}},
\qquad
\Delta_j^{\rm b}
=
\widetilde\omega_{q,j}
-
\widetilde\omega_m .
\end{align}
Here $\bar\omega_m$ is the magnon frequency after cavity dressing, and $\bar\omega_{q,j}$ is the qubit-$j$ frequency after both cavity dressing and the qubit--magnon Lamb shift generated by $G_j^{(0)}$~\cite{Blais2004CircuitQED,Gambetta2006,Boissonneault2009,Schuster2007PhotonNumber,LachanceQuirion2017,LachanceQuirion2020}. 
The coefficients $\bar\omega_{q,j}$ and $\chi_j$ are written separately for each qubit $j$, because the dispersive transformation is performed perturbatively for each off-resonant qubit--magnon pair. 
At the same perturbative order, the coupling of different qubits through the magnon mode gives an additional residual qubit--qubit exchange,
\begin{equation}
\label{eq:sm_Jres_definition}
J_{jk}^{({\rm res})}
\simeq
J_{jk}^{({\rm st})}
+
J_{jk}^{(m)},
\end{equation}
where
\begin{equation}
\label{eq:sm_Jst_definition}
J_{jk}^{({\rm st})}
=
J_{jk}^{(c)}
+
J_{jk}^{({\rm dir})}
\end{equation}
is the static qubit--qubit exchange already present after eliminating the cavity, and
\begin{equation}
\label{eq:sm_Jm}
J_{jk}^{(m)}
\simeq
\frac{G_j^{(0)}G_k^{(0)}}{2}
\left(
\frac{1}{\Delta_j^{\rm b}}
+
\frac{1}{\Delta_k^{\rm b}}
\right)
\end{equation}
is the magnon-mediated contribution generated by the dispersive transformation. 
Thus, to this order,
\begin{equation}
\label{eq:sm_Jres_explicit}
J_{jk}^{({\rm res})}
\simeq
J_{jk}^{(c)}
+
J_{jk}^{({\rm dir})}
+
\frac{G_j^{(0)}G_k^{(0)}}{2}
\left(
\frac{1}{\Delta_j^{\rm b}}
+
\frac{1}{\Delta_k^{\rm b}}
\right).
\end{equation}
The term $\hat H_{\rm small}$ collects corrections beyond this leading second-order description, including weak Kerr and cross-Kerr terms, higher-order dispersive shifts, and small frequency shifts induced by the far-off-resonant residual qubit--qubit exchange.

The residual exchange in Eq.\,\eqref{eq:sm_Jres_explicit} is suppressed by parking the qubits at sufficiently different dressed frequencies,
\begin{equation}
\label{eq:sm_residual_qubit_suppression}
|\Delta_{jk}^q|
\gg
|J_{jk}^{({\rm res})}|,
\qquad
\Delta_{jk}^q=\widetilde\omega_{q,j}-\widetilde\omega_{q,k}.
\end{equation}
Under this condition, the term proportional to $J_{jk}^{({\rm res})}$ produces only small perturbative frequency shifts and negligible real excitation exchange between different qubits. The important result of this static dispersive treatment is therefore that the same hardware coupling $G_j^{(0)}$ generates the magnon-number-dependent shift $\chi_j$, while real qubit--magnon exchange remains suppressed by the large static detuning $|\Delta_j^{\rm b}|$.
Consequently, after neglecting the off-resonant residual exchange terms, the relevant static Hamiltonian reduces to
\begin{equation}
\label{eq:sm_static_dispersive_reduced}
\hat H_{\rm disp}
\simeq
\bar\omega_m\hat n
+
\sum_{j=0}^{N_Q-1}
\frac{\bar\omega_{q,j}}{2}\hat\sigma_{z,j}
+
\sum_{j=0}^{N_Q-1}
\chi_j\hat n\hat\sigma_{z,j}.
\end{equation}

\subsection{Parametrically activated exchange coupling}
\label{subsec:parametric_exchange}

The exchange coefficient $g_j$ in Eq.\,\eqref{eq:sm_model_H} is introduced after the static dispersive structure has been established. 
One implementation is to modulate the same cavity-mediated qubit--magnon channel~\cite{ZakkaBajjani2011,Beaudoin2012,Strand2013,PhysRevLett.104.177004,McKay2016,Gu2017MicrowavePhotonics,RevModPhys.93.025005},
\begin{equation}
\label{eq:sm_Gt}
G_j(t)
=
G_j^{(0)}
+
G_j^{(1)}
\cos(\omega_{d,j}t+\phi_j).
\end{equation}
Here $G_j^{(1)}$ is the first Fourier amplitude of the modulation, $\omega_{d,j}$ is the modulation or sideband-drive frequency applied to qubit $j$, and $\phi_j$ is its phase. 
The static part $G_j^{(0)}$ remains off resonant and continues to produce $\chi_j$. 

Starting from the reduced static dispersive Hamiltonian in Eq.\,\eqref{eq:sm_static_dispersive_reduced}, the sideband modulation adds a time-dependent transverse coupling. The Hamiltonian before making the rotating-wave approximation is
\begin{equation}
\label{eq:sm_full_modulated_H}
\hat H(t)
=
\hat H_{\rm disp}
+
\sum_{j=0}^{N_Q-1}
G_j^{(1)}
\cos(\omega_{d,j}t+\phi_j)
\left(
\hat a^\dagger\hat\sigma_j^-
+
\hat a\hat\sigma_j^+
\right).
\end{equation}
In the interaction picture with respect to the free part of $\hat H_{\rm disp}$, the component proportional to $\hat a^\dagger\hat\sigma_j^-$ becomes
\begin{align}
\label{eq:sm_sideband_RWA}
\hat V_{j,1}^{(I)}(t)
&=
G_j^{(1)}
\cos(\omega_{d,j}t+\phi_j)
\hat a^\dagger\hat\sigma_j^-
e^{-\ii\Delta_j^{\rm b}t}
+
{\rm H.c.}
\notag\\
&=
\frac{G_j^{(1)}}{2}
e^{-\ii(\Delta_j^{\rm b}-\omega_{d,j})t+\ii\phi_j}
\hat a^\dagger\hat\sigma_j^-
\notag\\
&\quad+
\frac{G_j^{(1)}}{2}
e^{-\ii(\Delta_j^{\rm b}+\omega_{d,j})t-\ii\phi_j}
\hat a^\dagger\hat\sigma_j^-
+
{\rm H.c.},
\end{align}
where ${\rm H.c.}$ denotes the Hermitian conjugate. 
When $\omega_{d,j}$ is chosen close to the target qubit--magnon transition frequency ($\Delta_j^{\rm b}\simeq\omega_{d,j}$), the first term in the final expression of Eq.\,\eqref{eq:sm_sideband_RWA} is slowly varying, whereas the second term oscillates rapidly and is neglected under the rotating-wave approximation.
Keeping only the slowly varying term in Eq.\,\eqref{eq:sm_sideband_RWA} gives the sideband Hamiltonian
\begin{equation}
\label{eq:sm_g_from_G1}
\hat H_{g,j}
=
g_j
\left(
e^{\ii\phi_j}\hat a^\dagger\hat\sigma_j^-
+
e^{-\ii\phi_j}\hat a\hat\sigma_j^+
\right),
\qquad
g_j\simeq \frac{G_j^{(1)}}{2}.
\end{equation}
Thus the effective exchange $g_j$ is the sideband coupling obtained from the rotating-wave approximation applied to $\hat V_{j,1}^{(I)}(t)$.
Since $g_j$ is controlled by the modulation amplitude $G_j^{(1)}$, it can be made much smaller than the static hardware coupling $G_j^{(0)}$ without reducing the dispersive shift $\chi_j$~\cite{Beaudoin2012,Strand2013,McKay2016,Zhou2021}. Combining this driven exchange term with the reduced static dispersive Hamiltonian gives
\begin{equation}
\label{eq:sm_effective_sideband_H}
\hat H_{\rm eff}
=
\bar\omega_m\hat n
+
\sum_{j=0}^{N_Q-1}
\frac{\bar\omega_{q,j}}{2}\hat\sigma_{z,j}
+
\sum_{j=0}^{N_Q-1}
\chi_j\hat n\hat\sigma_{z,j}
+
\sum_{j=0}^{N_Q-1}
g_j
\left(
e^{\ii\phi_j}\hat a^\dagger\hat\sigma_j^-
+
e^{-\ii\phi_j}\hat a\hat\sigma_j^+
\right).
\end{equation}
After moving to the same rotating frame used in the main text and absorbing the drive phases into the definitions of the qubit operators, Eq.\,\eqref{eq:sm_effective_sideband_H} reduces to Eq.\,\eqref{eq:sm_model_H}.

An alternative way to obtain the same effective sideband coupling $g_j$ is to modulate the qubit frequency rather than the transverse qubit--magnon coupling directly~\cite{Beaudoin2012,Strand2013,ZakkaBajjani2011,McKay2016,Magnard2018PRL,Zhou2021}. 
To see this explicitly, consider the static transverse coupling $G_j^{(0)}$ together with a time-dependent qubit frequency, described by this time-dependent Hamiltonian
\begin{equation}
\label{eq:sm_frequency_modulated_H}
\hat H_j(t)
=
\omega_m \hat a^\dagger\hat a
+
\frac{\omega_{q,j}(t)}{2}\hat\sigma_{z,j}
+
G_j^{(0)}
\left(
\hat a^\dagger\hat\sigma_j^-
+
\hat a\hat\sigma_j^+
\right),
\end{equation}
where
\begin{equation}
\label{eq:sm_frequency_modulation}
\omega_{q,j}(t)
=
\omega_{q,j}
+
\varepsilon_j\cos(\omega_{d,j}t+\phi_j).
\end{equation}
In this reduced two-mode description, $\omega_m$ and $\omega_{q,j}$ denote the dressed effective magnon and qubit frequencies relevant to the sideband derivation.
Here $\varepsilon_j$ is the frequency-modulation amplitude. 
Moving to the interaction picture with respect to the time-dependent free Hamiltonian
\begin{equation}
\label{eq:sm_frequency_modulated_free_H}
\hat H_{0,j}(t)
=
\omega_m \hat a^\dagger\hat a
+
\frac{\omega_{q,j}(t)}{2}\hat\sigma_{z,j},
\end{equation}
the operator $\hat a^\dagger\hat\sigma_j^-$ acquires the phase
\begin{equation}
\label{eq:sm_modulated_phase}
\hat a^\dagger\hat\sigma_j^-
\rightarrow
\hat a^\dagger\hat\sigma_j^-
\exp\!\left[
-\ii\Delta_j^{\rm b}t
-\ii\beta_j\sin(\omega_{d,j}t+\phi_j)
\right],
\end{equation}
where $\Delta_j^{\rm b}=\omega_{q,j}-\omega_m$ is the static qubit--magnon detuning in this reduced description, and
\begin{equation}
\label{eq:sm_beta_definition}
\beta_j=\frac{\varepsilon_j}{\omega_{d,j}}
\end{equation}
is the dimensionless modulation index. 
Therefore the interaction-picture coupling becomes
\begin{equation}
\label{eq:sm_frequency_modulated_interaction}
\hat V_j^{(I)}(t)
=
G_j^{(0)}
\hat a^\dagger\hat\sigma_j^-
e^{-\ii\Delta_j^{\rm b}t}
e^{-\ii\beta_j\sin(\omega_{d,j}t+\phi_j)}
+
\Hc.
\end{equation}
Using the Jacobi--Anger expansion,
\begin{equation}
\label{eq:sm_jacobi_anger}
e^{-\ii\beta_j\sin(\omega_{d,j}t+\phi_j)}
=
\sum_{\ell=-\infty}^{\infty}
J_\ell(\beta_j)
e^{-\ii\ell(\omega_{d,j}t+\phi_j)},
\end{equation}
where $\ell$ labels the modulation sideband and $J_\ell(\beta_j)$ is the Bessel function of the first kind of integer order $\ell$, we obtain
\begin{equation}
\label{eq:sm_frequency_modulated_sidebands}
\hat V_j^{(I)}(t)
=
G_j^{(0)}
\sum_{\ell=-\infty}^{\infty}
J_\ell(\beta_j)
e^{-\ii(\Delta_j^{\rm b}+\ell\omega_{d,j})t}
e^{-\ii\ell\phi_j}
\hat a^\dagger\hat\sigma_j^-
+
\Hc .
\end{equation}
A near-resonant sideband is selected by choosing an integer $\ell$ such that $\Delta_j^{\rm b}+\ell\omega_{d,j}\simeq0$. 
For the first sideband, one may take $\ell=-1$, giving the slowly varying term
\begin{equation}
\label{eq:sm_frequency_modulated_first_sideband}
\hat H_{g,j}
\simeq
G_j^{(0)}J_{-1}(\beta_j)
e^{\ii\phi_j}
\hat a^\dagger\hat\sigma_j^-
+
\Hc .
\end{equation}
Since $J_{-1}(\beta_j)=-J_1(\beta_j)$, the minus sign can be absorbed into the drive phase. 
Thus the effective sideband coupling generated by frequency modulation is
\begin{equation}
\label{eq:sm_g_bessel}
g_j
\simeq
G_j^{(0)}J_1(\beta_j).
\end{equation}
For weak modulation, $\beta_j\ll1$, this reduces to
\begin{equation}
\label{eq:sm_g_weak_modulation}
g_j
\simeq
G_j^{(0)}\frac{\varepsilon_j}{2\omega_{d,j}}.
\end{equation}
This derivation shows explicitly that qubit-frequency modulation is a second experimental route to the same effective exchange coefficient $g_j$. 
In the transverse-coupling modulation scheme, $g_j$ comes from the Fourier component $G_j^{(1)}$ of $G_j(t)$; in the frequency-modulation scheme, $g_j$ comes from the first sideband of the static coupling $G_j^{(0)}$. 
In both cases, $G_j^{(0)}$ remains the static coupling responsible for the dispersive shift $\chi_j$, while the driven first sideband provides the controllable exchange term $g_j$.

\subsection{Connection to the rotating-frame detuning and number selectivity}
\label{subsec:number_selectivity}

The drive frequency $\omega_{d,j}$ determines the residual detuning $\Delta_j$ in Eq.\,\eqref{eq:sm_model_H}. A useful calibration relation is
\begin{equation}
\label{eq:sm_rotating_detuning}
\Delta_j
=
\Delta_j^{\rm b}
-
\omega_{d,j}
+
\delta_j^{\rm cal} .
\end{equation}
The term $\delta_j^{\rm cal}$ denotes all small calibrated shifts not written explicitly, including ac-Stark shifts, Lamb shifts beyond Eq.\,\eqref{eq:sm_dressed_frequencies}, and residual dispersive corrections. Thus tuning $\Delta_j$ in the effective model means tuning the sideband-drive frequency $\omega_{d,j}$, not bringing the static qubit and magnon into resonance.

The dispersive term shifts the sideband resonance according to the magnon number and the states of the spectator qubits~\cite{Schuster2007PhotonNumber,Gambetta2006,Boissonneault2009,LachanceQuirion2017,LachanceQuirion2020,PhysRevLett.125.117701}. 
Let $v_j$ denote a fixed configuration of all qubits except $j$. 
For each spectator qubit $k\neq j$, define
\begin{equation}
\label{eq:sm_zk_definition}
z_k(v_j)
=
\begin{cases}
+1, & \text{if qubit } k \text{ is in } \ket{e_k},\\
-1, & \text{if qubit } k \text{ is in } \ket{g_k}.
\end{cases}
\end{equation}
For the effective sideband Hamiltonian in Eq.\,\eqref{eq:sm_effective_sideband_H}, the number selectivity is determined by the static dispersive part,
\begin{equation}
\hat H_0
=
\bar\omega_m\hat n
+
\sum_{j=0}^{N_Q-1}
\frac{\bar\omega_{q,j}}{2}\hat\sigma_{z,j}
+
\sum_{j=0}^{N_Q-1}
\chi_j\hat n\hat\sigma_{z,j}.
\end{equation}
The drive frequency must match the energy difference between the two states connected by the exchange operator $\hat a^\dagger\hat\sigma_j^-$. Therefore, for the transition
\begin{equation}
\label{eq:sm_transition_definition}
\ket{n,e_j;v_j}
\leftrightarrow
\ket{n+1,g_j;v_j},
\end{equation}
the physical drive frequency is, to leading order,
\begin{equation}
\label{eq:sm_transition_frequency}
\Omega^{\rm sb}_j(n;v_j)
=
\Delta_j^{\rm b}
+
(2n+1)\chi_j
-
\sum_{k\neq j}
z_k(v_j)\chi_k
+
\delta_j^{\rm cal}.
\end{equation}
The symbol $\Omega^{\rm sb}_j(n;v_j)$ denotes a laboratory-frame sideband frequency. 
The integer $n$ is the magnon number before the transition, and $v_j$ specifies the spectator-qubit state. Driving at
\begin{equation}
\label{eq:sm_drive_choice}
\omega_{d,j}
=
\Omega^{\rm sb}_j(j;\bar v_j)
\end{equation}
makes the intended transition $\ket{j,e_j;\bar v_j}\leftrightarrow\ket{j+1,g_j;\bar v_j}$ resonant in the rotating frame. Equivalently, the rotating-frame compensation condition is
\begin{equation}
\label{eq:sm_compensation_general}
\Delta_j
=
-(2j+1)\chi_j
+
\sum_{k\neq j}
\bar z_k\chi_k,
\qquad
\bar z_k=z_k(\bar v_j).
\end{equation}
If the incoherent pump keeps the spectator qubits predominantly excited, one may take $\bar z_k=+1$. For homogeneous dispersive shifts $\chi_j=\chi$, Eq.\,\eqref{eq:sm_compensation_general} becomes
\begin{equation}
\label{eq:sm_compensation_homogeneous}
\Delta_j=(N_Q-2j-2)\chi .
\end{equation}

The transition is spectrally resolved when the dispersive separation is large compared with both the activated exchange coupling and the relevant dissipative rates. 
For the few-magnon transitions used in the scheme, a conservative condition is
\begin{equation}
\label{eq:sm_resolved_condition}
2|\chi_j|
\gg
g_j\sqrt{n+1},
\qquad
2|\chi_j|
\gg
\Gamma_{j,n} .
\end{equation}
Here $\Gamma_{j,n}$ is the coherence decay rate of the driven transition $\ket{n,e_j}\leftrightarrow\ket{n+1,g_j}$, including qubit relaxation, incoherent-pump broadening, magnon loss, and dephasing. 
This condition explains why the simulations use $\chi_j/g_j\gg1$: $g_j$ is deliberately chosen to be weak enough that only the desired number-selective sideband is activated.

\subsection{Dynamic crosstalk in a multi-qubit device}
\label{subsec:dynamic_crosstalk}

The suppression of the static qubit--qubit exchange in Eqs.\,\eqref{eq:sm_static_qubit_suppression} and \eqref{eq:sm_residual_qubit_suppression} does not by itself guarantee the absence of drive crosstalk. 
The sideband drive intended for qubit $j$ must also be off resonant from unwanted sidebands of qubit $k$~\cite{Beaudoin2012,Strand2013,McKay2016,RevModPhys.93.025005}. 
Using the definition of $\Omega^{\rm sb}_k(n;v_k)$ in Eq.\,\eqref{eq:sm_transition_frequency}, the selectivity condition is
\begin{equation}
\label{eq:sm_dynamic_crosstalk}
|\omega_{d,j}-\Omega^{\rm sb}_k(n;v_k)|
\gg
\max\!\left\{
g_j\sqrt{n+1},\; g_k\sqrt{n+1},\; \Gamma_{j,n},\; \Gamma_{k,n}\right\},
\qquad k\neq j .
\end{equation}
This condition can be satisfied by using distinct qubit parking frequencies, distinct sideband-drive frequencies, and calibrated local microwave or flux control. The implementation therefore does not require the qubits to be mutually resonant. The cavity supplies the virtual hardware coupling needed for $G_j^{(0)}$, while both static qubit--qubit exchange and dynamic sideband crosstalk are suppressed by detuning and frequency selectivity.

\subsection{Parameter estimates}
\label{subsec:parameter_estimates}

The simulations in the main text use homogeneous effective parameters $g_j=g$, $\chi_j=\chi$, $\gamma_j=\gamma$, and $\wp_j=\wp$ unless stated otherwise. Taking
\begin{equation}
\label{eq:sm_g_numerical}
g/2\pi=0.3\,{\rm MHz},
\end{equation}
the dimensionless parameters correspond to
\begin{align}
\label{eq:sm_parameter_conversion}
\chi/g=10\text{--}30
&\quad \Rightarrow \quad
\chi/2\pi=3\text{--}9\,{\rm MHz},
\notag\\
\kappa/g=0.01\text{--}0.1
&\quad \Rightarrow \quad
\kappa/2\pi=3\text{--}30\,{\rm kHz},
\notag\\
\gamma/g=0.05
&\quad \Rightarrow \quad
\gamma/2\pi=15\,{\rm kHz},
\notag\\
\wp/g=2
&\quad \Rightarrow \quad
\wp/2\pi=0.6\,{\rm MHz}.
\end{align}
Here $\kappa$ is the magnon energy-decay rate, $\gamma$ is the qubit relaxation rate, and $\wp$ is the engineered incoherent pumping rate from $\ket{g_j}$ to $\ket{e_j}$. The coupling $g$ in Eq.\,\eqref{eq:sm_g_numerical} is the activated sideband exchange, not the static hardware coupling $G_j^{(0)}$.

These values should be compared with two distinct experimental scales. 
The first is the static hardware qubit--magnon coupling $G_j^{(0)}$, which generates the dispersive shift $\chi_j$. 
The second is the driven sideband coupling $g_j$, which is a tunable effective exchange activated by microwave or flux modulation and should not be identified with the static coupling $G_j^{(0)}$.
For the static qubit--magnon coupling, coherent coupling between a Kittel mode and a superconducting qubit has been demonstrated with a coupling strength that corresponds, in our notation, to $G_j^{(0)}/2\pi=11.4\,{\rm MHz}$~\cite{Tabuchi2015}. 
Magnon-number-resolved spectroscopy further demonstrated the strong-dispersive regime with $G_j^{(0)}/2\pi=7.79\,{\rm MHz}$ and $\chi/2\pi=1.5\pm0.1\,{\rm MHz}$~\cite{LachanceQuirion2017}. 
Single-shot single-magnon detection reported $G_j^{(0)}/2\pi=7.13\,{\rm MHz}$ and a single-magnon qubit-frequency pull $2\chi/2\pi=-3.82\,{\rm MHz}$, corresponding to $|\chi|/2\pi=1.91\,{\rm MHz}$~\cite{LachanceQuirion2020}. 
These experiments directly support MHz-scale static qubit--magnon coupling and MHz-scale magnon-number-dependent dispersive shifts~\cite{Tabuchi2015,LachanceQuirion2017,LachanceQuirion2020,PhysRevLett.125.117701,PhysRevLett.130.193603,LachanceQuirion2019,ZareRameshti2022,Yuan2022}.
The lower part of the range used here is close to the demonstrated strong-dispersive regime, whereas the upper part should be regarded as an optimized extension enabled by increasing the effective static coupling or reducing the qubit--magnon detuning while maintaining dispersive selectivity.
The activated coupling $g_j$ in our model is a weaker, drive-controlled sideband coupling derived from $G_j^{(0)}$. 
This type of parametrically activated exchange is standard in circuit QED, where modulation of a qubit frequency or a coupling element generates sideband transitions between a qubit and a harmonic mode~\cite{Beaudoin2012,Strand2013,ZakkaBajjani2011,McKay2016}. 
In the present magnonic setting, the relevant estimate follows from the sideband relation
\begin{equation}
g_j\simeq G_j^{(0)}J_1(\beta_j)
\simeq
G_j^{(0)}\frac{\beta_j}{2},
\qquad
\beta_j\ll1 .
\end{equation}
Using the experimentally demonstrated static scale $G_j^{(0)}/2\pi\simeq 7$--$8\,{\rm MHz}$, the representative value $g/2\pi=0.3\,{\rm MHz}$ only requires a modest first-sideband weight $J_1(\beta_j)\simeq0.04$, corresponding to $\beta_j\simeq0.08$ in the weak-modulation limit. 
Thus the value of $g$ used in the simulations should be interpreted as a deliberately weak and spectrally selective driven coupling, not as a bare magnon--qubit coupling. 
However, to our knowledge, a driven sideband exchange rate between a superconducting qubit and a Kittel mode has not yet been directly established experimentally in the specific qubit--magnon setting considered here.
We therefore treat $g_j$ as an experimentally motivated estimate obtained by applying a modest sideband modulation to the available static coupling scale $G_j^{(0)}$, rather than as a directly measured magnon--qubit sideband benchmark.

A separate device-level challenge is the low effective magnon loss rate assumed in the simulations, $\kappa/2\pi=3$--$30\,{\rm kHz}$. 
Existing superconducting-qubit--magnon and integrated superconducting-magnonic devices often exhibit MHz-scale Kittel-mode loss rates~\cite{Tabuchi2015,LachanceQuirion2017,LachanceQuirion2020,Morris2017,Baity2021}.
However, high-quality yttrium iron garnet (YIG) is a benchmark low-damping ferrimagnetic insulator, as established by early ferromagnetic-resonance studies and subsequent magnonics experiments~\cite{PhysRev.73.155,LeCraw1958,Spencer1959,Spencer1961,Cherepanov1993,Serga2010,Chumak2015,Klingler2017APL,MaierFlaig2017}.
Measurements of magnetostatic modes in YIG spheres have reported a Gilbert damping parameter $\alpha_G=2.7(5)\times10^{-5}$ and identified mode-dependent inhomogeneous broadening and two-magnon scattering as important additional contributions~\cite{Klingler2017APL,MaierFlaig2017}.
This value corresponds to a Gilbert-limited contribution of order $\kappa/2\pi\simeq 2\alpha_G(\omega_m/2\pi)$; for example, for $\omega_m/2\pi=5\,\mathrm{GHz}$, one obtains $\kappa/2\pi\simeq0.27\,\mathrm{MHz}$~\cite{Gilbert2004,Klingler2017APL}.
In addition, a loss measurement based on the parallel-pumping method implied a much narrower YIG loss rate $\kappa/2\pi$ of about $45\,\mathrm{kHz}$ at $4.2\,\mathrm{K}$~\cite{PhysRevLett.113.083603,Spencer1961}.
This value is close to the upper end of the low-loss range considered here and indicates that the assumed $\kappa/2\pi=3$--$30\,\mathrm{kHz}$ regime should be regarded as an optimized low-extrinsic-loss target rather than as a routinely achieved loss rate in present integrated superconducting-qubit--magnon devices.
The difference between this intrinsic Gilbert contribution and the effective loss rate measured in a device can be strongly affected by inhomogeneous broadening, mode mixing, magnetic-field inhomogeneity, surface and volume defects, impurity relaxation, two-magnon scattering, and interface- or integration-induced loss~\cite{LeCraw1958,Spencer1959,Sparks1961,AriasMills1999,McMichael2004,Klingler2017APL,MaierFlaig2017,Jermain2017,Tserkovnyak2002,Kapelrud2013}.
Thus, reaching the low-$\kappa$ regime assumed in our simulations requires suppressing extrinsic broadening and integration-induced loss through high-quality YIG material, careful mode selection, homogeneous bias fields, and weakly perturbative coupling to the superconducting circuit.

The qubit relaxation requirement is more moderate.
The value $\gamma/2\pi=15\,\mathrm{kHz}$ corresponds to an energy-relaxation time $T_1=1/\gamma\simeq10.6\,\mu\mathrm{s}$, where $T_1$ denotes the lifetime of the excited state of the superconducting qubit.
This value is well within the range of modern superconducting qubits~\cite{Paik2011,Rigetti2012,Place2021NatCommun,Wang2022TaTransmon,Somoroff2023,Kjaergaard2020,RevModPhys.93.025005}.
For example, tantalum-based transmon qubits have demonstrated lifetimes and coherence times exceeding $0.3\,\mathrm{ms}$~\cite{Place2021NatCommun}, and transmon lifetimes approaching $0.5\,\mathrm{ms}$ have also been reported~\cite{Wang2022TaTransmon}.
In a qubit--magnon device, however, the Kittel mode usually requires a static bias magnetic field~\cite{Tabuchi2015,LachanceQuirion2017,LachanceQuirion2020,TABUCHI2016729,LachanceQuirion2019}, which can introduce additional loss or dephasing in superconducting circuits through trapped vortices, quasiparticles, and magnetic-field-induced degradation of superconducting microwave resonators~\cite{Song2009APL,Song2009PRB,Bothner2012,Nsanzineza2014,Samkharadze2016,Kroll2019}.
Thus, the main device-level challenge is to maintain a small qubit relaxation rate $\gamma$ while applying the magnetic field needed to tune the magnon mode.

The incoherent pump rate $\wp/2\pi\simeq0.6\,\mathrm{MHz}$ corresponds to an incoherent-pumping time scale $1/\wp\simeq0.27\,\mu{\rm s}$.
This rate is compatible with engineered qubit-transition rates in superconducting circuits, where driven reset, sideband-assisted cooling, and reservoir-engineered dissipation have demonstrated controllable transition rates on sub-microsecond or faster time scales~\cite{Valenzuela2006,Reed2010APL,Riste2012,PhysRevLett.110.120501,PhysRevLett.109.183602,Magnard2018PRL,Egger2018PRApplied,Sunada2022PRApplied,Zhou2021}.
For example, all-microwave reset of a transmon qubit has achieved sub-percent residual excitation in $280\,\mathrm{ns}$~\cite{Magnard2018PRL}, fast reset with an intrinsic Purcell filter has demonstrated a $100\,\mathrm{ns}$ reset time~\cite{Sunada2022PRApplied}, and rapid unconditional parametric reset of a tunable superconducting qubit has reported residual excitation below $0.1\%$ within $34\,\mathrm{ns}$~\cite{Zhou2021}.
These time scales correspond to characteristic rates comparable to, or larger than, the required $\wp/2\pi$, supporting the feasibility of engineering the incoherent transition rate assumed here.

\subsection{Implementation logic}
\label{subsec:implementation_logic}

The implementation logic can be summarized as follows. 
First, the cavity is eliminated in the dispersive regime, yielding the complete static Hamiltonian in Eq.\,\eqref{eq:sm_static_after_cavity}. 
Second, static qubit--qubit exchange is suppressed by parking the qubits at distinct dressed frequencies satisfying Eq.\,\eqref{eq:sm_static_qubit_suppression}. 
Third, each qubit is kept dispersively detuned from the magnon mode, so that the static hardware coupling $G_j^{(0)}$ produces the magnon-number-dependent shift $\chi_j\hat n\hat\sigma_{z,j}$ rather than resonant exchange. 
Fourth, a weak sideband modulation is used to generate the effective exchange $g_j$, which should be viewed as a controlled driven component derived from the static coupling scale rather than as a directly measured bare magnon--qubit coupling. 
Fifth, the sideband frequencies are chosen to satisfy the number-selective resonance condition in Eq.\,\eqref{eq:sm_compensation_general} while avoiding the crosstalk condition in Eq.\,\eqref{eq:sm_dynamic_crosstalk}. 
Under this interpretation, the static coupling and dispersive-shift scales are supported by existing superconducting-qubit--magnon experiments, while the driven coupling $g_j$ is treated as an experimentally motivated sideband estimate based on modest modulation of the available static coupling. 
The main device-level challenges are therefore the realization of a sufficiently low effective Kittel-mode loss rate in an integrated superconducting-qubit--magnon platform and the experimental implementation of the weak, spectrally selective magnon--qubit sideband modulation.

\end{document}